\documentclass[12pt,english]{article}
\usepackage{lmodern}

\usepackage[T1]{fontenc}
\usepackage[latin9]{inputenc}
\usepackage{color}
\usepackage{babel}
\usepackage{array}
\usepackage{booktabs}
\usepackage{amsmath}
\usepackage{amsthm}
\usepackage{amssymb}
\usepackage{graphicx}
\usepackage{geometry}
\usepackage{setspace}
\usepackage[authoryear]{natbib}
\usepackage[pdfusetitle,
 bookmarks=true,bookmarksnumbered=false,bookmarksopen=false,
 breaklinks=false,pdfborder={0 0 1},backref=false,colorlinks=true]
 {hyperref}
\hypersetup{
 linkcolor=red, citecolor=blue, unicode, psdextra, pdfencoding=auto}

\makeatletter

\providecommand{\tabularnewline}{\\}

\numberwithin{equation}{section}
\numberwithin{figure}{section}
\numberwithin{table}{section}
\theoremstyle{plain}
\newtheorem{assumption}{\protect\assumptionname}
\theoremstyle{plain}
\newtheorem{prop}{\protect\propositionname}[section]
\theoremstyle{definition}
\newtheorem{example}{\protect\examplename}[section]
\theoremstyle{plain}
\newtheorem{lem}{\protect\lemmaname}[section]
\theoremstyle{plain}
\newtheorem{thm}{\protect\theoremname}[section]
\theoremstyle{remark}
\newtheorem{rem}{\protect\remarkname}[section]
\theoremstyle{plain}
\newtheorem{cor}{\protect\corollaryname}[section]
\theoremstyle{plain}
\newtheorem{lyxalgorithm}{\protect\algorithmname}

\usepackage{mathtools}
\usepackage{amsmath}
\usepackage{amsthm}
\usepackage{amssymb}
\usepackage{babel}
\usepackage{xcolor,soul}
\usepackage{appendix}
\usepackage{booktabs}
\usepackage{tabularx}
\usepackage{threeparttable}
\usepackage{dcolumn}
\usepackage{fancyhdr}
\usepackage[T1]{fontenc}
\usepackage{color}
\usepackage{titlesec}

\makeatletter

\providecommand{\assumptionname}{Assumption}
\providecommand{\remarkname}{Remark}

\allowdisplaybreaks

\fancypagestyle{appendix}{%
  \fancyhf{}
  \fancyfoot[C]{Online Appendix \thepage}%
  
}

\makeatother

\providecommand{\algorithmname}{Algorithm}
\providecommand{\assumptionname}{Assumption}
\providecommand{\corollaryname}{Corollary}
\providecommand{\examplename}{Example}
\providecommand{\lemmaname}{Lemma}
\providecommand{\propositionname}{Proposition}
\providecommand{\remarkname}{Remark}
\providecommand{\theoremname}{Theorem}

\begin{document}
\title{Two-Step Estimation of a Strategic Network Formation Model with Clustering\thanks{\scriptsize We are grateful to Pablo Fajgelbaum, Bryan Graham, Bo Honore, Hiro Kasahara, Michael Leung, Rosa Matzkin, Konrad Menzel, Whitney Newey, Aureo de Paula, Andres Santos, Kevin Song, and seminar and conference participants at UC Berkeley, UC Davis, UC Riverside, Princeton, U Colorado, Northwestern, Maryland, Georgetown, Duke, UBC, Ohio State, TAMU, Florida State, Penn State, U Chicago, Simon Fraser, Tinbergen Institute, Groningen U, FGV Rio, INSPER, PUC Rio, Vanderbilt, the World Congress of the Econometric Society, the Winter Meeting and Summer Meeting of the Econometric Society, INET Conferences on Econometrics of Networks at Cambridge and at USC, the Berkeley/CeMMAP Conference on Networks, the NYU CRATE Conference, and the Cowles Foundation Conference. We are particularly grateful to Terence Tao for his insightful suggestions. All errors are our own. This paper supersedes the earlier working paper \textit{Estimation of Large Network Formation Games}.}}
\author{Geert Ridder\thanks{\scriptsize Department of Economics, University of Southern California, Los Angeles, CA 90089. E-mail: ridder@usc.edu.}\and
Shuyang Sheng\thanks{\scriptsize Shenzhen Finance Institute, School of Management and Economics, The Chinese University of Hong Kong, Shenzhen, China. Email: shengshuyang@cuhk.edu.cn.}}
\maketitle
\begin{abstract}
This paper explores strategic network formation under incomplete information
using data from a single large network. We allow the utility function
to be nonseparable in an individual's link choices to capture the
spillover effects from friends in common. In a network with $n$ individuals,
an individual with a nonseparable utility function chooses between
$2^{n-1}$ overlapping portfolios of links. We develop a novel approach
that applies the Legendre transform to the utility function so that
the optimal link choices can be represented as a sequence of correlated
binary choices. The link dependence that results from the preference
for friends in common is captured by an auxiliary variable introduced
by the Legendre transform. We propose a two-step estimator that is
consistent and asymptotically normal. We also derive a limiting approximation
of the game as $n$ grows large that simplifies the computation in
large networks. We apply these methods to favor exchange networks
in rural India and find that the direction of support from a mutual
link matters in facilitating favor provision.


\textsc{Keywoods}: network formation, strategic interactions, clustering,
two-step estimation, limiting game, favor exchange
\end{abstract}

\section{\label{sec:intro}Introduction}

Network formation has attracted considerable interest from economists
due to its many applications to phenomena such as job referrals \citep{Beaman2012},
favor exchange \citep{Jackson_2012}, interbank lending \citep{Elliott2014},
and production networks \citep{Acemoglu2020}. A challenge in modeling
the formation of social and economic networks is that the formation
of a link may be influenced by the presence of other links (\citealp{Jackson_2008};
\citealp{Jackson_2017}). For example, individuals get to know each
other as friends of friends \citep{Jackson_Rogers_2007}. Having a
friend in common can be an incentive for establishing a relationship
\citep{Jackson_2012}. These externalities from indirect connections
can create strategic interactions between links that complicate the
empirical analysis of network formation.

To account for this link dependence, empirical models of strategic
network formation typically exploit a game-theoretic framework in
which the latent utility from forming a link depends on the other
links in a network, and the network that individuals form is an equilibrium
outcome (see \citet{Graham_2020} and \citet{Paula_2020} for surveys).
Depending on how we specify the information individuals possess, and
the strategies they take, significant challenges can arise in the
identification, estimation, and computation of model parameters. \citet{Miyauchi_2016},
\citet{Paula_2018}, and \citet{Sheng_2020} assumed that individuals
form links simultaneously under complete information. Because of the
prevalence of multiple equilibria, the parameters in general are partially
identified.\footnote{These papers considered undirected networks and used pairwise stability
\citep{Jackson_Wolinsky_1996} as the equilibrium solution.} \citet{Mele_2017} and \citet{Christakis_2020} circumvented multiplicity
by assuming that links in a network are formed in a random sequence.
This evolutionary process of network formation provides a particular
equilibrium selection mechanism that yields either a unique network
or a unique stationary distribution over networks (\citealp{Jackson_Watts_2002}).
Strategic interactions under complete information also generate a
complex dependence structure, which makes it difficult to establish
asymptotic results if one only observes a single large network. \citet{Leung_2019}
and \citet{Menzel_2017} proved a Law of Large Numbers, and \citet{Leung_Moon_2021}
proved a Central Limit Theorem with further restrictions on sparsity
and preferences.

In this paper, we develop a model of strategic network formation under
incomplete information. We assume that individuals know the unobserved
(by the researcher) utility shocks for their own potential links,
but not the unobserved utility shocks for the potential links of the
other individuals. Individuals simultaneously choose the links they
wish to form, and the directed network they form is a Bayesian Nash
equilibrium. The Bayesian Nash equilibrium has been widely used in
other network-related models.\footnote{Examples include \citet{Blume_2015}, who developed an incomplete
information game of social interactions where individuals do not observe
the utility shocks of other individuals in a network, and \citet{Galeotti_2010}
and \citet{Jackson_Yariv_2007}, who explored more general games played
on a network where individuals do not observe the private costs or
degrees of other individuals.} For network formation, we provide evolutionary results resembling
those under complete information (\citealp{Jackson_Watts_2002}; \citealp{Mele_2017}).
We argue that a Bayesian Nash equilibrium can be regarded as a long-term
equilibrium in a dynamic process of network formation (\citealp{Myatt_Wallace_2003};
\citealp{Myatt_Wallace_2004}; \citealp{Jackson_Yariv_2007}). Incomplete
information can offer an advantage over complete information in the
econometric analysis. The microfounded assumption of independent private
information yields conditional independence between links formed by
different individuals, thereby simplifying the asymptotic analysis
in a single large network. \citet{Leung_2015} pioneered the study
of strategic network formation under incomplete information. He assumed
that the utility function is additively separable in one's own links.
We extend his work to a more general utility function that is nonseparable
in one's own links.

Our extension to \citet{Leung_2015} is motivated by the empirical
regularity that social and economic networks typically present a high
degree of clustering (\citealp{Jackson_2008}; \citealp{Jackson_2017};
\citealp{Graham_2016}). This phenomenon occurs in part because two
individuals who have a mutual friend may have an increased chance
of knowing each other or have stronger incentives to build a cooperative
relationship to share risks or exchange favors \citep{Jackson_2012}.
To capture the preference for friends in common, we allow the utility
function to depend on the interaction between an individual's two
link choices.\footnote{The clustering considered here is different from that in the statistical
literature on community detection, which typically assumes that there
is a latent community structure in the data (\citealp{Abbe_2018};
\citealp{Mele_2022}).} In a network with $n$ individuals and nonseparable utility, an individual
chooses between $2^{n-1}$ overlapping portfolios of links, a seemingly
intractable discrete choice problem, as we assume that $n$ grows
large. We propose a novel approach that applies the Legendre transform
\citep{Rockafellar_1970} to the utility function so that the intractable
discrete choice problem is transformed into an equivalent tractable
sequence of correlated binary choice problems.\footnote{We are grateful to Terence Tao for suggesting this approach.}
The dependence between an individual's link choices that results from
the preference for friends in common is captured by an auxiliary variable
introduced by the Legendre transform. After the transformation we
can derive the optimal link choices of an individual explicitly.

We propose a two-step estimation procedure where we estimate the link
choice probabilities in the first step, and estimate the model parameters
in the second step. Two-step estimation has been widely used in dynamic
discrete choice models and games of incomplete information.\footnote{Seminal papers on two-step estimation include \citet{Hotz_1993},
\citet{BBL_2007}, \citet{Aguirregabiria2007}, and \citet{BHKN_2010}.} We extend this approach to network formation using data from a single
large network. Our framework requires that the network is dense, so
that the probability of forming a link does not vanish as $n\rightarrow\infty$.
The asymptotic analysis is complicated by the fact that the preference
for friends in common leads to dependence between an individual's
link choices. The auxiliary variable in the Legendre transform provides
a useful tool for investigating how the link dependence affects the
asymptotic properties of the estimator. We show that the two-step
estimator is consistent and asymptotically normal. The link dependence
does not affect the rate of convergence but increases the asymptotic
variance of the estimator.

While the two-step estimation facilitates computation, accounting
for link dependence when computing a link choice probability can be
computationally costly in a large network. To make the estimation
procedure practical, we show that a link choice probability in a finite-$n$
network converges to a limiting link probability as $n\rightarrow\infty$.
The limiting link probability has a closed form that is simple to
compute. We provide simulation evidence that using the limiting approximation
in the second step yields estimates that are similar to those from
the finite-$n$ model, provided that the networks are sufficiently
large. In addition, we also discuss how to extend our approach to
undirected networks.

We apply our approach to favor exchange networks in rural India. We
extend \citet{Jackson_2012} by investigating the directed links in
favor exchange, that is, who offers a favor to whom. We find that
indirect links have significant effects on favor provision; ignoring
these spillover effects will overestimate the homophily effects. More
strikingly, we find that the effect of support from a mutual link
depends critically on the direction of the support from the provider's
perspective. Individual $i$ is more likely to offer a favor to individual
$j$ if $i$ offers a favor to $j$'s favor-exchange companion $k$
instead of being offered a favor by $k$. These directional results
complement the findings in \citet{Jackson_2012} and \citet{Leung_2015}
and shed further light on the outcomes of policy interventions.

The remainder of the paper is organized as follows. Section \ref{sec:model}
introduces the model, including the utility function, the information
structure, and the equilibrium. Section \ref{sec:legendre} derives
an explicit expression for the optimal link choices of an individual.
Section \ref{sec:estimation} presents the two-step estimator and
its asymptotic properties. Section \ref{sec:extension} explores extensions
to our approach, including undirected networks and the limiting approximation.
Section \ref{sec:empirical} discusses the empirical application.
Section \ref{sec:conclusion} concludes the paper. Additional results
are presented in the Online Appendix.

\section{\label{sec:model}Model}

Consider a set of $n$ individuals who choose to form a network. Each
individual $i$ is endowed with a vector of observed characteristics
$X_{i}$ with support $\mathcal{X}$, and a vector of unobserved link-specific
utility shocks $\epsilon_{i}=(\epsilon_{i1},\ldots,\epsilon_{i,i-1},\epsilon_{i,i+1},\ldots,\epsilon_{in})'\in\mathbb{R}^{n-1}$,
where $\epsilon_{ij}$ is the utility shock for link $ij$. Let $X=(X_{1}',\ldots,X_{n}')'\in\mathcal{X}^{n}$
denote the characteristic profile and $\epsilon=(\epsilon_{1}',\ldots,\epsilon_{n}')'\in\mathbb{R}^{n(n-1)}$
denote the utility shock profile.

The network formed is denoted by an $n\times n$ binary matrix $G\in\mathcal{G}$,
where the $ij$th entry $G_{ij}=1$ if individual $i$ forms a link
to individual $j$ and $G_{ij}=0$ otherwise. The diagonal elements
$G_{ii}$ are set to $0$ for all $i$, so there are no self-links.
In this paper, we focus on directed networks, that is, $G_{ij}$ and
$G_{ji}$ can be different. While relationships such as friendships
and collaborations are typically undirected, many economic networks
are in fact formed as a result of directed individual decisions. Examples
include a village resident lending money to another resident, a buyer
purchasing a product from a seller, and an employee referring a candidate
for a job. Following \citet{Bala_2000}, \citet{Mele_2017}, and \citet{Leung_2015},
we consider a noncooperative framework where individual $i$ unilaterally
decides to form the link $ij$.\footnote{Equivalently, we can characterize the formation of a link as a bilateral
decision between the provider and the recipient in which the recipient
always prefers the link. For example, a village resident always likes
to receive a favor, a seller always wants to sell a product (given
the price), and a job candidate always wishes to get a job referral.} In Section \ref{sec:undirected}, we extend our analysis to undirected
networks.

\paragraph*{Utility.}

For a given characteristic profile $X$ and utility shock vector $\epsilon_{i}$,
individual $i$'s utility in a network $G$ is given by
\begin{equation}
U_{i}(G,X,\epsilon_{i};\theta_{u})=\frac{1}{n-1}\sum_{j\neq i}G_{ij}\left(u_{ij}(G_{-i},X;\beta)+\frac{1}{2(n-2)}\sum_{k\neq i,j}G_{ik}v_{i,jk}(G_{-i},X;\gamma)-\epsilon_{ij}\right),\label{eq:U}
\end{equation}
where $G_{-i}$ is the submatrix of $G$ with the $i$th row deleted,
that is, the links formed by individuals other than $i$. We assume
that the utility function is known up to the parameter $\theta_{u}=(\beta',\gamma')'$
in a compact set $\Theta_{u}\subset\mathbb{R}^{d_{u}}$.

In this specification, the term $u_{ij}(G_{-i},X;\beta)$ represents
individual $i$'s incremental utility from linking to individual $j$
that does not depend on the other links that $i$ forms. A typical
specification of $u_{ij}(G_{-i},X;\beta)$ is
\begin{eqnarray}
u_{ij}(G_{-i},X;\beta) & = & \beta_{1}+X_{i}^{\prime}\beta_{2}+d(X_{i},X_{j})^{\prime}\beta_{3}+G_{ji}\beta_{4}\nonumber \\
 &  & +\frac{1}{n-2}\sum_{k\neq i,j}G_{kj}\beta_{5}+\frac{1}{n-2}\sum_{k\neq i,j}G_{jk}\beta_{6},\label{eq:U.sep}
\end{eqnarray}
where $d(X_{i},X_{j})$ represents a vector of known functions of
$X_{i}$ and $X_{j}$ that measure the social proximity between $i$
and $j$; for example, whether they have the same gender, age, education,
and caste. This term captures the homophily effect (\citealp{Jackson_2008};
\citealp{Jackson_2017}). The last three terms in equation (\ref{eq:U.sep})
capture the spillover effects from other links that involve $j$,
including the reciprocity effect of link $ji$ ($\beta_{4}$) and
the effects of $j$'s in-degree ($\beta_{5}$) and out-degree ($\beta_{6}$).
Note that we normalize $j$'s in-degree and out-degree by $n-2$ to
ensure that these terms remain bounded as $n\rightarrow\infty$, so
that they do not dominate when $n$ is large. The specification in
equation (\ref{eq:U.sep}) is similar to that in \citet{Leung_2015}.

In addition to utility that is separable in individual $i$'s links,
we also allow for utility that is nonseparable in individual $i$'s
links. The term $v_{i,jk}(G_{-i},X;\gamma)$ represents the incremental
utility that $i$ derives from linking to both individual $j$ and
individual $k$, if $j$ and $k$ link to each other. An important
example is
\begin{eqnarray}
v_{i,jk}(G_{-i},X;\gamma) & = & (G_{jk}+G_{kj})\gamma_{1}(X_{j},X_{k})\nonumber \\
 &  & +\frac{1}{n-3}\sum_{l\neq i,j,k}(G_{jl}G_{lk}+G_{kl}G_{lj})\gamma_{2}(X_{j},X_{k}).\label{eq:U.nonsep}
\end{eqnarray}
The two terms are motivated by the prevalence of triadic closure ($\gamma_{1}>0$)
and cyclic closure ($\gamma_{2}>0$), which mean that individual $i$
is more likely to link to individual $j$ if $i$ links to a third
individual $k$ who is connected to $j$ directly or indirectly (\citealp{Kossinets_Watts_2006};
\citealp{Jackson_2008}; \citealp{Jackson_2017}).\footnote{The terms $G_{jk}+G_{kj}$ and $G_{jl}G_{lk}+G_{kl}G_{lj}$ in equation
(\ref{eq:U.nonsep}) can be replaced by other functions that are symmetric
in $j$ and $k$. For example, we can replace $G_{jk}+G_{kj}$ by
$G_{jk}G_{kj}$ (i.e., both $j$ links to $k$ and $k$ links to $j$)
or $1\{G_{jk}+G_{kj}\geq1\}$ (i.e., either $j$ links to $k$ or
$k$ links to $j$).} One possible reason for triadic/cyclic closure is that via a mutual
friend $k$, individual $i$ may have an increased chance to know
$j$. Another reason---one relevant particularly in favor exchange
and risk sharing networks---is that linking to a third individual
$k$ whom $i$ trusts and who trusts $j$ may give $i$ the basis
to trust $j$ (\citealp{Easley_2010}; \citealp{Karlan2009}; \citealp{Jackson_2012}).\footnote{In the context of directed links, while certain variants of triadic
and cyclic closure statistics can be specified through the separable
component $u_{ij}$ (e.g., the supported trust in \citet{Leung_2015}),
we demonstrate in the empirical application in Section \ref{sec:empirical}
that the empirically relevant variants of these statistics may inevitably
require the nonseparable component $v_{i,jk}$.} These mutual-friend effects can also depend on the social proximity
between $j$ and $k$, as captured by $\gamma_{1}(X_{j},X_{k})$ and
$\gamma_{2}(X_{j},X_{k})$, which consist of known nonnegative functions
of $X_{j}$ and $X_{k}$, such as whether $j$ and $k$ share certain
characteristics and a vector of parameters.\footnote{One example is $\gamma_{1}(X_{j},X_{k})=d_{1}(X_{j},X_{k})'\gamma_{1}$
and $\gamma_{2}(X_{j},X_{k})=d_{2}(X_{j},X_{k})'\gamma_{2}$, where
$d_{1}(X_{j},X_{k})$ and $d_{2}(X_{j},X_{k})$ are vectors that measure
the social distance between $j$ and $k$.} Note that $v_{i,jk}(G_{-i},X;\gamma)$ is symmetric in $j$ and $k$
so that the utility function does not depend on how we label the individuals.\footnote{This requires that $\gamma_{1}(X_{j},X_{k})$ and $\gamma_{2}(X_{j},X_{k})$
are symmetric in $j$ and $k$.} The second term in equation (\ref{eq:U.nonsep}) is also normalized
to guarantee its boundedness for large $n$.

\paragraph*{Information.}

Most literature on network formation games assumes that individuals
have complete information about the game \citep{Jackson_Wolinsky_1996,Bala_2000,Christakis_2020,Mele_2017,Miyauchi_2016,Paula_2018,Sheng_2020,Menzel_2017}.
While this is appropriate in small networks, in a large network an
individual may not observe every aspect of the other individuals.
In this paper, we follow \citet{Leung_2015} and assume that each
individual only has partial information about the other individuals.
In particular, we assume that the characteristic profile $X$ is observed
by all the individuals, but the utility shock vector $\epsilon_{i}$
is observed by individual $i$ only.\footnote{This is a standard setup for games of incomplete information (e.g.,
\citealp{BHKN_2010}).} We also assume that the utility shocks are i.i.d. and are independent
of the characteristics. Formally,
\begin{assumption}
\label{ass:e=000026x}(i) $\epsilon_{ij}$ is i.i.d. with cdf $F_{\epsilon}(\epsilon_{ij};\theta_{\epsilon})$
known up to the parameter $\theta_{\epsilon}\in\Theta_{\epsilon}\subset\mathbb{R}^{d_{\epsilon}}$.
(ii) The distribution of $\epsilon_{ij}$ has a density function $f_{\epsilon}(\epsilon_{ij};\theta_{\epsilon})$
with respect to the Lebesgue measure, which is continuously differentiable
in $\theta_{\epsilon}$, strictly positive, and bounded on $\mathbb{R}$.
(iii) $\epsilon$ and $X$ are independent.
\end{assumption}
The independence of $\epsilon_{i}$ across $i$ is a crucial assumption.
It enables us to break the link dependence across individuals and
reduce the complexity of the model. The independence of $\epsilon_{ij}$
and $\epsilon_{ik}$ is imposed for simplicity.\footnote{\citet{Leung_2015} allows $\epsilon_{ij}$ and $\epsilon_{ik}$ to
be arbitrarily correlated, which generates persistent correlation
between $G_{ij}$ and $G_{ik}$, leading to a rate of convergence
slower than ours.} Assumptions \ref{ass:e=000026x}(ii)--(iii) are standard regularity
assumptions.

\paragraph*{Equilibrium.}

We assume that individuals form links simultaneously. Let $G_{i}$
be the $i$th row of network $G$, that is, the links formed by individual
$i$, and $\mathcal{G}_{i}=\{0,1\}^{n-1}$ the set of all possible
$G_{i}$. A strategy of individual $i$ is a function $G_{i}(X,\epsilon_{i}):\mathcal{X}^{n}\times\mathbb{R}^{n-1}\rightarrow\mathcal{G}_{i}$
that maps $i$'s information $(X,\epsilon_{i})$ to a row vector of
links $G_{i}$. Denote the strategy profile of all individuals by
$G(X,\epsilon)=(G_{1}(X,\epsilon_{1})',\ldots,G_{n}(X,\epsilon_{n})')'$.
A Bayesian Nash equilibrium (or an equilibrium for short) of the game
is a strategy profile $G(X,\epsilon)$ such that each $G_{i}(X,\epsilon_{i})$
maximizes the expected utility $\mathbb{E}[U_{i}(G_{i},G_{-i},X,\epsilon_{i})|X,\epsilon_{i}]$,
where the expectation is taken with respect to the strategies of individuals
other than $i$, $G_{-i}$.

For the utility function in (\ref{eq:U}), the expected utility of
individual $i$ is
\begin{eqnarray}
 &  & \mathbb{E}[U_{i}(G_{i},G_{-i},X,\epsilon_{i})|X,\epsilon_{i}]\nonumber \\
 & = & \frac{1}{n-1}\sum_{j\neq i}G_{ij}\left(\mathbb{E}[u_{ij}(G_{-i},X)|X]+\frac{1}{2(n-2)}\sum_{k\neq i,j}G_{ik}\mathbb{E}[v_{i,jk}(G_{-i},X)|X]-\epsilon_{ij}\right).\label{eq:EU}
\end{eqnarray}
Under the specifications in (\ref{eq:U.sep})--(\ref{eq:U.nonsep}),
we have
\begin{eqnarray}
\mathbb{E}[u_{ij}(G_{-i},X)|X] & = & \beta_{1}+X_{i}^{\prime}\beta_{2}+d(X_{i},X_{j})^{\prime}\beta_{3}+\sigma_{ji}(X)\beta_{4}\nonumber \\
 &  & +\frac{1}{n-2}\sum_{k\neq i,j}\sigma_{kj}(X)\beta_{5}+\frac{1}{n-2}\sum_{k\neq i,j}\sigma_{jk}(X)\beta_{6},\label{eq:Eu}
\end{eqnarray}
and
\begin{eqnarray}
\mathbb{E}[v_{i,jk}(G_{-i},X)|X] & = & (\sigma_{jk}(X)+\sigma_{kj}(X))\gamma_{1}(X_{j},X_{k})\nonumber \\
 &  & +\frac{1}{n-3}\sum_{l\neq i,j,k}(\sigma_{jl}(X)\sigma_{lk}(X)+\sigma_{kl}(X)\sigma_{lj}(X))\gamma_{2}(X_{j},X_{k}),\label{eq:Ev}
\end{eqnarray}
where $\sigma_{ij}(X)=\mathbb{E}[G_{ij}|X]$. The expressions for
$\mathbb{E}[u_{ij}(G_{-i},X)|X]$ and $\mathbb{E}[v_{i,jk}(G_{-i},X)|X]$
follow because the independence of $\epsilon_{i}$ across $i$ implies
that $\epsilon_{i}$ is independent of the strategies of others $G_{-i}$
conditional on $X$ and hence $\mathbb{E}[u_{ij}(G_{-i},X)|X]$ and
$\mathbb{E}[v_{i,jk}(G_{-i},X)|X]$ only depend on the public information
$X$. Equation (\ref{eq:Ev}) holds also because, conditional on $X,$
the strategies $G_{j}$ (or $G_{k}$) and $G_{l}$ are independent.

Following the literature on incomplete information games \citep{BHKN_2010},
we can represent an equilibrium in the space of conditional choice
probabilities. Given $X$, let $\sigma_{i}(g_{i}|X)$ denote the conditional
probability that individual $i$ chooses link vector $g_{i}$
\begin{eqnarray}
\sigma_{i}(g_{i}|X) & = & \Pr(G_{i}=g_{i}|X)\nonumber \\
 & = & \Pr\left(\left.\mathbb{E}[U_{i}(g_{i},G_{-i},X,\epsilon_{i})|X,\epsilon_{i}]\geq\max_{\tilde{g}_{i}\in\mathcal{G}_{i}}\mathbb{E}[U_{i}(\tilde{g}_{i},G_{-i},X,\epsilon_{i})|X,\epsilon_{i}]\right\vert X\right)\label{eq:CCP}
\end{eqnarray}
and $\sigma(X)=\{\sigma_{i}(g_{i}|X),g_{i}\in\mathcal{G}_{i},i=1,\ldots,n\}$
the conditional choice probability (CCP) profile. The right-hand side
of equation (\ref{eq:CCP}) defines a mapping from $\sigma_{-i}(X)=\{\sigma_{j}(g_{j}|X),g_{j}\in\mathcal{G}_{j},j\neq i\}$
to $\sigma_{i}(g_{i}|X)$ that we denote by $\mathcal{P}_{i}(g_{i}|X,\sigma_{-i}(X))$.
An equilibrium CCP profile $\sigma^{\ast}(X)$ is a fixed point of
the equation
\begin{equation}
\sigma_{i}^{\ast}(g_{i}|X)=\mathcal{P}_{i}(g_{i}|X,\sigma_{-i}^{\ast}(X))\label{eq:BNE}
\end{equation}
for all $g_{i}\in\mathcal{G}_{i}$ and all $i=1,\ldots,n$. From an
equilibrium CCP profile, we can derive the equilibrium strategy profile
from the optimal decision of each individual $i$ for a given $X$
and $\epsilon_{i}$. Therefore, we can represent an equilibrium equivalently
by the CCP profile.

The Bayesian Nash equilibrium, though less common in network formation,
has been widely used in other network-related models. For example,
\citet{Blume_2015} developed an incomplete information game of social
interactions where individuals do not observe the private utility
shocks of other individuals in a network. They show that Bayesian
Nash equilibria of the game provide a microfoundation that can nest
the standard social interaction models such as \citet{Manski_1993}.
\citet{Galeotti_2010} and \citet{Jackson_Yariv_2007} considered
more general games played on a network, where an individual's payoff
depends on the actions taken by neighbors. In such a game, individuals
do not observe the private costs or degrees of other individuals.
They form beliefs about the degrees of their neighbors based on the
degree distribution in the network. Both \citet{Galeotti_2010} and
\citet{Jackson_Yariv_2007} investigated Bayesian Nash equilibria
of the game.

One concern regarding the application of the Bayesian Nash equilibrium
to network formation is that links may not be formed simultaneously.
In network formation games with complete information, equilibrium
solutions that assume simultaneous move (e.g., the Nash equilibrium
for directed networks and pairwise stability for undirected networks)
are usually justified by an evolutionary process that converges to
equilibria in the static game. For example, \citet{Jackson_Watts_2002}
developed a dynamic process of network formation that converges to
pairwise stable networks if cycles are ruled out. \citet{Mele_2017}
considered a similar dynamic process for directed networks that converges
to Nash equilibria in the static game. In Online Appendix \ref{online:dynamic},
we show that similar evolutionary results can be established for the
Bayesian Nash equilibrium. Specifically, we construct two dynamic
processes of network formation where links are formed over time. The
first process assumes that one link is updated in each period as in
\citet{Myatt_Wallace_2004}, and the second process assumes that all
links are updated in each period as in \citet{Myatt_Wallace_2003}.
Unlike the dynamic processes in \citet{Jackson_Watts_2002} and \citet{Mele_2017},
where an active individual observes all the links formed in previous
periods, we assume that an active individual only observes the distribution
of the links formed previously. Following \citet{Myatt_Wallace_2004}
and \citet{Myatt_Wallace_2003}, we show that for a sufficiently large
network, the first process generates a Markov chain of networks that
has a unique limiting distribution with local modes coinciding with
stable Bayesian Nash equilibria, and the second process converges
to a Bayesian Nash equilibrium in probability. The result in our second
process is in line with \citet{Jackson_Yariv_2007}, who also showed
that Bayesian Nash equilibria in a static game are equivalent to steady
states of a dynamic process. These evolutionary results suggest that
a Bayesian Nash equilibrium can be regarded as a long-term equilibrium
in a dynamic process of network formation.

\paragraph*{Symmetric Equilibrium.}

In this paper, we focus on symmetric equilibria where observationally
identical individuals have the same choice probabilities. In a symmetric
equilibrium, the CCP profile $\sigma(X)$ satisfies that for any individuals
$i$ and $j$ with $X_{i}=X_{j}$, we have $\sigma_{i}(g_{i}|X)=\sigma_{j}(g_{j}|X)$
for all $g_{i}\in\mathcal{G}_{i}$ and $g_{j}\in\mathcal{G}_{j}$,
with $g_{j}$ obtained from $g_{i}$ by swapping its $i$th and $j$th
components $g_{ii}$ and $g_{ij}$. Simply put, individuals with the
same observed characteristics choose their links with the same probability.\footnote{This restriction does not rule out the possibility that two observationally
equivalent individuals form different links in an observed network
because they can have different unobserved utility shocks.} This restriction is motivated by the observation that the utility
function is the same for all individuals, so if they hold symmetric
beliefs about the decisions of others (as specified in a symmetric
CCP profile), then individuals of any given $X_{i}$ and $\epsilon_{i}$
face the same decision problem. The optimal decision in it is unique
with probability one, leading to symmetry in the choice probabilities.
The symmetry of an equilibrium guarantees that the conditional choice
probabilities of an individual do not depend on how we label the individuals,
a desirable feature in most networks where the identities of individuals
do not play any role and individuals are labeled arbitrarily.

In Proposition \ref{prop:existsym}, we establish the existence of
a symmetric equilibrium. Our proof is similar to that in \citet[Thereom 1]{Leung_2015}.
We assume that in observed data, individuals coordinate on a symmetric
equilibrium, independently of the utility shocks $\epsilon$.\footnote{The symmetry implies that the expected utility terms in (\ref{eq:Eu})
and (\ref{eq:Ev}) depend on $i$, $j$ and $k$ only through $X_{i}$,
$X_{j}$, and $X_{k}$.} There may be multiple symmetric equilibria that satisfy condition
(\ref{eq:BNE}).
\begin{prop}
\label{prop:existsym}Suppose that Assumption \ref{ass:e=000026x}
is satisfied. For any $X$, there exists a symmetric equilibrium CCP
profile $\sigma(X)$.
\end{prop}
\begin{proof}
See Appendix \ref{app:proof.model}.
\end{proof}
\begin{assumption}
\label{ass:sym}The equilibrium selection mechanism selects a symmetric
equilibrium $\sigma$ independently of the utility shocks $\epsilon$.
\end{assumption}
A game with multiple equilibria is considered incomplete unless restrictions
are imposed on the equilibrium selection mechanism. In data scenarios
with many markets, it is often assumed that the selection mechanism
is degenerate, meaning that any market with observationally equivalent
individuals must select the same equilibrium \citep{BHKN_2010}. This
assumption ensures that the CCPs can be estimated by pooling observations
across markets \citep{Paula_2013}. In contrast, we build on the insight
of \citet{Leung_2015} and restrict the selection mechanism to select
only symmetric equilibria. This approach allows us to estimate the
CCPs by pooling pairs of individuals in a single large network.

The main challenge in analyzing the model involves characterizing
the optimal decision of an individual. Because the expected utility
depends on the interaction $G_{ij}G_{ik}$, an individual no longer
chooses between separable links as in \citet{Leung_2015}, but between
portfolios of links. This is a multinomial discrete choice problem
with $2^{n-1}$ overlapping alternatives. Note that links $G_{ij}$
and $G_{ik}$ are strategic complements (substitutes) if $\gamma_{1},\gamma_{2}>0$
($<0$). The nonseparable decision over links naturally leads to the
links being dependent on one another. In the subsequent section,
we develop a novel method to derive the optimal decision of an individual
and characterize the link dependence.

In addition, this challenge may arise in other applications where
individuals select a set of binary choices that are complements or
substitutes for one another. Formally, suppose individual $i$ has
a set of binary choices $D_{ij},j\in\mathcal{C}=\{1,\dots,n_{c}\}$,
where the number of choices $n_{c}$ is large. The utility of individual
$i$ is nonseparable in $D_{ij}$ and includes the term $\frac{1}{n_{c}(n_{c}-1)}\sum_{j}\sum_{k\neq j}D_{ij}D_{ik}v_{i,jk}$,
where $v_{i,jk}$ captures the complementarity or substitutability
between $D_{ij}$ and $D_{ik}$.\footnote{$v_{i,jk}$ does not need to depend on an equilibrium, as it does
in our model. Instead, it can be flexibly specified according to the
context.} We illustrate this setting with several examples. Our method can
be used to derive the optimal choices in these cases.

\begin{example}[Discrete choice]
Consider the discrete choice models for bundles in \citet{Gentzkow_2007}
and \citet{Fox_2017}, where $D_{ij}$ indicates whether consumer
$i$ purchases product $j$. These studies focus on two products ($1$
and $2$) and allow the utility of $i$ to depend on the interaction
term $D_{i1}D_{i2}$, which captures the complementarity between the
products when purchased as a bundle. Our setting extends theirs to
accommodate a large number of products (e.g., shopping on Amazon).
\end{example}

\begin{example}[Trade]
\citet{Morales_2019_extgravity} develop a model in which firms are
more likely to export to foreign countries that are similar to their
prior export destinations (extended gravity) due to lower entry costs.
We can extend their insight to a static setting by allowing firm $i$'s
export decision to country $j$ to depend on its export decision to
country $k$. Specifically, let $D_{ij}$ indicate whether firm $i$
exports to country $j$. The utility of $i$ depends on the average
of $D_{ij}D_{ik}v_{i,jk}$, where $v_{i,jk}$ captures the extended
gravity between countries $j$ and $k$. 
\end{example}

\begin{example}[Multinational production]
Consider a multinational production problem, where $D_{ij}$ indicates
whether firm $i$ selects foreign location $j$ for production. \citet{Arkolakis_2025}
allow for cross-location dependence to account for complementarity,
which leads to a combinatorial discrete choice problem. They propose
an iterative algorithm to solve for the optimal locations under a
single crossing condition. We provide an alternative approach to derive
the optimal locations if the utility is quadratic in location choices.

\end{example}

\section{\label{sec:legendre}Optimal Link Choices}

In this section, we develop an approach that yields an explicit expression
for the optimal link choices of an individual. The idea is to find
an auxiliary variable that captures the strategic interactions between
an individual's link choices, so that after the inclusion of this
auxiliary variable the link choices become correlated binary choices,
with the correlation captured by the auxiliary variable.

Recall that the incremental utility $v_{i,jk}(G_{-i},X)$ is symmetric
in $j$ and $k$. Moreover, in a symmetric equilibrium $\sigma$ the
expected incremental utility $\mathbb{E}[v_{i,jk}(G_{-i},X)|X,\sigma]$
depends on $j$ and $k$ only through the values of $X_{j}$ and $X_{k}$.\footnote{The inclusion of $\sigma$ in the notation indicates that the expectation
is taken according to $\sigma$. } These symmetry properties imply that $\mathbb{E}[v_{i,jk}(G_{-i},X)|X,\sigma]$
is a symmetric function of $X_{j}$ and $X_{k}$.

To facilitate the exposition, we focus on the case where $X_{i}$
is discrete. Assume that $X_{i}$ takes a finite number of values,
which we refer to as the types of an individual.\footnote{It is more complicated to derive the optimal link choices when $X_{i}$
is continuous, as the matrix notation must be replaced with linear
operators. Exploring the theoretical results for continuous $X_{i}$
is beyond the scope of this paper. In practice, our approach can be
applied by discretizing continuous covariates, as demonstrated in
the empirical application in Section \ref{sec:empirical}.}
\begin{assumption}
\label{ass:discX}$X_{i}$ takes $T<\infty$ distinct values $x_{1},\ldots,x_{T}$.
\end{assumption}
Under Assumption \ref{ass:discX}, we can represent the expected utility
in (\ref{eq:EU}) in matrix form. For $1\leq s,t\leq T$, let $V_{i,st}(X,\sigma)$
denote the value of $\mathbb{E}[v_{i,jk}(G_{-i},X)|X,\sigma]$ if
individuals $j$ and $k$ are of types $x_{s}$ and $x_{t}$ respectively;
that is, $V_{i,st}(X,\sigma)=\mathbb{E}[v_{i,jk}(G_{-i},X)|X_{j}=x_{s},X_{k}=x_{t},X,\sigma]$.
Arrange the $T^{2}$ type-specific expected incremental utilities
$V_{i,st}(X,\sigma)$ in a $T\times T$ matrix $V_{i}(X,\sigma)=(V_{i,st}(X,\sigma))\in\mathbb{R}^{T\times T}$.
Because $V_{i,st}(X,\sigma)$ is symmetric in $s$ and $t$, $V_{i}(X,\sigma)$
is a symmetric matrix. Using the matrix notation, we can represent
the expected utility in (\ref{eq:EU}) as
\begin{eqnarray}
\mathbb{E}[U_{i}(G_{i},G_{-i},X,\epsilon_{i})|X,\epsilon_{i},\sigma] & = & \frac{1}{n-1}\sum_{j\neq i}G_{ij}(U_{ij}(X,\sigma)-\epsilon_{ij})\nonumber \\
 &  & +\frac{1}{2(n-1)(n-2)}\sum_{j\neq i}\sum_{k\neq i}G_{ij}G_{ik}Z'_{j}V_{i}(X,\sigma)Z_{k},\label{eq:EU.mat}
\end{eqnarray}
where $Z_{j}=(1\{X_{j}=x_{1}\},\ldots,1\{X_{j}=x_{T}\})'$ is a $T\times1$
vector of binary variables that indicates the type of individual $j$,
and $U_{ij}(X,\sigma)=\mathbb{E}[u_{ij}(G_{-i},X)|X,\sigma]-\frac{1}{2(n-2)}Z'_{j}V_{i}(X,\sigma)Z_{j}$.
The term $Z'_{j}V_{i}(X,\sigma)Z_{k}$ represents the expected incremental
utility that individual $i$ receives from linking to both $j$ and
$k$.

To derive the optimal decision of individual $i$, we \textquotedbl linearize\textquotedbl{}
the quadratic term in (\ref{eq:EU.mat}) using the Legendre transform
\citep{Rockafellar_1970}. Observe that $V_{i}(X,\sigma)$ is real
and symmetric and thus has a real spectral decomposition 
\begin{equation}
V_{i}(X,\sigma)=\Phi_{i}(X,\sigma)\Lambda_{i}(X,\sigma)\Phi'_{i}(X,\sigma),\label{eq:spectral}
\end{equation}
where $\Lambda_{i}(X,\sigma)=\text{diag}(\lambda_{i1}(X,\sigma),\ldots,\lambda_{iT}(X,\sigma))$
denotes the $T\times T$ diagonal matrix of eigenvalues in $\mathbb{R}$
and $\Phi_{i}(X,\sigma)=(\phi_{i1}(X,\sigma),\ldots,\phi_{iT}(X,\sigma))$
denotes the $T\times T$ orthogonal matrix of eigenvectors in $\mathbb{R}^{T}$.
Using the spectral decomposition, we can express the quadratic term
in (\ref{eq:EU.mat}) as a function of the squares of $\frac{1}{n-1}\sum_{j\neq i}G_{ij}Z'_{j}\phi_{it}(X,\sigma)$,
$t=1,\ldots,T$, which are linear in link choices $G_{ij}$. 

Next, we ``linearize'' these squares of linear functions using a
special case of the Legendre transform. In particular, for any scalar
$y\in\mathbb{R}$, we have
\begin{equation}
\frac{1}{2}y^{2}=\max_{\omega\in\mathbb{R}}\left\{ y\omega-\frac{1}{2}\omega^{2}\right\} ,\label{eq:legendre.scalar}
\end{equation}
where $\omega\in\mathbb{R}$ is a scalar auxiliary variable. By choosing
$y=\frac{1}{n-1}\sum_{j\neq i}G_{ij}Z'_{j}\phi_{it}(X,\sigma)$, we
can replace its square by the maximization on the right-hand-side
of (\ref{eq:legendre.scalar}). This maximization has an objective
function that is linear in $y$ and thus linear in the link choices
$G_{ij}$. The transformation of the expected utility is presented
in Lemma \ref{lem:EU.legendre}.
\begin{lem}
\label{lem:EU.legendre}Suppose that Assumptions \ref{ass:e=000026x}--\ref{ass:discX}
are satisfied. The expected utility in (\ref{eq:EU.mat}) satisfies
\begin{eqnarray}
 &  & \mathbb{E}[U_{i}(G_{i},G_{-i},X,\epsilon_{i})|X,\epsilon_{i},\sigma]\nonumber \\
 & = & \frac{1}{n-1}\sum_{j\neq i}G_{ij}(U_{ij}(X,\sigma)-\epsilon_{ij})+\frac{n-1}{2(n-2)}\sum_{t=1}^{T}\lambda_{it}(X,\sigma)\left(\frac{1}{n-1}\sum_{j\neq i}G_{ij}Z'_{j}\phi_{it}(X,\sigma)\right)^{2}\nonumber \\
 & = & \frac{1}{n-1}\sum_{j\neq i}G_{ij}(U_{ij}(X,\sigma)-\epsilon_{ij})\nonumber \\
 &  & +\frac{n-1}{n-2}\sum_{t=1}^{T}\lambda_{it}(X,\sigma)\max_{\omega_{t}\in\mathbb{R}}\left\{ \frac{1}{n-1}\sum_{j\neq i}G_{ij}Z'_{j}\phi_{it}(X,\sigma)\omega_{t}-\frac{1}{2}\omega_{t}^{2}\right\} .\label{eq:EU.legendre}
\end{eqnarray}
\end{lem}
\begin{proof}
See Appendix \ref{app:proof.legendre}.
\end{proof}
The optimal decision of individual $i$ is a link vector $G_{i}\in\mathcal{G}_{i}$
that maximizes her expected utility. By Lemma \ref{lem:EU.legendre},
the expected utility can be expressed as the optimal value obtained
from optimizing over the components of the auxiliary variable $\omega=(\omega_{1},\dots,\omega_{T})'\in\mathbb{R}^{T}$.
Note that the transformed expected utility in (\ref{eq:EU.legendre})
is separable in each maximization. Therefore, if we move $\lambda_{it}(X,\sigma)$
inside the maximization over $\omega_{t}$, the maximization remains
unchanged if $\lambda_{it}(X,\sigma)>0$ and switches to a minimization
if $\lambda_{it}(X,\sigma)<0$, leading to a maximin problem over
$\omega$ in general. The separability also implies that the order
of the maximizations and minimizations does not matter. If we can
further interchange the maximization over $G_{i}$ and the maximin
over $\omega$, we can solve for the optimal $G_{i}$ first from a
simple maximization with an objective function linear in $G_{i}$.
This optimal $G_{i}$ is evidently a function of $\omega$. By solving
for the optimal $\omega$ next and evaluating the optimal $G_{i}$
at the optimal $\omega$, we can derive the optimal decision that
maximizes the expected utility. The validity of this approach and
the derivation of the optimal decision are demonstrated in Theorem
\ref{thm:optG}.
\begin{thm}
\label{thm:optG}Suppose that Assumptions \ref{ass:e=000026x}--\ref{ass:discX}
are satisfied. For each $i$, the optimal link choices $G_{i}(\epsilon_{i},X,\sigma)=(G_{ij}(\epsilon_{i},X,\sigma),j\neq i)\in\mathcal{G}_{i}$
are given by
\begin{equation}
G_{ij}(\epsilon_{i},X,\sigma)=1\left\{ U_{ij}(X,\sigma)+\frac{n-1}{n-2}Z'_{j}\Phi_{i}(X,\sigma)\Lambda_{i}(X,\sigma)\omega_{i}(\epsilon_{i},X,\sigma)\geq\epsilon_{ij}\right\} ,\forall j\neq i,\label{eq:gij}
\end{equation}
almost surely. The $T\times1$ vector $\omega_{i}(\epsilon_{i},X,\sigma)\in\mathbb{R}^{T}$
in (\ref{eq:gij}) is an optimal solution to the maximin problem
\begin{equation}
\max_{\omega_{t},t\in\mathcal{T}_{i+}}\min_{\omega_{t},t\in\mathcal{T}_{i-}}\frac{1}{n-1}\sum_{j\neq i}\left[U_{ij}(\ensuremath{X,\sigma)+\frac{n-1}{n-2}Z'_{j}\Phi_{i}(X,\sigma)\Lambda_{i}(X,\sigma)\omega-\epsilon_{ij}}\right]_{+}-\frac{n-1}{2(n-2)}\omega'\Lambda_{i}(X,\sigma)\omega,\label{eq:maxmin}
\end{equation}
where $\mathcal{T}_{i+}=\{1\leq t\leq T:\lambda_{it}\left(X,\sigma\right)>0\}$
and $\mathcal{T}_{i-}=\{1\leq t\leq T:\lambda_{it}\left(X,\sigma\right)<0\}$.
We set $\omega_{it}(\epsilon_{i},X,\sigma)=0$ if $\lambda_{it}(\epsilon_{i},X,\sigma)=0$.\footnote{If $\lambda_{it}(X,\sigma)=0$, the objective function does not depend
on $\omega_{t}$, so we set $\omega_{it}(\epsilon_{i},X,\sigma)=0$.} Moreover, both $G_{i}(\epsilon_{i},X,\sigma)$ and $\omega_{i}(\epsilon_{i},X,\sigma)$
are unique almost surely.
\end{thm}
\begin{proof}
See Appendix \ref{app:proof.legendre}.
\end{proof}
To gain some intuition about the auxiliary variable $\omega_{i}(\epsilon_{i},X,\sigma)$
in (\ref{eq:gij}), multiplying the first-order condition of problem
(\ref{eq:maxmin}) (see Lemma \ref{lem:foc.w}) by $\frac{n-1}{n-2}Z'_{j}\Phi_{i}(X,\sigma)$,
we derive that
\begin{equation}
\frac{n-1}{n-2}Z'_{j}\Phi_{i}(X,\sigma)\Lambda_{i}(X,\sigma)\omega_{i}(\epsilon_{i},X,\sigma)=\frac{1}{n-2}\sum_{k\neq i}G_{ik}(\epsilon_{i},X,\sigma)Z'_{j}V_{i}(X,\sigma)Z_{k}\label{eq:w.interp}
\end{equation}
almost surely, where $G_{ik}(\epsilon_{i},X,\sigma)$ is defined in
(\ref{eq:gij}). The left-hand side of (\ref{eq:w.interp}) is the
component added to the latent utility in (\ref{eq:gij}). The right-hand
side of (\ref{eq:w.interp}) interprets this component as the expected
incremental utility from friends in common. We interpret it as such
because if individual $i$ contemplates a link to individual $j$,
she anticipates that her friend $k$ can potentially become a mutual
friend with $j$. If individual $j$ is of type $x_{s}$ and individual
$i$'s friend $k$ is of type $x_{t}$, then $i$'s expected utility
from this potential friend in common is $V_{i,st}(X,\sigma)$. Taking
the average over all friends of individual $i$, we obtain the expected
incremental utility from friends in common if individual $i$ links
to individual $j$. By adding this component to the latent utility,
we internalize the strategic interactions between the link choices
due to the preference for friends in common, so that the optimal decision
breaks down into a collection of binary choices.

The auxiliary variable $\omega_{i}(\epsilon_{i},X,\sigma)$ provides
an explicit expression for the dependence of the links formed by individual
$i$. Note that $\omega_{i}(\epsilon_{i},X,\sigma)$ is a function
of $\epsilon_{i}$ because it is an optimal solution to problem (\ref{eq:maxmin})
whose objective function depends on $\epsilon_{i}$. The randomness
in $\omega_{i}(\epsilon_{i},X,\sigma)$ leads to dependence between
the link choices. From (\ref{eq:gij}) we can see that two link choices
$G_{ij}$ and $G_{ik}$ are dependent either through the presence
of $\omega_{i}(\epsilon_{i},X,\sigma)$ in both links, or through
the dependence between the utility shock $\epsilon_{ij}$ and $\omega_{i}(\epsilon_{i},X,\sigma)$
in $G_{ik}$ or symmetrically between the utility shock $\epsilon_{ik}$
and $\omega_{i}(\epsilon_{i},X,\sigma)$ in $G_{ij}$. This explicit
characterization of the link dependence is useful for studying the
asymptotic properties of an estimator.
\begin{rem}
In the special case where there is no effect from friends in common
($\gamma_{1},\gamma_{2}=0$), we have $V_{i}(X,\sigma)=0$ and the
utility function is separable in one's own links. In this case, the
optimal decision in (\ref{eq:gij}) reduces to $G_{ij}=1\{\mathbb{E}[u_{ij}(G_{-i},X)|X,\sigma]\geq\epsilon_{ij}\}$
for $j\neq i$. Because there are no strategic interactions between
the link choices of an individual, each link choice is a separate
binary choice \citep{Leung_2015}.
\end{rem}

\section{\label{sec:estimation}Estimation}

We now turn our attention to estimating the parameter $\theta=(\theta'_{u},\theta'_{\epsilon})'$.
We propose a two-step estimation procedure, where we estimate conditional
link choice probabilities in the first step and estimate the parameter
$\theta$ in the second step \citep{Leung_2015}. The asymptotic analysis
of the estimator is complicated by the fact that link choices of an
individual are correlated due to the preference for friends in common.
We exploit the optimal link choices in Theorem \ref{thm:optG} to
investigate the link dependence and derive the asymptotic properties
of the estimator.

We start with the data generating process. We consider the scenario
in which a single large network is observed. In the asymptotic analysis,
we assume that the number of individuals in the network $n$ goes
to infinity. Because the network depends on $n$, we denote it by
$G_{n}=(G_{n,ij})$ hereafter. We assume that links in a network are
generated as follows. First, we draw a vector of characteristics $X=(X_{1}^{\prime},\ldots X_{n}^{\prime})^{\prime}$
from a joint discrete distribution, where $X_{i}$ represents the
observed characteristics of individual $i$. Because $X$ is ancillary,
we treat it as deterministic. Note that $X_{i}$ can be dependent
across $i$. Next, we draw an $(n-1)\times1$ vector of unobserved
preferences $\epsilon_{i}\in\mathbb{R}^{n-1}$ for each $i$, independently
across $i$. After that, each individual chooses to form links, and
an equilibrium (a fixed point of (\ref{eq:BNE})) emerges. There can
be multiple equilibria, and nature selects one equilibrium $\sigma_{n}$
among the equilibria. The network $G_{n}$ observed in the data is
obtained from the optimal links chosen under $\sigma_{n}$.

\paragraph*{First step.}

The optimal link choices in (\ref{eq:gij}) depend on the equilibrium
$\sigma_{n}$ only through the conditional probabilities of forming
each link, denoted by $p_{n,ij}=\mathbb{E}[G_{n,ij}|X]$, $1\leq i\neq j\leq n$.
Moreover, the symmetry of the equilibrium (Assumption \ref{ass:sym})
implies that each $p_{n,ij}$ depends on $i$ and $j$ only through
their types $X_{i}$ and $X_{j}$. Under Assumption \ref{ass:discX},
it is thus sufficient to consider the type-specific conditional link
probabilities $p_{n,(st)}=\mathbb{E}[G_{n,ij}|X_{i}=x_{s},X_{j}=x_{t},X]$
for $1\leq s,t\leq T$. Denote $p_{n}=(p_{n,(st)},1\leq s,t\leq T)'$.
This is the parameter we need to estimate in the first step.

Specifically, for each $1\leq s,t\leq T$, we estimate $p_{n,(st)}$
by the relative frequency of forming a link among the pairs of individuals
that are of types $x_{s}$ and $x_{t}$
\begin{equation}
\hat{p}_{n,(st)}=\frac{\sum_{i}\sum_{j\neq i}G_{n,ij}1\{X_{i}=x_{s},X_{j}=x_{t}\}}{\sum_{i}\sum_{j\neq i}1\{X_{i}=x_{s},X_{j}=x_{t}\}}.\label{eq:p_hat}
\end{equation}
Let $\hat{p}_{n}=(\hat{p}_{n,(st)},1\leq s,t\leq T)'$ denote the
first-step estimator.

\paragraph*{Second step.}

To estimate $\theta_{0}$, we define $P_{n,ij}(\theta,p)=\mathbb{E}[G_{n,ij}(\epsilon_{i},\theta,p)|X]$
as the model-implied probability that $i$ forms a link to $j$, where
$G_{n,ij}(\epsilon_{i},\theta,p)$ represents the optimal link choice
in (\ref{eq:gij}) given $\theta$ and $p$. The equilibrium condition
in (\ref{eq:BNE}) yields a set of conditional moment restrictions
\begin{equation}
\mathbb{E}[G_{n,ij}-P_{n,ij}(\theta_{0},p_{n})|X]=0,\label{eq:m_cond}
\end{equation}
where each value of $(X_{i},X_{j})$ gives one moment restriction.
Based on (\ref{eq:m_cond}), we can construct a GMM estimator for
$\theta_{0}$. Let $q_{n,ij}=q_{n}(X_{i},X_{j})$ denote a $d_{\theta}\times1$
vector of instruments that can depend on $X$ as well as $\theta_{0}$
and $p_{n}$, and $\hat{q}_{n,ij}$ denote an estimator of $q_{n,ij}$.
Define
\begin{equation}
\hat{m}_{n}(\theta,\hat{p}_{n})=\frac{1}{n(n-1)}\sum_{i}\sum_{j\neq i}\hat{q}_{n,ij}(G_{n,ij}-P_{n,ij}(\theta,\hat{p}_{n}))\label{eq:m_hat}
\end{equation}
to be the sample moment, where $\hat{p}_{n}$ is the first-step estimator.
The minimizer of the function $\hat{m}_{n}(\theta,\hat{p}_{n})'\hat{m}_{n}(\theta,\hat{p}_{n})$
gives a GMM estimator $\hat{\theta}_{n}$.\footnote{We formulate the instrument in a way so that the weighting matrix
is absorbed into the instrument. See \citet[Section 5.4]{Newey_McFadden_1994}
for justification of this general formulation.} Suppose that the estimator $\hat{\theta}_{n}$ satisfies $\hat{m}_{n}(\hat{\theta}_{n},\hat{p}_{n})=o_{p}(n^{-1})$.

\paragraph*{Asymptotic analysis.}

We now investigate the asymptotic properties of the estimator $\hat{\theta}_{n}$.
Given $X$, define the population counterpart of $\hat{m}_{n}(\theta,p)$
by
\begin{equation}
m_{n}(\theta,p)=\frac{1}{n(n-1)}\sum_{i}\sum_{j\neq i}q_{n,ij}(\mathbb{E}[G_{n,ij}|X]-P_{n,ij}(\theta,p)).\label{eq:m}
\end{equation}
Equation (\ref{eq:m_cond}) implies that $m_{n}(\theta_{0},p_{n})=0$.\footnote{The population moment $m_{n}(\theta,p)$ has the subscript $n$ because
the probability of forming a link depends on the network size.} Because $p_{n}$ is uniquely determined in the first step,\footnote{The first step can be characterized by the moment restrictions $\mathbb{E}[G_{n,ij}-p_{n,ij}|X]=0$;
or equivalently, $\mathbb{E}[G_{n,ij}-p_{n,(st)}|X_{i}=x_{s},X_{j}=x_{t},X]=0$
for all $1\leq s,t\leq T$, where $p_{n}$ is a unique solution.} we assume that $m_{n}(\theta,p_{n})=0$ has a unique solution at
$\theta_{0}$. Stacking the moments in the first and second steps
then uniquely identifies $\theta_{0}$ and $p_{n}$. With abuse of
notation, we write $(\theta,p)$ for $(\theta',p')'$.

With the addition of Assumption \ref{ass:theta.consist}, we show
that $(\hat{\theta}_{n},\hat{p}_{n})$ is consistent for $(\theta_{0},p_{n})$.
\begin{assumption}
\label{ass:theta.consist}(i) The parameter $\theta$ lies in a compact
set $\Theta\subseteq\mathbb{R}^{d_{\theta}}$. (ii) For any $\delta>0$,
there is $\xi>0$ such that for $n$ sufficiently large, $\|m_{n}(\theta,p_{n})\|>\xi$
for all $\|\theta-\theta_{0}\|>\delta$. (iii) The instrument $q_{n,ij}$
and its estimator $\hat{q}_{n,ij}$ satisfy $\max_{1\leq i,j\leq n}\max\{\Vert q_{n,ij}\Vert,\|\hat{q}_{n,ij}\|\}\leq C_{q}<\infty$
and $\max_{1\leq i,j\leq n}\Vert\hat{q}_{n,ij}-q_{n,ij}\Vert=o_{p}(1)$.
(iv) $\lim\inf_{n\rightarrow\infty}\frac{1}{n(n-1)}\sum_{i}\sum_{j\neq i}1\{X_{i}=x_{s},X_{j}=x_{t}\}>0$
for all $1\leq s,t\leq T$.
\end{assumption}
Assumption \ref{ass:theta.consist}(i) is a standard regularity condition.
Assumption \ref{ass:theta.consist}(ii) is the identification condition
previously discussed. This assumption requires that $\mathbb{E}[G_{n,ij}-P_{n,ij}(\theta,p_{n})|X]=0$
has a unique solution at $\theta_{0}$. Equivalently, we can consider
the model-implied type-specific link choice probabilities $P_{n,(st)}(\theta,p)=\mathbb{E}[G_{n,ij}(\epsilon_{i},\theta,p)|X_{i}=x_{s},X_{j}=x_{t},X]$,
$1\leq s,t\leq T$. The assumption requires that for any $\theta\neq\theta_{0}$,
there exist $1\leq s,t\leq T$ such that $P_{n,(st)}(\theta,p_{n})\neq P_{n,(st)}(\theta_{0},p_{n})$.
When there is no effect from friends in common ($\gamma_{1},\gamma_{2}=0$),
this assumption reduces to a standard rank condition that the regressors
in (\ref{eq:Eu}) evaluated at $p_{n}$ are linearly independent \citep{Leung_2015}.
Moreover, note that the auxiliary term $\Phi_{ni}\Lambda_{ni}\omega_{ni}(\epsilon_{i})$
in (\ref{eq:gij}) can be viewed as a solution to the first-order
condition in (\ref{eq:foc.maxmin}) (multiplied by $\Phi_{ni}$).
The assumption requires that the solution to this first-order condition
under $\theta\neq\theta_{0}$ must differ from the solution under
$\theta_{0}$, so that $\gamma_{1}$ and $\gamma_{2}$ can be identified.\footnote{A necessary condition is that the two network statistic terms in (\ref{eq:Ev})
must be linearly independent, so that $V_{ni}$ does not remain the
same for different values of $\gamma_{1}$ and $\gamma_{2}$.} Assumption \ref{ass:theta.consist}(iii) is a standard assumption
that the instrument is bounded and its estimator is consistent, both
uniformly over $i$ and $j$. This assumption ensures that estimating
the instrument has no impact on the asymptotic distribution of $\hat{\theta}_{n}$,
as the terms involving the estimation error $\hat{q}_{n,ij}-q_{n,ij}$
are of a smaller order compared to those involving the true instrument
$q_{n,ij}$. Assumption \ref{ass:theta.consist}(iv) imposes a mild
restriction on $X$. It requires that the fractions of pairs of each
type remain positive as $n\rightarrow\infty$, so that the numbers
of pairs of each type grow without bounds, and we can identify and
estimate each $p_{n,(st)}$. If $X_{i}$ is i.i.d. or has limited
dependence across $i$ such that $\frac{1}{n(n-1)}\sum_{i}\sum_{j\neq i}1\{X_{i}=x_{s},X_{j}=x_{t}\}$
converges to $\Pr(X_{i}=x_{s},X_{j}=x_{t})$ almost surely, and if
$\Pr(X_{i}=x_{s},X_{j}=x_{t})>0$ for all $1\leq s,t\leq T$, then
Assumption \ref{ass:theta.consist}(iv) holds for almost every realization
of $X$.
\begin{thm}[Consistency]
\label{thm:theta.consist}Suppose that Assumptions \ref{ass:e=000026x}--\ref{ass:theta.consist}
are satisfied. Conditional on $X$, we have $\hat{\theta}_{n}-\theta_{0}=o_{p}(1)$
and $\hat{p}_{n}-p_{n}=o_{p}(1)$.
\end{thm}
\begin{proof}
See Appendix \ref{app:proof.estimation}.
\end{proof}
The consistency is established as a result of the fact that given
$X$ links formed by different individuals are independent, although
links formed by the same individual are correlated. The conditional
independence allows us to establish a uniform LLN for the stacked
sample moment, which together with the identification condition yields
consistency.

Analyzing the asymptotic distribution of $\hat{\theta}_{n}$ is more
complicated because links formed by an individual are correlated.
Theorem \ref{thm:optG} shows that link choices $G_{n,ij}$ and $G_{n,ik}$
are correlated because they both depend on the auxiliary variable
$\omega_{ni}(\epsilon_{i})$, which is a maximin solution of the function
\begin{equation}
\Pi_{ni}(\omega,\epsilon_{i})=\frac{1}{n-1}\sum_{j\neq i}\left[U_{n,ij}+\frac{n-1}{n-2}Z_{j}^{\prime}\Phi_{ni}\Lambda_{ni}\omega-\varepsilon_{ij}\right]_{+}-\frac{n-1}{2(n-2)}\omega^{\prime}\Lambda_{ni}\omega.\label{eq:Pai}
\end{equation}
In the expression, we add subscript $n$ to $\omega_{ni}$, $\Pi_{ni}$,
$U_{n,ij}$ and $V_{ni}$ to indicate their dependence on $n$, and
all of the terms are evaluated at $(\theta_{0},p_{n})$, abbreviated
for simplicity. To investigate how the link dependence will affect
the asymptotic distribution of $\hat{\theta}_{n}$, we represent $\omega_{ni}(\epsilon_{i})$
in an asymptotically linear form. Specifically, let $\Pi_{ni}^{\ast}(\omega)$
denote the population counterpart of $\Pi_{ni}(\omega,\epsilon_{i})$
given $X$
\begin{equation}
\Pi_{ni}^{\ast}(\omega)=\frac{1}{n-1}\sum_{j\neq i}\mathbb{E}\left[\left.\left[U_{n,ij}+\frac{n-1}{n-2}Z_{j}^{\prime}\Phi_{ni}\Lambda_{ni}\omega-\varepsilon_{ij}\right]_{+}\right\vert X\right]-\frac{n-1}{2(n-2)}\omega^{\prime}\Lambda_{ni}\omega,\label{eq:Pai*}
\end{equation}
and $\omega_{ni}^{\ast}$ denote a maximin solution of $\Pi_{ni}^{\ast}(\omega)$.
Under the regularity conditions in Assumption \ref{ass:w}, we show
in Lemma \ref{lem:w.asymlin} that $\omega_{ni}(\epsilon_{i})$ has
an asymptotically linear representation
\begin{equation}
\Lambda_{ni}(\omega_{ni}(\epsilon_{i})-\omega_{ni}^{\ast})=\frac{1}{n-1}\sum_{j\neq i}\phi_{n,ij}^{\omega}(\omega_{ni}^{\ast},\epsilon_{ij})+o_{p}(n^{-1/2}),\label{eq:Vw.alr}
\end{equation}
where $\phi_{n,ij}^{\omega}(\omega_{ni}^{\ast},\epsilon_{ij})$ is
an influence function defined in the lemma. Observe that $\omega_{ni}^{\ast}$
is deterministic, so link choices evaluated at $\omega_{ni}^{\ast}$
are independent. The representation indicates that the link dependence
due to $\omega_{ni}(\epsilon_{i})$ vanishes at the rate of $n^{-1/2}$,
which is crucial in determining the asymptotic distribution of $\hat{\theta}_{n}$.

With the addition of Assumptions \ref{ass:theta.asymdist} and \ref{ass:w},
we show that $\hat{\theta}_{n}$ is asymptotically normal.
\begin{assumption}
\label{ass:theta.asymdist}(i) For $n$ sufficiently large, $P_{n,ij}(\theta,p)$
is continuously differentiable with respect to $\theta$ and $p$
in a neighborhood of $(\theta_{0},p_{n})$, $1\leq i\neq j\leq n$.
(ii) For $n$ sufficiently large, the $d_{\theta}\times d_{\theta}$
matrix $J_{n}=\frac{1}{n(n-1)}\sum_{i}\sum_{j\neq i}q_{n,ij}\nabla_{\theta^{\prime}}P_{n,ij}(\theta_{0},p_{n})$
is nonsingular.
\end{assumption}
Assumption \ref{ass:theta.asymdist}(i) imposes a smoothness restriction
on $P_{n,ij}(\theta,p)$. We show in Lemma \ref{lem:ccp.cont.esti}
that $P_{n,ij}(\theta,p)$ is continuous in $\theta$ and $p$, by
exploiting the fact that there is a one-to-one mapping between the
optimal link choices of an individual and a partition of the $\epsilon_{i}$
space $\mathbb{R}^{n-1}$ (Online Appendix \ref{online:partition}),
where the function that defines the boundary of each set in the partition
is continuous in $\theta$ and $p$. The proof suggests that $P_{n,ij}(\theta,p)$
can have kinks if the binding inequalities that define the partition
vary with $\theta$ and $p$. This assumption requires that there
is a neighborhood of $(\theta_{0},p_{n})$ that has no kinks. In fact,
we show in Proposition \ref{prop:CCP.limit.consist} that $P_{n,ij}(\theta,p)$
converges (pointwise in $\theta$ and $p$) to a limit as $n\rightarrow\infty$,
which is continuously differentiable in $\theta$ and $p$. Therefore,
the assumption is less of a concern for larger $n$. Assumption \ref{ass:theta.asymdist}(ii)
is a standard regularity condition for $\hat{\theta}_{n}$ to have
a well-behaved asymptotic distribution. It also ensures that $(\theta_{0},p_{n})$
is locally identified in a small neighborhood of $(\theta_{0},p_{n})$.
Assumption \ref{ass:w} imposes additional regularity conditions on
the auxiliary variable $\omega$ so that we can derive the asymptotically
linear representation in (\ref{eq:Vw.alr}) as well as other needed
asymptotic properties of $\omega_{ni}(\epsilon_{i})$. 
\begin{thm}[Asymptotic Distribution]
\label{thm:theta.asymdist}Suppose that Assumptions \ref{ass:e=000026x}--\ref{ass:theta.asymdist}
and \ref{ass:w} are satisfied. Conditional on $X$, we have $\sqrt{n(n-1)}\Sigma_{n}^{-1/2}J_{n}(\hat{\theta}_{n}-\theta_{0})\overset{d}{\rightarrow}N(0,I_{d_{\theta}})$,
where $I_{d_{\theta}}$ is the $d_{\theta}\times d_{\theta}$ identity
matrix, $\Sigma_{n}=\frac{1}{n(n-1)}\sum_{i}\sum_{j\neq i}\mathbb{E}[\phi_{n,ij}^{\theta}\phi_{n,ij}^{\theta\prime}|X]$,
and $\phi_{n,ij}^{\theta}$ is a $d_{\theta}\times1$ vector defined
by (\ref{eq:theta.influence}) in the proof.
\end{thm}
\begin{proof}
See Appendix \ref{app:proof.estimation}.
\end{proof}
We derive the asymptotic distribution by decomposing the sample moment
into two leading terms, corresponding to the two components in the
influence function $\phi_{n,ij}^{\theta}$. The first captures the
sampling variation in link choices that does not account for the link
dependence due to $\omega_{ni}(\epsilon_{i})$. The second captures
the contribution of the link dependence to the asymptotic distribution.
Note that $\hat{\theta}_{n}$ converges to $\theta_{0}$ at the rate
of $n$ (the square root of the sample size). The link dependence
vanishes sufficiently fast so that it does not slow down the rate
at which $\hat{\theta}_{n}$ converges, but increases its asymptotic
variance.

The asymptotic variance of $\hat{\theta}_{n}$ can be calculated as
$\frac{1}{n(n-1)}J_{n}^{-1}\Sigma_{n}(J'_{n})^{-1}$. We can estimate
the asymptotic variance consistently by a plug-in estimator, where
we replace $\theta_{0}$ and $p_{n}$ by their estimators $\hat{\theta}_{n}$
and $\hat{p}_{n}$. In Online Appendix \ref{online:implementation},
we provide detailed guidance on how to estimate $\theta$ and compute
the standard errors in practice. 
\begin{rem}
\label{rem:P_sim}For each $\theta$ to be evaluated, the link choice
probability $P_{n,ij}(\theta,\hat{p}_{n})$ is an $n-1$ dimensional
integral that has no closed form and must be computed by simulation.
Specifically, we draw $\epsilon_{i}$ independently $R$ times, and
for each simulated $\epsilon_{i,r}$, $r=1,\ldots,R$, we compute
$\omega_{ni}(\epsilon_{i,r},\theta,\hat{p}_{n})$ and $G_{n,ij}(\epsilon_{i,r},\theta,\hat{p}_{n})$
in (\ref{eq:gij}). The sample average of the simulated $G_{n,ij}(\epsilon_{i,r},\theta,\hat{p}_{n})$
gives a simulated link choice probability (Online Appendix \ref{online:implementation}).
The simulation does not affect the consistency and asymptotic normality
of the estimator, but increases the asymptotic variance by $1+R^{-1}$
fold \citep{Pakes_Pollard_1989}.
\end{rem}

\paragraph*{Instrument.}

In practice, we need to choose an instrument. We suggest using the
instrument derived from quasi-maximum likelihood estimation (QMLE).\footnote{This is also the optimal instrument given the conditional moment restrictions
in (\ref{eq:m_cond}) \citep{Chamberlain_1987}.} Let $\mathcal{L}_{n}(\theta,\hat{p}_{n})$ denote the log of the
quasi-likelihood function evaluated at the first-step estimator $\hat{p}_{n}$\footnote{The quasi-likelihood function does not take into account the joint
distribution of link choices $G_{n,ij}$ and $G_{n,ik}$, which can
be informative about $\theta$.}
\begin{equation}
\mathcal{L}_{n}(\theta,\hat{p}_{n})=\sum_{i}\sum_{j\neq i}G_{n,ij}\ln P_{n,ij}(\theta,\hat{p}_{n})+(1-G_{n,ij})\ln(1-P_{n,ij}(\theta,\hat{p}_{n})).\label{eq:logL}
\end{equation}
Taking the derivative with respect to $\theta$, we obtain the quasi-likelihood
equation
\[
\frac{1}{n(n-1)}\sum_{i}\sum_{j\neq i}\frac{\nabla_{\theta}P_{n,ij}(\theta,\hat{p}_{n})}{P_{n,ij}(\theta,\hat{p}_{n})(1-P_{n,ij}(\theta,\hat{p}_{n}))}(G_{n,ij}-P_{n,ij}(\theta,\hat{p}_{n}))=0.
\]
Comparing this equation with the sample moment in (\ref{eq:m_hat})
suggests the instrument
\begin{equation}
\hat{q}_{n,ij}(\theta)=\frac{\nabla_{\theta}P_{n,ij}(\theta,\hat{p}_{n})}{P_{n,ij}(\theta,\hat{p}_{n})(1-P_{n,ij}(\theta,\hat{p}_{n}))}.\label{eq:instrument}
\end{equation}
Note that the instrument depends on $\theta$. We can either construct
a preliminary estimator of $\theta$ using an initial instrument\footnote{For example, we can use powers and interactions of $X_{i}$ and $X_{j}$
to construct an initial instrument.} or use continuous updating, as in \citet{Hansen_1996}.\footnote{Using instrument (\ref{eq:instrument}) with continuous mapping is
equivalent to QMLE based on (\ref{eq:logL}). However, GMM provides
more flexibility in choosing the instrument, especially when the link
choice probability $P_{n,ij}(\theta,p)$ is not fully differentiable
in $\theta$ and $p$. See Section \ref{sec:limiting} for more discussions.}

\section{\label{sec:extension}Extensions}

\subsection{\label{sec:undirected}Undirected Networks}

In some applications, the networks observed by researchers are undirected,
such as friendship and coauthorship networks \citep{Jackson_2008}.
In this section, we demonstrate that the approach developed in Section
\ref{sec:legendre} can be extended to undirected networks. However,
because the underlying decisions that induce undirected links are
not observed, the identification and estimation of parameters become
more challenging.

Let $G_{ij}$ denote an undirected link between individuals $i$ and
$j$, and thus $G_{ij}=G_{ji}$. We assume that individuals propose
the links they wish to form as in a link announcement game \citep{Myerson_1991}.
The undirected link $G_{ij}$ is formed if both $i$ and $j$ propose
to form it. Specifically, let $D_{ij}$ indicate whether $i$ proposes
a link to $j$. We have $G_{ij}=D_{ij}D_{ji}$. 

We adapt the utility specifications in (\ref{eq:U.sep}) and (\ref{eq:U.nonsep})
to accommodate undirected links. By removing the reciprocity effect
and replacing the spillover effects of directed links with their undirected
counterparts, we specify
\[
u_{ij}(G_{-i},X;\beta)=\beta_{1}+X_{i}^{\prime}\beta_{2}+d(X_{i},X_{j})'\beta_{3}+\frac{1}{n-2}\sum_{k\neq i,j}G_{jk}\beta_{4}
\]
and 
\[
v_{i,jk}(G_{-i},X;\gamma)=G_{jk}\gamma_{1}(X_{j},X_{k})+\frac{1}{n-3}\sum_{l\neq i,j,k}G_{jl}G_{kl}\gamma_{2}(X_{j},X_{k}).
\]
This utility specification is more general than that of \citet{Comola_Dekel_2023},
who also extend \citet{Leung_2015}'s approach to undirected networks.
\citet{Comola_Dekel_2023} maintain \citet{Leung_2015}'s assumption
of separable utility, which is more restrictive in the undirected
context because it excludes any triadic closure statistic.\footnote{For example, \citet{Leung_2015} allows for the supported trust $\frac{1}{n-2}\sum_{k\neq i,j}G_{ki}G_{kj}$
in $u_{ij}$. In an undirected setting, however, this statistic takes
the form $\frac{1}{n-2}\sum_{k\neq i,j}D_{ik}D_{ki}D_{jk}G_{kj}$,
which depends on $i$'s decision $D_{ik}$ and therefore violates
the separability assumption. } 

Because $G_{ij}=D_{ij}D_{ji}$, we write $G=G(D_{i},D_{-i})$, where
$D_{i}=(D_{ij},j\neq i)\in\mathcal{D}_{i}=\{0,1\}^{n-1}$ represents
the links proposed by individual $i$, and $D_{-i}=(D_{j},j\neq i)$
the links proposed by individuals other than $i$. By taking the expectation
with respect to $D_{-i}$, we can calculate the expected utility of
individual $i$ as follows:
\begin{align}
 & \mathbb{E}[U_{i}(G(D_{i},D_{-i}),X,\epsilon_{i})|X,\epsilon_{i}]\nonumber \\
= & \frac{1}{n-1}\sum_{j\neq i}D_{ij}\left(\mathbb{E}[D_{ji}u_{ij}(G_{-i},X)|X]+\frac{1}{2(n-2)}\sum_{k\neq i,j}D_{ik}\mathbb{E}[D_{ji}D_{ki}v_{i,jk}(G_{-i},X)|X]-\sigma_{ji}(X)\epsilon_{ij}\right),\label{eq:EU.undi}
\end{align}
where
\begin{equation}
\mathbb{E}[D_{ji}u_{ij}(G_{-i},X)|X]=\sigma_{ji}(X)(\beta_{1}+X_{i}^{\prime}\beta_{2}+d(X_{i},X_{j})'\beta_{3})+\frac{1}{n-2}\sum_{k\neq i,j}\sigma_{j,ik}(X)\sigma_{kj}(X)\beta_{4},\label{eq:Eu.undi}
\end{equation}
and
\begin{eqnarray}
\mathbb{E}[D_{ji}D_{ki}v_{i,jk}(G_{-i},X)|X] & = & \sigma_{j,ik}(X)\sigma_{k,ij}(X)\gamma_{1}(X_{j},X_{k})\nonumber \\
 &  & +\frac{1}{n-3}\sum_{l\neq i,j,k}\sigma_{j,il}(X)\sigma_{k,il}(X)\sigma_{l,jk}(X)\gamma_{2}(X_{j},X_{k}).\label{eq:Ev.undi}
\end{eqnarray}
In these expressions, we denote $\sigma_{ij}(X)=\mathbb{E}[D_{ij}|X]$
and $\sigma_{i,jk}(X)=\mathbb{E}[D_{ij}D_{ik}|X]$. Equations (\ref{eq:Eu.undi})
and (\ref{eq:Ev.undi}) hold because, conditional on $X$, the proposals
$D_{j}$ and $D_{k}$ are independent. Note that (\ref{eq:Eu.undi})
and (\ref{eq:Ev.undi}) involve the probability of an individual proposing
two links (e.g., $\sigma_{j,ik}(X)$).

The expected utility in (\ref{eq:EU.undi}) is similar to that in
(\ref{eq:EU}) when viewed as a function of proposals. Therefore,
we can apply the approach in Section \ref{sec:legendre} to derive
the optimal proposals. For $1\leq s,t\leq T$, let $V_{i,st}^{u}(X,\sigma)$
denote the value of $\mathbb{E}[D_{ji}D_{ki}v_{i,jk}(G_{-i},X)|X]$
if individuals $j$ and $k$ are of types $s$ and $t$, respectively;
that is, $V_{i,st}^{u}(X,\sigma)=\mathbb{E}[D_{ji}D_{ki}v_{i,jk}(G_{-i},X)|X_{j}=x_{s},X_{k}=x_{t},X,\sigma]$.
The superscript $u$ indicates an undirected network. Arrange the
$T^{2}$ type-specific expected incremental utilities $V_{i,st}^{u}(X,\sigma)$
in a $T\times T$ matrix $V_{i}^{u}(X,\sigma)=(V_{i,st}^{u}(X,\sigma))\in\mathbb{R}^{T\times T}$.
Because $V_{i,st}^{u}(X,\sigma)$ is symmetric in $s$ and $t$, $V_{i}^{u}(X,\sigma)$
is a symmetric matrix. Hence, it has a real spectral decomposition
\[
V_{i}^{u}(X,\sigma)=\Phi_{i}^{u}(X,\sigma)\Lambda_{i}^{u}(X,\sigma)\Phi_{i}^{u\prime}(X,\sigma),
\]
where $\Lambda_{i}^{u}(X,\sigma)=\text{diag}(\lambda_{i1}^{u}(X,\sigma),\ldots,\lambda_{iT}^{u}(X,\sigma))$
denotes the $T\times T$ diagonal matrix of eigenvalues in $\mathbb{R}$,
and $\Phi_{i}^{u}(X,\sigma)=(\phi_{i1}^{u}(X,\sigma),\ldots,\phi_{iT}^{u}(X,\sigma))$
denotes the $T\times T$ orthogonal matrix of eigenvectors in $\mathbb{R}^{T}$.
Define $U_{ij}^{u}(X,\sigma)=\mathbb{E}[D_{ji}u_{ij}(G_{-i},X)|X]-\frac{1}{2(n-2)}Z_{j}^{\prime}V_{i}^{u}(X,\sigma)Z_{j}$.

Following Theorem \ref{thm:optG}, we derive the optimal proposals
in Corollary \ref{cor:optD}.
\begin{cor}
\label{cor:optD}Suppose that Assumptions \ref{ass:e=000026x}-\ref{ass:discX}
are satisfied. For each $i$, the optimal proposals $D_{i}(\epsilon_{i},X,\sigma)=(D_{ij}(\epsilon_{i},X,\sigma),j\neq i)\in\mathcal{D}_{i}$
are given by
\begin{equation}
D_{ij}(\epsilon_{i},X,\sigma)=1\{U_{ij}^{u}(X,\sigma)+\frac{n-1}{n-2}Z_{j}^{\prime}\Phi_{i}^{u}(X,\sigma)\Lambda_{i}^{u}(X,\sigma)\omega_{i}^{u}(\epsilon_{i},X,\sigma)\geq\sigma_{ji}\epsilon_{ij}\},\forall j\neq i,\label{eq:dij}
\end{equation}
almost surely. The $T\times1$ vector $\omega_{i}^{u}(\epsilon_{i},X,\sigma)\in\mathbb{R}^{T}$
is an optimal solution to the maximin problem

\begin{align*}
\max_{\omega_{t},t\in\mathcal{T}_{i+}}\min_{\omega_{t},t\in\mathcal{T}_{i-}} & \frac{1}{n-1}\sum_{j\neq i}\left[U_{ij}^{u}(X,\sigma)+\frac{n-1}{n-2}Z_{j}^{\prime}\Phi_{i}^{u}(X,\sigma)\Lambda_{i}^{u}(X,\sigma)\omega-\sigma_{ji}\epsilon_{ij}\right]_{+}-\frac{n-1}{2(n-2)}\omega^{\prime}\Lambda_{i}^{u}(X,\sigma)\omega
\end{align*}
where $\mathcal{T}_{i+}=\{1\leq t\leq T:\lambda_{it}^{u}(X,\sigma)>0\}$
and $\mathcal{T}_{i-}=\{1\leq t\leq T:\lambda_{it}^{u}(X,\sigma)<0\}$.
We set $\omega_{it}^{u}(\epsilon_{i},X,\sigma)=0$ if $\lambda_{it}^{u}(X,\sigma)=0$.
Moreover, both $D_{i}(\epsilon_{i},X,\sigma)$ and $\omega_{i}^{u}(\epsilon_{i},X,\sigma)$
are unique almost surely.
\end{cor}
Corollary \ref{cor:optD} shows that the optimal proposals can be
expressed as binary choices, with the addition of an auxillary variable
$\omega_{i}^{u}(\epsilon_{i},X,\sigma)$, which serves the same role
as $\omega_{i}(\varepsilon_{i},X,\sigma)$ in directed networks. We
anticipate that proposals in an undirected network exhibit a dependence
structure analogous to that of links in a directed network. However,
since we observe links rather than proposals, the estimation method
must be adapted. In Online Appendix \ref{online:undi.estimation},
we discuss how to extend the estimation procedure in Section \ref{sec:estimation}
to undirected networks. A complete econometric analysis for undirected
networks is left for future research.

\subsection{\label{sec:limiting}Limiting Approximation}

In this section, we demonstrate that under certain conditions, a
link choice probability in the finite-$n$ game converges to a limit
as $n\rightarrow\infty$. In contrast to its finite-$n$ counterpart,
the limiting link probability is continuously differentiable in the
parameters and can be calculated analytically. It provides a useful
approximation to facilitate the estimation and computation of the
parameters.

Given a characteristic profile $X$ and equilibrium $p=(p_{(st)},1\leq s,t\leq T)'$,
recall that the probability that individual $i$ forms a link to individual
$j$ is given by
\begin{equation}
P_{n,ij}(X,p)=\Pr\left(\left.U_{n,ij}(X,p)+\frac{n-1}{n-2}Z_{j}^{\prime}\Phi_{ni}(X,p)\Lambda_{ni}(X,p)\omega_{ni}(\epsilon_{i},X,p)\geq\epsilon_{ij}\right\vert X\right),\label{eq:CCP.n}
\end{equation}
where the auxiliary variable $\omega_{ni}(\epsilon_{i},X,p)$ is a
maximin solution of the objective function $\Pi_{ni}(\omega,\epsilon_{i},X,p)$
in problem (\ref{eq:maxmin}). Suppose that $U_{n,ij}(X,p)$ and $V_{ni}(X,p)$
converge to some limites $U^{*}(X_{i},X_{j},p)$ and $V^{*}(X_{i},p)$
as $n\rightarrow\infty$, and $V^{*}(X_{i},p)$ has a real spectral
decomposition $V^{*}(X_{i},p)=\Phi^{*}(X_{i},p)\Lambda^{*}(X_{i},p)\Phi^{*\prime}(X_{i},p)$,
where $\Lambda^{*}(X_{i},p)=\text{diag}(\lambda_{1}^{*}(X_{i},p),\dots,\lambda_{T}^{*}(X_{i},p))$.
Let $\omega^{*}(X_{i},p)\in\mathbb{R}^{T}$ denote an optimal solution
to the maximin problem
\begin{equation}
\max_{\omega_{t},t\in\mathcal{T}_{+}}\min_{\omega_{t},t\in\mathcal{T}_{-}}\mathbb{E}[[U^{*}(X_{i},X_{j},p)+Z_{j}^{\prime}\Phi^{*}(X_{i},p)\Lambda^{*}(X_{i},p)\omega-\varepsilon_{ij}]_{+}|X_{i}]-\frac{1}{2}\omega^{\prime}\Lambda^{*}(X_{i},p)\omega,\label{eq:maxmin.lim}
\end{equation}
where $\mathcal{T}_{+}=\{1\leq t\le T:\lambda_{t}^{*}(X_{i},p)>0\}$
and $\mathcal{T}_{-}=\{1\leq t\leq T:\lambda_{t}^{*}(X_{i},p)>0\}$.
We set $\omega_{t}^{*}(X_{i},p)=0$ if $\lambda_{t}^{*}(X_{i},p)=0$.
Let $\Pi^{*}(\omega,X_{i},p)$ denote the objective function in (\ref{eq:maxmin.lim}),
where we condition on $X_{i}$ and take expectation with respect to
$X_{j}$ and $\epsilon_{ij}$, $j\neq i$. We can regard problem (\ref{eq:maxmin.lim})
as the limiting counterpart of problem (\ref{eq:maxmin}) and show
that the finite-$n$ solution $\omega_{ni}(\epsilon_{i},X,p)$ converges
to the the limiting solution $\omega^{*}(X_{i},p)$ as a result. From
these results, we can derive that the finite-$n$ link probability
$P_{n,ij}(X,p)$ converges to a limit defined by
\begin{equation}
P^{*}(X_{i},X_{j},p)=\Pr(U^{*}(X_{i},X_{j},p)+Z_{j}^{\prime}\Phi^{*}(X_{i},p)\Lambda^{*}(X_{i},p)\omega^{*}(X_{i},p)\geq\epsilon_{ij}|X_{i},X_{j}).\label{eq:CCP.lim}
\end{equation}
We refer to $P^{*}(X_{i},X_{j},p)$ as the limiting link probability.

To formally establish the convergence result, we impose the following
assumptions.
\begin{assumption}
\label{ass:lim}(i) The auxiliary variable $\omega$ lies in a compact
set $\Omega\subseteq\mathbb{R}^{T}$. (ii) For any $p$, any $\omega^{*}(X_{i},p)$
that solves problem (\ref{eq:maxmin.lim}) yields a unique $\Lambda^{*}(X_{i},p)\omega^{*}(X_{i},p)$.
(iii) $X_{i}$ is i.i.d. across $i$. (iv) For any $p$, there exist
$U^{*}(X_{i},X_{j},p)\in\mathbb{R}$ and $V^{*}(X_{i},p)\in\mathbb{R}^{T\times T}$
such that $\max_{j\neq i}|U_{n,ij}(X,p)-U^{*}(X_{i},X_{j},p)|=o_{p}(1)$
conditional on $X_{i}$ and $X_{j}$, and $\|V_{ni}(X,p)-V^{*}(X_{i},p)\|=o_{p}(1)$
conditional on $X_{i}$.
\end{assumption}
Because $\frac{\partial}{\partial c}\mathbb{E}[c-\epsilon]_{+}=\frac{\partial}{\partial c}\int_{-\infty}^{c}(c-\epsilon)f_{\epsilon}(\epsilon)d\epsilon=F_{\epsilon}(c)$,
(\ref{eq:maxmin.lim}) has the first-order condition
\begin{equation}
\Lambda^{*}(X_{i},p)\Phi^{*\prime}(X_{i},p)\mathbb{E}[Z_{j}F_{\epsilon}(U^{*}(X_{i},X_{j},p)+Z_{j}^{\prime}\Phi^{*}(X_{i},p)\Lambda^{*}(X_{i},p)\omega)|X_{i}]=\Lambda^{*}(X_{i},p)\omega.\label{eq:foc.maxmin.lim}
\end{equation}
Any solution to this first-order condition must be bounded. Therefore,
it is reasonable to assume that $\omega$ lies in a compact set $\Omega\subseteq\mathbb{R}^{T}$
as in Assumption \ref{ass:lim}(i).\footnote{This assumption resembles Assumption \ref{ass:w}(i), which is imposed
to derive the asymptotic properties of $\omega_{ni}(\epsilon_{i})$
conditional on $X$. } Assumption \ref{ass:lim}(ii) is an identification condition.\footnote{This assumption resembles Assumption \ref{ass:w}(ii) for fixed $X$.}
It is imposed to derive the consistency result for $\omega_{ni}(\epsilon_{i},X,p)$
in Lemma \ref{lem:wlim.consist}.\footnote{We require $\Lambda^{*}(X_{i},p)\omega^{*}(X_{i},p)$, rather than
$\omega^{*}(X_{i},p)$, to be unique because $V^{*}(X_{i},p)$ may
be singular, but it is $\Lambda^{*}(X_{i},p)\omega^{*}(X_{i},p)$
that affects the link choices.} Our analysis in the previous sections avoided making assumptions
about how $X_{i}$ are generated. In Assumption \ref{ass:lim}(iii),
we assume that $X_{i}$ is i.i.d. in order to establish the limiting
approximation. Assumption \ref{ass:lim}(iv) posits the convergence
of the expected utilities. In Example \ref{ex:U=000026V.consist},
we demonstrate that Assumption \ref{ass:lim}(iv) holds under Assumption
\ref{ass:lim}(iii) for the expected utility specified in (\ref{eq:Eu})--(\ref{eq:Ev}).
Intuitively, the spillover effects in (\ref{eq:Eu})--(\ref{eq:Ev})
take the form of sample averages over functions of $X_{k}$, $k\neq i,j$,
(e.g., $p(X_{k},X_{j})$). These averages converge to population means
when $X_{i}$ is i.i.d..
\begin{example}
\label{ex:U=000026V.consist}Consider the expected utility in (\ref{eq:Eu})--(\ref{eq:Ev}).
Fix $X_{i}$ and $X_{j}$. Let $p=(p(x_{s},x_{t}),1\leq s,t\leq T)'$
be an equilibrium. Define
\begin{eqnarray}
U^{*}(X_{i},X_{j},p) & = & \beta_{1}+X_{i}^{\prime}\beta_{2}+d(X_{i},X_{j})^{\prime}\beta_{3}+p(X_{j},X_{i})\beta_{4}\nonumber \\
 &  & +\mathbb{E}[p(X_{k},X_{j})|X_{j}]\beta_{5}+\mathbb{E}[p(X_{j},X_{k})|X_{j}]\beta_{6},\label{eq:Eu.lim}
\end{eqnarray}
where the expectations in (\ref{eq:Eu.lim}) are taken with respect
to $X_{k}$. Further, define the $T\times T$ matrix $V^{*}(p)=(V_{st}^{*}(p))$,
where the $st$th entry is given by
\begin{eqnarray}
V_{st}^{*}(p) & = & (p(x_{s},x_{t})+p(x_{t},x_{s}))\gamma_{1}(x_{s},x_{t})\nonumber \\
 &  & +\mathbb{E}[p(x_{s},X_{l})p(X_{l},x_{t})+p(x_{t},X_{l})p(X_{l},x_{s})]\gamma_{2}(x_{s},x_{t}).\label{eq:Ev.lim}
\end{eqnarray}
The expectation in (\ref{eq:Ev.lim}) is taken with respect to $X_{l}$.
We verify in Lemma \ref{lem:U=000026V.consist} that $U_{n,ij}(X,p)$
and $V_{ni}(X,p)$ converge to $U^{*}(X_{i},X_{j},p)$ and $V^{*}(p)$
respectively under Assumption \ref{ass:lim}(iii).
\end{example}
Proposition \ref{prop:CCP.limit.consist} shows that the finite-$n$
link probabilities converge in probability to the limiting link probabilities
as $n\rightarrow\infty$. 
\begin{prop}
\label{prop:CCP.limit.consist}Under Assumptions \ref{ass:e=000026x}--\ref{ass:discX}
and \ref{ass:lim}, for any equilibrium $p$, we have $P_{n,ij}(X,p)-P^{*}(X_{i},X_{j},p)=o_{p}(1)$
conditional on $X_{i}$ and $X_{j}$.
\end{prop}
\begin{proof}
See Appendix \ref{app:proof.extension}.
\end{proof}

We establish the result in Proposition \ref{prop:CCP.limit.consist}
by first noting that the finite-$n$ first-order condition in (\ref{eq:foc.maxmin})
takes the form of a sample average over functions of $X_{j}$, $j\neq i$.
Under the assumption of i.i.d. $X_{i}$ and converging expected utilities,
we can show that the finite-$n$ first-order condition converges to
the limiting first-order condition in (\ref{eq:foc.maxmin.lim}).
Consequently, the solution to the finite-$n$ first-order condition
also converges to the solution to the limiting counterpart. While
the finite-$n$ auxiliary variable $\omega_{ni}(\epsilon_{i},X,p)$
depends on both $\epsilon_{i}$ and the entire $X$, its limiting
counterpart $\omega^{*}(X_{i},p)$ depends on $X_{i}$ only. Therefore,
conditional on $X_{i}$, individual $i$'s link choices in the limit
become independent.\footnote{In equation (\ref{eq:CCP.lim}), the latent utility of forming a link
depends on the equilibrium $p$, indicating that strategic interactions
among link choices do not vanish in the limit. The presence of $\omega^{*}(X_{i},p)$
further suggests that, even in the limit, strategic interactions due
to the preference for friends in common persist. By incorporating
$\omega^{*}(X_{i},p)$ in the latent utility, we internalize the limiting
approximation of the spillover effects caused by this preference.} Our result aligns with the literature that employs large-market approximations
as a simplification for finite-$n$ markets, which are often challenging
to analyze due to complex equilibria.\footnote{For example, \citet{Menzel_2015} discovered the large-market approximation
for a one-to-one matching model under non-transferable utility. \citet{Azevedo2016}
established the convergence of equilibrium cutoffs for a many-to-one
matching model under non-transferable utility as the market size grows
large.} In our context, we derive the limiting approximation to simplify
the link dependence arising from the preference for friends in common,
thereby yielding simpler link choice probabilities.

\paragraph{Advantage of the limiting approximation.}

The two-step estimator proposed in Section \ref{sec:estimation} requires
an instrument in the second stage. We suggested using the instrument
derived from quasi-maximum likelihood (equation (\ref{eq:instrument}));
however, this instrument involves the derivative of a link choice
probability. Because the limiting auxiliary variable $\omega^{*}(X_{i},p)$
does not depend on $\epsilon_{i}$, the limiting link probability
$P^{*}(X_{i},X_{j},p)$ is continuously differentiable in the parameters.
Therefore, we can use the derivative of a limiting link probability
to construct the instrument, addressing the concern that finite-$n$
link probabilities may have kinks. Given that finite-$n$ and limiting
link probabilities are asymptotically close (Proposition \ref{prop:CCP.limit.consist}),
the instrument based on limiting link probabilities should achieve
asymptotic efficiency similar to that of the instrument based on finite-$n$
link probabilities.\footnote{Because we only approximate the instrument, the consistency of the
estimator remains unaffected.}

We can further simplify the moment condition by replacing the finite-$n$
link probabilities in the moment function with their limiting counterparts.
This approximation improves computational efficiency, as limiting
link probabilities can be computed without simulation. Although the
approximated moment function yields a misspecified model, the misspecification
vanishes asymptotically.\footnote{To analyze the asymptotic properties of such an estimator, we must
examine the extent to which the limiting link probabilities evaluated
at a finite-$n$ equilibrium differ from that equilibrium. In the
presence of multiple equilibria, additional assumptions ensuring the
convergence of a sequence of equilibrium selection mechanisms would
be needed to achieve the consistency of the estimator. This issue
is related to the convergence of equilibria explored in \citet{Menzel_2016}.}

\paragraph{Simulation evidence.}

Given the scope of this paper, we do not investigate the theoretical
properties of the limiting approximation. However, we provide simulation
evidence on its performance. In Online Appendix \ref{online:simulation},
we evaluate our approach in a simulation study, where limiting link
probabilities are used to approximate the instrument and/or the moment
function. The estimates that use limiting link probabilities for the
instrument (Table \ref{tab:simulation} Case (ii)) are similar to
those that use the finite-$n$ counterparts (Table \ref{tab:simulation}
Case (i)), although they are biased and have larger root MSEs in small
networks ($n\leq25$). The estimates that use limiting link probabilities
for both the moment function and the instrument (Table \ref{tab:simulation}
Case (iii)) are the most biased and have the largest root MSEs in
small networks, but once networks become moderately large ($n\geq100$),
they perform similarly to the other estimates -- remaining unbiased
with comparable root MSEs. These results suggest that limiting link
probabilities provide a useful approximation in sufficiently large
networks.

\section{\label{sec:empirical}Empirical Application}

\paragraph*{Data and setup.}

We apply our approach to investigate favor exchange networks in rural
India. The dataset was collected from 75 rural villages in southern
India as part of a study on a microfinance program (see \citet{Jackson_2012}
and \citet{Banerjee_2013} for detailed descriptions of the data).
Respondents in the survey were asked whether they provided monetary,
in-kind (kerorice), advisory, or medical help to -- or received such
favors from -- other individuals surveyed in the same village. Because
providing and receiving favors represent distinct decisions, we keep
the directed relationships and construct a directed network of favor
exchange in each village.

In particular, we say that individual $i$ lends money or kerorice
to individual $j$ if either $i$ reports lending money or kerorice
to $j$ or $j$ reports borrowing money or kerorice from $i$. Similarly,
we say that individual $i$ gives advice or medical help to individual
$j$ if either $i$ reports providing such help to $j$ or $j$ reports
receiving such help from $i$.\footnote{These directed relationships are constructed using the variables Borrow-money,
Lend-money, Borrow-kerorice, Lend-kerorice, Advice-come, Advice-go
and Medical-help in the data. For detailed descriptions of these variables,
see \citet{Jackson_2012}. } We say that individual $i$ does a favor for individual $j$ if $i$
lends money or kerorice, or gives advice or medical help to $j$.
This creates a directed link from $i$ to $j$ in a favor exchange
network.

Our empirical study is motivated by \citet{Jackson_2012}, who found
that the provision of a favor is supported by mutual relationships
with other individuals. We aim to provide further evidence on self-support
within a directed network of favor exchange. Given the intrinsic nature
of a favor, we assume that whoever receives a favor accepts it, so
that the presence of a favor is determined unilaterally by the provider.
Inspired by the findings of \citet{Jackson_2012}, we allow individual
$i$'s marginal utility from providing a favor to individual $j$
to depend on the support from the connections that $i$ and $j$ have
with another individual $k$. From individual $i$'s perspective,
her incentive to provide a favor to $j$ may differ depending on whether
she provides a favor to $k$ or receives a favor from $k$. Therefore,
we distinguish the supporting connections based on the direction of
the link $ik$. We define the \textit{inward support} for the link
$ij$ as $\frac{1}{n-2}\sum_{k\neq i,j}G_{ki}G_{kj}$, where $i$
receives a favor from a supporting individual $k$.\footnote{There are other possible variants of inward support, such as $\frac{1}{n-2}\sum_{k\neq i,j}G_{ki}(G_{jk}+G_{kj})$,
which aligns with our definition of outward support. However, we choose
$\frac{1}{n-2}\sum_{k\neq i,j}G_{ki}G_{kj}$, as it coincides with
the supported trust defined in \citet{Leung_2015} and thus facilitates
comparison.} In contrast, we define the \textit{outward support} for the link
$ij$ as $\frac{1}{n-2}\sum_{k\neq i,j}G_{ik}(G_{jk}+G_{kj})$, where
$i$ provides a favor to a supporting individual $k$.

Specifically, we consider the utility function in (\ref{eq:U}), where
the unobservable $\epsilon_{ij}$ is assumed to follow a logistic
distribution. Our specification of the separable utility $u_{ij}$
in (\ref{eq:U.sep}) includes the provider $i$'s characteristics
(gender, age, education, caste), homophily measures (same gender,
same age, same education, same caste), and spillover effects that
are separable in $i$'s links: reciprocity ($G_{ji}$), recipient's
in-degree ($\frac{1}{n-2}\sum_{k\neq i,j}G_{kj}$), recipient's out-degree
($\frac{1}{n-2}\sum_{k\neq i,j}G_{jk}$), and inward support ($\frac{1}{n-2}\sum_{k\neq i,j}G_{ki}G_{kj}$).
Our specification of the nonseparable utility includes outward support
($\frac{1}{n-2}\sum_{k\neq i,j}G_{ik}(G_{jk}+G_{kj})$), with $v_{i,jk}=(G_{jk}+G_{kj})\gamma_{1}$,
where $\gamma_{1}$ is constant.\footnote{We do no consider the second term in (\ref{eq:U.nonsep}) and set
$\gamma_{2}=0$.} While the spillover effects in $u_{ij}$ can be estimated using the
approach in \citet{Leung_2015}, estimating the effect of outward
support requires our approach.

Our approach requires discrete types. We discretize age into three
categories (under 29, 30--49 and over 50) and education into two
categories (below and above the median).\footnote{The median number of years of schooling in the dataset is 5.}
Castes are classified into three categories: scheduled (including
scheduled castes and scheduled tribes), other backward class (OBC),
and general. This discretization and categorization result in a type
space of $36$ types ($T=36$). 

To align with the asymptotic framework in the paper, we use only one
village from the dataset for our empirical analysis. Our sample consists
of $n=395$ individuals in the village. Among these individuals, there
are $n(n-1)=155,630$ potential directed links. 

\paragraph*{Estimation and inference.}

We estimate the utility parameters in two steps. In the first step,
we estimate the probability that individual $i$ provides a favor
to individual $j$ given the characteristics of $i$ and $j$. In
our sample, certain pair types are absent.\footnote{Our sample consists of 1,084 pair types. In fact, no village in the
dataset contains all 1,296 pair types.} Therefore, rather than using a frequency estimator, as discussed
in Section \ref{sec:estimation}, we use a series logit estimator
\citep{Hirano_2003}. Specifically, we run a logit regression of favor
provision on a second-order polynomial series of provider and recipient
characteristics. The predicted link choice probabilities for each
pair type yield our first-step estimates.  

In the second step, we estimate the utility parameters in $u_{ij}$
and $v_{i,jk}$ by GMM. We use the moment in (\ref{eq:m_hat}), with
the instrument given by (\ref{eq:instrument}).\footnote{In practice, we implement GMM by weighted nonlinear least squares
(NLS), where we use the optimal weight $1/(P_{n,ij}(1-P_{n,ij}))$
for link $G_{ij}$. The first-order condition of the weighted NLS
coincide with that of GMM, so the estimates should be equivalent.
An advantage of weighted NLS is that we can use the built-in command
in MATLAB $\textit{nlinfit}$ to calculate the estimates.} To reduce the computational burden, we approximate the finite-$n$
link choice probability $P_{n,ij}$ in (\ref{eq:m_hat}) using a variant
of the limiting approximation developed in Section \ref{sec:limiting}.
This approximation retains all terms from $P_{n,ij}$, except that
the auxiliary variable $\omega_{ni}(\epsilon_{i})$, a maximin solution
of (\ref{eq:Pai}), is replaced by its population counterpart $\omega_{ni}^{\ast}$,
a maximin solution of (\ref{eq:Pai*}). Unlike $\omega_{ni}(\epsilon_{i})$,
which must be calculated for each individual by simulation (Online
Appendix \ref{online:implementation}), $\omega_{ni}^{\ast}$ depends
on $i$ only through her type and does not involve unobservables.
Consequently, it can be calculated for each type without simulation.\footnote{On an 8-core CPU, a single evaluation of the approximated link probabilities
for all the 1296 pair types in our sample takes 0.03 seconds.} While $\omega_{ni}^{\ast}$ is a maximin solution, we compute it
by solving the first-order condition of (\ref{eq:Pai*}) using a standard
fixed-point algorithm.

In practice, to get an educated guess for the initial values of the
parameters, we first estimate the parameters approximately by logit,
where we use the right-hand side of (\ref{eq:w.interp})---calculated
based on the observed links---to approximate the auxiliary term on
the left-hand side. Given that the logit approximation runs quite
fast, it is also useful to researchers who want to experiment with
specifications.\footnote{For our specification with all spillover effects---the specification
that is most costly in computation---on an 8-core CPU, the logit
approximation takes 0.4 seconds, while the GMM estimation, implemented
by weighted NLS, takes 32.7 seconds.}

The standard errors are calculated using the asymptotic variance in
Theorem \ref{thm:theta.asymdist} with modifications. First, the use
of the population $\omega_{ni}^{\ast}$ implies that the links are
conditional independent. This property simplifies the asymptotic distribution
in Theorem \ref{thm:theta.asymdist}, leaving only the first term
in the influence function $\ensuremath{\phi_{n,ij}^{\theta}}$. Moreover,
recall that the first step is estimated by series logit. Following
\citet{Ackerberg2012}, we can account for the contribution of a nonparametric
first step to the asymptotic variance in the same way as in two-step
estimation with a parametric first step.\footnote{\citet{Ackerberg2012} established the numerical equivalence result
when the first step is estimated using sieves. This result can be
extended to series logit by applying the approach in \citet{Chernozhukov2021}.} 

\paragraph*{Results.}

\begin{table}
\centering \begin{threeparttable}\caption{\label{tab:emp.2s}Two-Step GMM Estimation of Favor Provision}

\begin{tabular}{lr@{\extracolsep{0pt}.}lr@{\extracolsep{0pt}.}lr@{\extracolsep{0pt}.}lr@{\extracolsep{0pt}.}l}
\toprule 
 & \multicolumn{2}{c}{(1)} & \multicolumn{2}{c}{(2)} & \multicolumn{2}{c}{(3)} & \multicolumn{2}{c}{(4)}\tabularnewline
\midrule 
Homophily effects & \multicolumn{2}{c}{} & \multicolumn{2}{c}{} & \multicolumn{2}{c}{} & \multicolumn{2}{c}{}\tabularnewline
\quad{}Same gender & 1&553 & 1&259 & 1&151 & 1&125\tabularnewline
 & (0&054) & (0&096) & (0&046) & (0&059)\tabularnewline
\quad{}Same age & 0&244 & 0&172 & 0&241 & 0&217\tabularnewline
 & (0&045) & (0&047) & (0&040) & (0&045)\tabularnewline
\quad{}Same education & 0&057 & 0&067 & 0&041 & 0&086\tabularnewline
 & (0&047) & (0&042) & (0&036) & (0&040)\tabularnewline
\quad{}Same caste & 1&738 & 1&396 & 1&281 & 1&232\tabularnewline
 & (0&049) & (0&107) & (0&045) & (0&064)\tabularnewline
Provider's characteristics & \multicolumn{2}{c}{} & \multicolumn{2}{c}{} & \multicolumn{2}{c}{} & \multicolumn{2}{c}{}\tabularnewline
\quad{}Female & -0&242 & -0&128 & -0&096 & -0&245\tabularnewline
 & (0&047) & (0&063) & (0&028) & (0&048)\tabularnewline
\quad{}Age 30--49 & 0&056 & 0&045 & -0&034 & -0&033\tabularnewline
 & (0&066) & (0&068) & (0&028) & (0&047)\tabularnewline
\quad{}Age 50+ & 0&217 & 0&203 & 0&074 & 0&143\tabularnewline
 & (0&077) & (0&077) & (0&029) & (0&051)\tabularnewline
\quad{}Education > Median & 0&045 & 0&056 & -0&016 & 0&019\tabularnewline
 & (0&055) & (0&056) & (0&024) & (0&043)\tabularnewline
\quad{}Scheduled & 0&265 & 0&048 & 0&181 & 0&007\tabularnewline
 & (0&067) & (0&073) & (0&044) & (0&073)\tabularnewline
\quad{}OBC & -0&240 & -0&215 & -0&161 & -0&117\tabularnewline
 & (0&054) & (0&064) & (0&031) & (0&039)\tabularnewline
Spillover effects & \multicolumn{2}{c}{} & \multicolumn{2}{c}{} & \multicolumn{2}{c}{} & \multicolumn{2}{c}{}\tabularnewline
\quad{}Reciprocity & \multicolumn{2}{c}{} & 3&717 & \multicolumn{2}{c}{} & 0&447\tabularnewline
 & \multicolumn{2}{c}{} & (3&141) & \multicolumn{2}{c}{} & (2&680)\tabularnewline
\quad{}Recipient's in-degree & \multicolumn{2}{c}{} & 89&603 & \multicolumn{2}{c}{} & 102&496\tabularnewline
 & \multicolumn{2}{c}{} & (21&315) & \multicolumn{2}{c}{} & (5&921)\tabularnewline
\quad{}Recipient's out-degree & \multicolumn{2}{c}{} & -57&518 & \multicolumn{2}{c}{} & -55&226\tabularnewline
 & \multicolumn{2}{c}{} & (13&235) & \multicolumn{2}{c}{} & (7&923)\tabularnewline
\quad{}Inward support & \multicolumn{2}{c}{} & 635&047 & \multicolumn{2}{c}{} & 368&108\tabularnewline
 & \multicolumn{2}{c}{} & (502&789) & \multicolumn{2}{c}{} & (211&205)\tabularnewline
\quad{}Outward support & \multicolumn{2}{c}{} & \multicolumn{2}{c}{} & 769&731 & 884&305\tabularnewline
 & \multicolumn{2}{c}{} & \multicolumn{2}{c}{} & (61&699) & (52&171)\tabularnewline
\midrule 
Observations & \multicolumn{2}{c}{155,630} & \multicolumn{2}{c}{155,630} & \multicolumn{2}{c}{155,630} & \multicolumn{2}{c}{155,630}\tabularnewline
\bottomrule
\end{tabular} 

\begin{tablenotes}[flushleft]\small \item 
Note: Standard errors are in parentheses. The dependent variable is an indicator for whether resident $i$ provides a favor to resident $j$. The moment function and instrument in the second step are constructed using limiting link probabilities.
\end{tablenotes}\end{threeparttable}
\end{table}
Table \ref{tab:emp.2s} presents the second-step GMM estimates and
their standard errors. We consider four specifications of spillover
effects. Column 1 assumes no spillover effects. Column 2 allows for
four separable spillover effects (reciprocity, recipient's in-degree,
recipient's out-degree, and inward support).\footnote{To estimate the effect of inward support, we need a first-step estimate
for $\mathbb{E}[G_{ki}G_{kj}|X]$, that is, the conditional probability
that individual $k$ forms a link with both $i$ and $j$. Under the
limiting approximation, the two links are conditional independent.
Therefore, we estimate $\mathbb{E}[G_{ki}G_{kj}|X]$ approximately
by the products of the estimated $\mathbb{E}[G_{ki}|X]$ and $\mathbb{E}[G_{kj}|X]$.} Column 3 allows for outward support only. Column 4 considers all
five spillover effects. In all the four specifications, we control
for homophily measures and the provider's characteristics. Across
specifications, we find that individuals sharing the same gender,
age, and caste are significantly more likely to exchange favors, with
caste and gender similarity having the greatest impacts. Moreover,
individuals are significantly more likely to provide a favor if they
are male, over the age of 50, and belong to a higher caste. These
findings are consistent with evidence documented in the literature
\citep{Jackson_2012}. Additionally, we observe that homophily effects
tend to be smaller in the specifications with spillover (Column 1
vs. Columns 2-4). This suggests that ignoring spillover effects and
estimating a dyadic model may overestimate homophily effects.

In addition to the dyadic factors, Table \ref{tab:emp.2s} provides
evidence of spillover effects. Columns 2 and 4 show a positive reciprocity
effect, suggesting that individuals are more willing to do favors
for those who also do favors for them, although this effect is insignificant.
Furthermore, individuals are significantly more likely to provide
favors to recipients with higher in-degrees and less likely to do
so for those with higher out-degrees. This indicates that providers
interpret these network metrics as signals of need: a high in-degree
implies greater reliance on others (and thus more need), while a high
out-degree reflects the capacity to help others (and thus less need).
In short, favors tend to flow toward those perceived as needing help
and away from those already helping others. 

More importantly, Table \ref{tab:emp.2s} highlights the distinct
effects of inward and outward support on favor provision. Inward support
shows a positive but insignificant effect (Columns 2 and 4), suggesting
that $i$'s decision to help $j$ is unaffected by receiving favors
from a third party $k$ connected to $j$. This finding is consistent
with the results of \citet{Leung_2015}. In contrast, outward support
has a positive and significant effect (Columns 3 and 4), indicating
that $i$ is more likely to help $j$ if $i$ provides favors to a
third party $k$ connected to $j$. While mutual connections with
a third party matter, as shown in \citet{Jackson_2012}, the direction
of these connections is crucial: receiving a favor from $k$ has no
impoact on $i$'s decision to help $j$, whereas providing a favor
to $k$ increases the likelihood of $i$ helping $j$. These findings
suggest that policies prioritizing outward support over inward support
are more effective in promoting favor exchange. For example, targeting
active providers with high out-degrees can amplify support and strengthen
favor provision throughout the network.

\paragraph*{Variance decomposition of predicted log odds ratio.}

In our framework, favor provision is influenced by three types of
factors: (i) dyadic attributes (homophily measures and provider's
characteristics), (ii) separable spillover (reciprocity, recipient's
degrees, and inward support), and (iii) nonseparable spillover (outward
support). Using the estimates in Column 4 of Table \ref{tab:emp.2s},
we calculate the predicted values of these components for each link.
The sum of the three gives the predicted log odds ratio of the link
($\log(\hat{P}_{n,ij}/(1-\hat{P}_{n,ij}))$).\footnote{We calculate the impact of dyadic attributes as $\hat{\beta}_{1}+X_{i}^{\prime}\hat{\beta}_{2}+d(X_{i},X_{j})^{\prime}\hat{\beta}_{3}$,
separable spillover as $\hat{p}_{ji}\hat{\beta}_{4}+\frac{1}{n-2}\sum_{k\neq i,j}\hat{p}_{kj}\hat{\beta}_{5}+\frac{1}{n-2}\sum_{k\neq i,j}\hat{p}_{jk}\hat{\beta}_{6}+\frac{1}{n-2}\sum_{k\neq i,j}\hat{p}_{ki}\hat{p}_{kj}\hat{\beta}_{7}-\frac{1}{2(n-2)}Z'_{j}\hat{V}_{ni}Z_{j}$,
and nonseparable spillover as $\frac{n-1}{n-2}Z'_{j}\hat{\Phi}_{ni}\hat{\Lambda}_{ni}\hat{\omega}_{ni}^{\ast}$.}

\begin{table}
\centering \begin{threeparttable}\caption{\label{tab:var_dcp}Variance Decomposition of Predicted Log Odds Ratio
of Favor Provision}

\begin{tabular}{l>{\centering}p{0.08\textwidth}>{\centering}p{0.08\textwidth}}
\hline 
 &  & Percentage\tabularnewline
 & (1) & (2)\tabularnewline
\hline 
Total variance of predicted log odds ratio & 1.205 & 100.0\%\tabularnewline
Variance of dyadic attributes & 0.698 & 57.9\%\tabularnewline
Variance of separable spillover & 0.024 & 2.0\%\tabularnewline
Variance of nonseparable spillover & 0.049 & 4.1\%\tabularnewline
2{*}Cov(dyadic attr., separable spil.) & 0.088 & 7.3\%\tabularnewline
2{*}Cov(dyadic attr., nonseparable spil.) & 0.324 & 26.9\%\tabularnewline
2{*}Cov(separable spil., nonseparable spil.) & 0.022 & 1.8\%\tabularnewline
\hline 
\end{tabular}

\begin{tablenotes}[flushleft]\small \item 
Note: Variance decomposition of the predicted log odds ratio of favor provision into three components: dyadic attributes, separable spillover, and nonseparable spillover. Dyadic attributes include homophily measures and provider's characteristics. Separable spillover includes reciprocity, recipient's in-degree and out-degree, and inward support. Nonseparable spillover includes outward support.
\end{tablenotes}\end{threeparttable}
\end{table}
Table \ref{tab:var_dcp} decomposes the total variance of the predicted
log odds ratio into the variances of dyadic attributes, separable
spillover, and nonseparable spillover, along with their covariances.
Dyadic attributes alone account for only 58\% of the total variance
in the predicted log odds ratio. Including separable spillover increases
this proportion to 67\%. However, 33\% of the total variance remains
unexplained without nonseparable spillover. These results highlight
the importance of accounting for nonseparable spillover.

\paragraph*{Predicting support distribution.}

Next we investigate our model's performance in predicting support
measures under different specifications of spillover effects. Using
the estimates in Column 4 of Table \ref{tab:emp.2s}, we simulate
three directed networks. In the first network, all spillover effects
are set to zero. In the second network, only the effect of outward
support is set to zero. In the third network, all spillover effects
are included.\footnote{We fix the first-step estimates when predicting a network under alternative
parameter values. The predicted links do not reflect the potential
change in the equilibrium.} The support measure introduced by \citet{Jackson_2012} is defined
at the network level for undirected networks. We adapt it to the individual
level for directed networks. Specifically, we calculate the support
measure of individual $i$ in network $G$ as
\[
\text{Supp}_{i}(G)=\frac{\sum_{j\neq i}G_{ij}\max_{k\neq i,j}((G_{ik}\lor G_{ki})\land(G_{jk}\lor G_{kj}))}{\sum_{j\neq i}G_{ij}},
\]
where $x\lor y=\max\{x,y\}$ and $x\land y=\min\{x,y\}$. A link $G_{ij}$
is supported in network $G$ if there exists a third party $k$ that
is connected (in any direction) to both $i$ and $j$. The support
measure of individual $i$ in network $G$ is calculated as the ratio
of the number of supported links $i$ forms to the total number of
links $i$ forms.

\begin{figure}[t]
\centering

\caption{\label{fig:Support}Support Distributions in Observed and Predicted
Networks}
\includegraphics[scale=0.7]{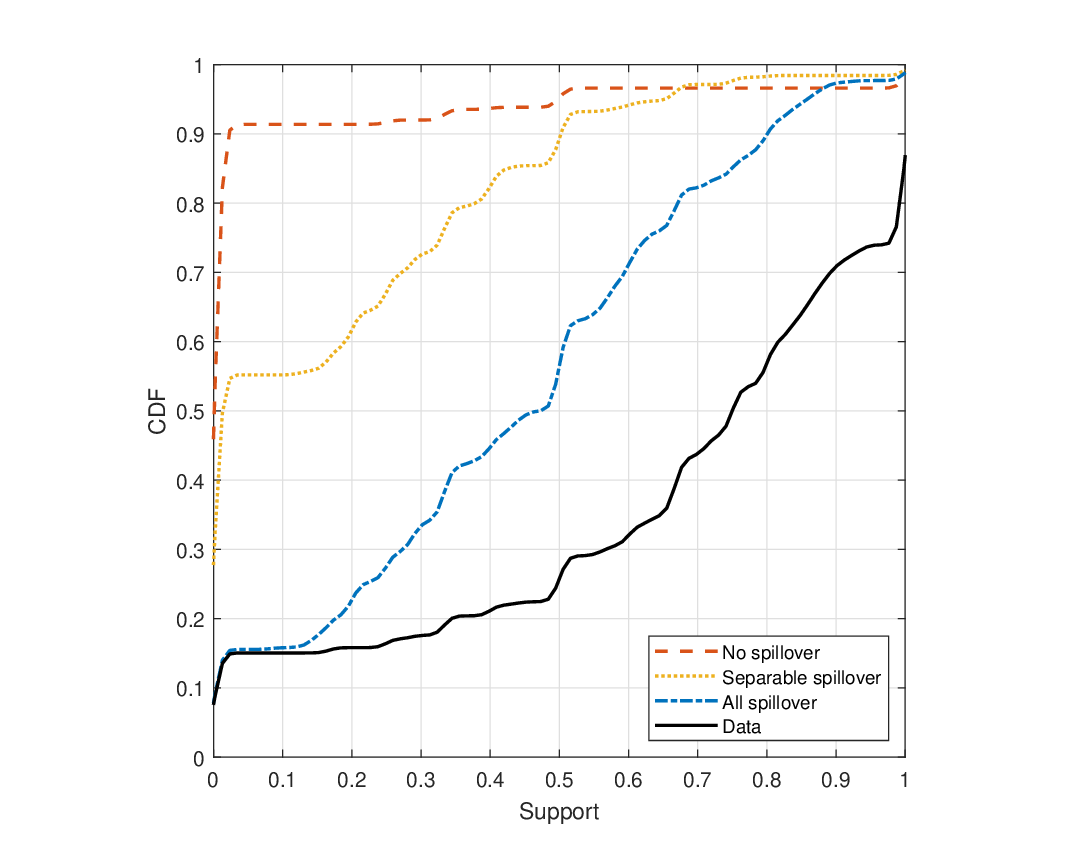}
\end{figure}
Figure \ref{fig:Support} plots the cumulative distribution function
(CDF) of the support measure across individuals in the observed network
and three predicted networks. Predictions from the specification with
no spillover severely understate the support distribution in the data.
Including separable spillover (reciprocity, recipient's degrees, and
inward support) improves the predicted support distribution to some
extent. However, the specification that yields predictions best matching
the data is the one that includes both separable and nonseparable
spillover. These findings underscore the importance of spillover effects,
in particular nonseparable spillover effects (outward support), in
predicting the support distribution.

\section{\label{sec:conclusion}Conclusion}

In this paper, we develop an econometric methodology for strategic
network formation under incomplete information using data from a single
large network. The utility function can be nonseparable in an individual's
link choices because of the spillover effects from friends in common.
We develop a novel approach that applies the Legendre transform to
the utility function so that the optimal decision of an individual
can be represented equivalently as a sequence of correlated binary
choices. We propose a two-step estimation procedure, where we estimate
the link choice probabilities in the first step and estimate the model
parameters in the second step. We show that the two-step estimator
is consistent and asymptotically normal. The link dependence due to
the preference for friends in common does not affect the rate of convergence,
but increases the asymptotic variance of the estimator. We also explore
a scenario of undirected networks and derive a limiting approximation
of the game that simplifies the computation in large networks.

There are a few more extensions of our approach that might be of interest.
We may relax the i.i.d. assumption on the utility shocks by adding
an individual-invariant heterogeneity \citep{Graham_2017}. Both the
individual heterogeneity and the strategic interactions considered
in this paper can generate link dependence. It would be valuable to
investigate the extent to which each of them accounts for the link
dependence in network data. A recent strand of literature explores
social interactions in endogenous networks where the endogeneity of
a network is characterized through a network formation model (\citealp{Goldsmith_Imbens_2013};
\citealp{Hsieh_Lee_2016}; \citealp{Johnsson_Moon_2021}; \citealp{Auerbach2022}).
These studies typically model network formation by a dyadic regression
or a sequential process. Our paper provides an alternative model of
network formation that is simple to analyze and allows for strategic
interactions.\footnote{Other related studies along this line include \citet{Badev_2021},
who developed a joint model of network formation and individual outcomes,
and \citet{Battaglini_2021}, who developed a model of network formation
to recover unobserved social networks using only observable outcomes.}

\renewcommand*{\thesection}{\Alph{section}}
\setcounter{section}{0}

\section{Appendix}

\paragraph{Notation}

We use $\|\cdot\|$ to denote the Euclidean norm. For an $n\times1$
vector $x\in\mathbb{R}^{n}$ and an $n\times n$ matrix $A\in\mathbb{R}^{n^{2}}$,
we have $\|x\|=(\sum_{i=1}^{n}x_{i}^{2})^{1/2}$ and $\|A\|=(\text{tr}(AA'))^{1/2}=(\text{\ensuremath{\sum}}_{i=1}^{n}\sum_{j=1}^{n}a_{ij}^{2})^{1/2}$.
$I_{T}$ denotes the $T\times T$ identity matrix. The notation $o_{p}(1)$
and $O_{p}(1)$ are defined conditionally on $X$ or certain components
of $X$, depending on the context. For example, the statement ``$Y_{n}=o_{p}(1)$
conditional on $X$'' means that for any $\delta>0$, $\lim_{n\rightarrow\infty}\Pr(\|Y_{n}\|>\delta|X)=0$.

\subsection{\label{app:proof.model}Proofs in Section \ref{sec:model}}
\begin{proof}[Proof of Proposition \ref{prop:existsym}]
We follow the proof in \citet[Theorem 1]{Leung_2015}. Organize the
conditional choice probabilities in an $n\times2^{n-1}$ matrix $\sigma(X)$.
The $i$th row consists of individual $i$'s conditional choice probabilities
$\sigma_{i}(X)=\{\sigma_{i}(g_{i}|X),g_{i}\in\mathcal{G}_{i}\}$.
The elements in the row sum to $1$. Denote the set of all such matrices
by $\Sigma(X)$. With row $i$ of $\sigma(X)$ we associate $X_{i}$.
Let $\Sigma^{s}(X)\subset\Sigma(X)$ denote the subset of matrices
of conditional choice probabilities such that if $X_{i}=X_{j}$ then
$\sigma_{i}(X)=\sigma_{j}(X)$, that is, $\sigma_{i}(g_{i}|X)=\sigma_{i}(g_{j}|X)$
for $g_{i}\in\mathcal{G}_{i}$ and $g_{j}\in\mathcal{G}_{j}$ where
$g_{j}$ is obtained from $g_{i}$ by swapping the $i$th and $j$th
components of $g_{i}$. If we organize the conditional choice probabilities
in (\ref{eq:CCP}) in an $n\times2^{n-1}$ matrix $P(X,\sigma)$,
it maps the matrix $\sigma$ to a matrix of conditional choice probabilities
in $\Sigma(X)$. An equilibrium is a fixed point of this mapping.
Because we focus on the symmetric equilibria in $\Sigma^{s}(X)$,
we must show that $P(X,\sigma)$ is a continuous mapping from $\Sigma^{s}(X)$
to $\Sigma^{s}(X)$ and that the set $\Sigma^{s}(X)$ is convex and
compact.

First, the mapping $P(X,\sigma)$ maps $\Sigma^{s}(X)$ to itself.
Let $\sigma(X)\in\Sigma^{s}(X)$. If $X_{i}=X_{j}$, then individuals
$i$ and $j$ have the same expected incremental utilities in (\ref{eq:Eu})
and (\ref{eq:Ev}). Because $\epsilon_{i}$ and $\epsilon_{j}$ follow
the same distribution, rows $i$ and $j$ of $P(X,\sigma(X))$ are
identical, so indeed $P(X,\sigma(X))\in\Sigma^{s}(X)$. Second, a
convex combination of matrices $\sigma(X)$,$\tilde{\sigma}(X)\in\Sigma^{s}(X)$
is a matrix with rows that sum to $1$ and that rows $i$ and $j$
are identical if $X_{i}=X_{j}$. The convex combination is therefore
in $\Sigma^{s}(X)$. Third, $\Sigma^{s}(X)$ is bounded. It is also
closed. Let $\{\sigma^{k}(X),k=1,2,\ldots\}$ be a sequence in $\Sigma^{s}(X)$
that converges to a limit. Then for all $k$ the rows of $\sigma^{k}(X)$
sum to $1$ and rows $i$ and $j$ are identical if $X_{i}=X_{j}$.
So the limit has the same properties and is therefore in $\Sigma^{s}(X)$.
Finally, the mapping $P(X,\sigma)$ is continuous on $\Sigma^{s}(X)$,
which is proved in Lemma \ref{lem:ccp.cont}. We conclude that by
Brouwer's fixed point theorem, $P(X,\sigma)$ has a fixed point in
$\Sigma^{s}(X)$.
\end{proof}

\subsection{\label{app:proof.legendre}Proofs in Section \ref{sec:legendre}}
\begin{proof}[Proof of Lemma \ref{lem:EU.legendre}]
The first equality follows because by the spectral decomposition
(\ref{eq:spectral}) we have
\begin{eqnarray*}
\sum_{j\neq i}\sum_{k\neq i}G_{ij}G_{ik}Z'_{j}V_{i}(X,\sigma)Z_{k} & = & \left(\sum_{j\neq i}G_{ij}Z'_{j}\Phi_{i}(X,\sigma)\right)\Lambda_{i}(X,\sigma)\left(\sum_{k\neq i}G_{ik}\Phi'_{i}(X,\sigma)Z_{k}\right)\\
 & = & (n-1)^{2}\sum_{t=1}^{T}\lambda_{it}(X,\sigma)\left(\frac{1}{n-1}\sum_{j\neq i}G_{ij}Z'_{j}\phi_{it}(X,\sigma)\right)^{2}.
\end{eqnarray*}
The second equality follows from (\ref{eq:legendre.scalar}).
\end{proof}

\begin{proof}[Proof of Theorem \ref{thm:optG}]
We prove the theorem for the general case where both $\mathcal{T}_{i+}$
and $\mathcal{T}_{i-}$ are nonempty. The special cases where $\mathcal{T}_{i+}$
is empty (negative semi-definite $V_{i}(X,\sigma)$) or $\mathcal{T}_{i-}$
is empty (positive semi-definite $V_{i}(X,\sigma)$) can be proved
similarly and thus omitted.

By Lemma \ref{lem:EU.legendre}, the expected utility satisfies
\begin{eqnarray}
 &  & \mathbb{E}[U_{i}(G_{i},G_{-i},X,\epsilon_{i})|X,\epsilon_{i},\sigma]\nonumber \\
 & = & \max_{\omega_{t},t\in\mathcal{T}_{i+}}\min_{\omega_{t},t\in\mathcal{T}_{i-}}\frac{1}{n-1}\sum_{j\neq i}G_{ij}\left(U_{ij}(X,\sigma)+\frac{n-1}{n-2}Z_{j}^{\prime}\sum_{t=1}^{T}\phi_{it}(X,\sigma)\lambda_{it}(X,\sigma)\omega_{t}-\epsilon_{ij}\right)\nonumber \\
 &  & -\frac{n-1}{2(n-2)}\sum_{t=1}^{T}\lambda_{it}(X,\sigma)\omega_{t}^{2}\nonumber \\
 & = & \max_{\omega_{t},t\in\mathcal{T}_{i+}}\min_{\omega_{t},t\in\mathcal{T}_{i-}}\frac{1}{n-1}\sum_{j\neq i}G_{ij}\left(U_{ij}(X,\sigma)+\frac{n-1}{n-2}Z_{j}^{\prime}\Phi_{i}(X,\sigma)\Lambda_{i}(X,\sigma)\omega-\epsilon_{ij}\right)\nonumber \\
 &  & -\frac{n-1}{2(n-2)}\omega^{\prime}\Lambda_{i}(X,\sigma)\omega.\label{eq:EU.maxmin}
\end{eqnarray}
The first equality follows because if we move an eigenvalue $\lambda_{it}$
inside a maximization, it remains a maximization if $\lambda_{it}>0$
and switches to a minimization if $\lambda_{it}<0$.

Let $\tilde{\Pi}(G_{i},\omega,\epsilon_{i},X,\sigma)$ denote the
objective function of the last maximin problem in (\ref{eq:EU.maxmin}).
We have 
\begin{eqnarray}
\max_{G_{i}}\mathbb{E}[U_{i}(G_{i},G_{-i},X,\epsilon_{i})|X,\epsilon_{i},\sigma] & = & \max_{G_{i}}\max_{\omega_{t},t\in\mathcal{T}_{i+}}\min_{\omega_{t},t\in\mathcal{T}_{i-}}\tilde{\Pi}_{i}(G_{i},\omega,\epsilon_{i},X,\sigma)\nonumber \\
 & \leq & \max_{\omega_{t},t\in\mathcal{T}_{i+}}\min_{\omega_{t},t\in\mathcal{T}_{i-}}\max_{G_{i}}\tilde{\Pi}_{i}(G_{i},\omega,\epsilon_{i},X,\sigma)\nonumber \\
 & = & \max_{\omega_{t},t\in\mathcal{T}_{i+}}\min_{\omega_{t},t\in\mathcal{T}_{i-}}\Pi_{i}(\omega,\epsilon_{i},X,\sigma),\label{eq:EU.G*}
\end{eqnarray}
where $\Pi_{i}(\omega,\epsilon_{i},X,\sigma)$ denotes the objective
function in (\ref{eq:maxmin}). The inequality follows because $\text{\ensuremath{\max_{\omega_{t},t\in\mathcal{T}_{i+}}\min_{\omega_{t},t\in\mathcal{T}_{i-}}}}\tilde{\Pi}_{i}(G_{i},\omega,\cdot)\leq\max_{\omega_{t},t\in\mathcal{T}_{i+}}\min_{\omega_{t},t\in\mathcal{T}_{i-}}\max_{G_{i}}\tilde{\Pi}_{i}(G_{i},\omega,\cdot)$
for all $G_{i}$ and hence the maximum of the left-hand side over
$G_{i}$ is bounded by the right-hand side. The last equality in (\ref{eq:EU.G*})
holds because for any $\omega$, $\tilde{\Pi}_{i}(G_{i},\omega,\cdot)$
is separable in each $G_{ij}$ so the optimal $G_{ij}$ is given by
(\ref{eq:gij}) with $\omega_{i}(\epsilon_{i},X,\sigma)$ replaced
by $\omega$ and $\max_{G_{i}}\tilde{\Pi}_{i}(G_{i},\omega,\cdot)=\Pi_{i}(\omega,\cdot)$.

Next we show that the inequality in (\ref{eq:EU.G*}) is an equality.
By Lemma \ref{lem:foc.w}, $\omega_{i}(\epsilon_{i},X,\sigma)$ satisfies
the first-order condition in (\ref{eq:foc.maxmin}). Multiplying both
sides by $\Phi_{i}(X,\sigma)$ gives 
\begin{equation}
\Phi_{i}(X,\sigma)\Lambda_{i}(X,\sigma)\omega_{i}(\epsilon_{i},X,\sigma)=\frac{1}{n-1}V_{i}(X,\sigma)\sum_{j\neq i}G_{ij}(\epsilon_{i},X,\sigma)Z_{j},\text{ a.s.,}\label{eq:foc.maxmin.phi}
\end{equation}
where $G_{ij}(\epsilon_{i},X,\sigma)$ is given in (\ref{eq:gij}).
By the definition of $G_{i}(\epsilon_{i},X,\sigma)$ and $\omega_{i}(\epsilon_{i},X,\sigma)$,
the maximin value of $\Pi(\omega,\epsilon_{i},X,\sigma)$ is given
by
\begin{eqnarray}
 &  & \max_{\omega_{t},t\in\mathcal{T}_{i+}}\min_{\omega_{t},t\in\mathcal{T}_{i-}}\Pi_{i}(\omega,\epsilon_{i},X,\sigma)\nonumber \\
 & = & \frac{1}{n-1}\sum_{j\neq i}G_{ij}(\epsilon_{i},X,\sigma)(U_{ij}(X,\sigma)-\epsilon_{ij})\nonumber \\
 &  & +\frac{1}{n-2}\sum_{j\neq i}G_{ij}(\epsilon_{i},X,\sigma)Z_{j}^{\prime}\Phi_{i}(X,\sigma)\Lambda_{i}(X,\sigma)\omega_{i}(\epsilon_{i},X,\sigma)\nonumber \\
 &  & -\frac{n-1}{2(n-2)}\omega_{i}(\epsilon_{i},X,\sigma)'\Lambda_{i}(X,\sigma)\omega_{i}(\epsilon_{i},X,\sigma)\nonumber \\
 & = & \frac{1}{n-1}\sum_{j\neq i}G_{ij}(\epsilon_{i},X,\sigma)(U_{ij}(X,\sigma)-\epsilon_{ij})\nonumber \\
 &  & +\frac{1}{2(n-2)}\sum_{j\neq i}G_{ij}(\epsilon_{i},X,\sigma)Z_{j}^{\prime}\Phi_{i}(X,\sigma)\Lambda_{i}(X,\sigma)\omega_{i}(\epsilon_{i},X,\sigma),\text{ a.s.}\nonumber \\
 & = & \frac{1}{n-1}\sum_{j\neq i}G_{ij}(\epsilon_{i},X,\sigma)(U_{ij}(X,\sigma)-\epsilon_{ij})\nonumber \\
 &  & +\frac{1}{2(n-1)(n-2)}\sum_{j\neq i}\sum_{k\neq i}G_{ij}(\epsilon_{i},X,\sigma)G_{ik}(\epsilon_{i},X,\sigma)Z_{j}^{\prime}V_{i}(X,\sigma)Z_{k},\text{ a.s.}\nonumber \\
 & = & \mathbb{E}[U_{i}(G_{i}(\epsilon_{i},X,\sigma),G_{-i},X,\epsilon_{i})|X,\epsilon_{i},\sigma],\label{eq:EU.gij}
\end{eqnarray}
where the second equality follows from the first-order condition (\ref{eq:foc.maxmin})
and the third equality follows from equation (\ref{eq:foc.maxmin.phi}). 

Combining (\ref{eq:EU.G*}) and (\ref{eq:EU.gij}) yields 
\begin{eqnarray*}
\max_{G_{i}}\mathbb{E}[U_{i}(G_{i},G_{-i},X,\epsilon_{i})|X,\epsilon_{i},\sigma] & \leq & \max_{\omega_{t},t\in\mathcal{T}_{i+}}\min_{\omega_{t},t\in\mathcal{T}_{i-}}\Pi_{i}(\omega,\epsilon_{i},X,\sigma)\\
 & = & \mathbb{E}[U_{i}(G_{i}(\epsilon_{i},X,\sigma),G_{-i},X,\epsilon_{i})|X,\epsilon_{i},\sigma]\text{, a.s..}
\end{eqnarray*}
Because $\max_{G_{i}}\mathbb{E}[U_{i}(G_{i},G_{-i},X,\epsilon_{i})|X,\epsilon_{i},\sigma]\geq\mathbb{E}[U_{i}(G_{i}(\epsilon_{i},X,\sigma),G_{-i},X,\epsilon_{i})|X,\epsilon_{i},\sigma]$,
the inequality becomes an equality, and all the terms are equal almost
surely. Hence, $G_{i}(\epsilon_{i},X,\sigma)$ is an optimal solution
almost surely.

As for the uniqueness, $G_{i}(\epsilon_{i},X,\sigma)$ is unique almost
surely because $\epsilon_{i}$ has a continuous distribution, so two
decisions achieve the same utility with probability zero. The uniqueness
of $\Lambda_{i}(X,\sigma)\omega_{i}(\epsilon_{i},X,\sigma)$ follows
from the uniqueness of $G_{i}(\epsilon_{i},X,\sigma)$, equation (\ref{eq:foc.maxmin.phi})
and the invertibility of $\Phi_{i}(X,\sigma)$. 
\end{proof}

\subsection{\label{app:proof.estimation}Proofs in Section \ref{sec:estimation}}
\begin{proof}[Proof of Theorem \ref{thm:theta.consist}]
For $p=(p_{(st)},1\leq s,t\leq T)\in\mathcal{P}=[0,1]^{T^{2}}$,
define the $T^{2}\times1$ vector function $\hat{h}_{n}(p)=(\hat{h}_{n,st}(p),1\leq s,t\leq T)'$,
where
\[
\hat{h}_{n,st}(p)=\frac{1}{n(n-1)}\sum_{i}\sum_{j\neq i}1\{X_{i}=x_{s},X_{j}=x_{t}\}(G_{n,ij}-p_{(st)}).
\]
Provided that $\frac{1}{n(n-1)}\sum_{i}\sum_{j\neq i}1\{X_{i}=x_{s},X_{j}=x_{t}\}>0$,
which holds for $n$ sufficiently large by Assumption \ref{ass:theta.consist}(iv),
the first-step estimator $\hat{p}_{n}$ satisfies that $\hat{h}_{n}(\hat{p}_{n})=0$.
Define the population counterpart of $\hat{h}_{n}(p)$ by $h_{n}(p)=(h_{n,st}(p),1\leq s,t\leq T)'$,
where
\[
h_{n,st}(p)=\frac{1}{n(n-1)}\sum_{i}\sum_{j\neq i}1\{X_{i}=x_{s},X_{j}=x_{t}\}(\mathbb{E}[G_{n,ij}|X]-p_{(st)}).
\]
We can write the moment restrictions in the first step as $h_{n}(p_{n})=0$.
Stack the moments in the first and second steps and define $\tilde{m}_{n}(\theta,p)=[m_{n}(\theta,p)',h_{n}(p)']'$
and $\hat{\tilde{m}}_{n}(\theta,p)=[\hat{m}_{n}(\theta,p)',\hat{h}_{n}(p)']'$.
The optimality of $\hat{\theta}_{n}$ and $\hat{h}_{n}(\hat{p}_{n})=0$
imply that $\hat{\tilde{m}}_{n}(\hat{\theta}_{n},\hat{p}_{n})=o_{p}(1)$.

We prove consistency following \citet{Newey_McFadden_1994}. Fix $\delta>0$.
Let $\mathcal{B}_{0}(\delta)=\{(\theta,p)\in\Theta\times\mathcal{P}:\Vert(\theta,p)-(\theta{}_{0},p_{n})\Vert<\delta\}$
be an open $\delta$-ball centered at $(\theta_{0},p_{n})$. We have
\begin{equation}
\Pr(\Vert(\hat{\theta}_{n},\hat{p}_{n})-(\theta{}_{0},p_{n})\Vert<\delta|X)\geq\Pr\left(\left.\Vert\tilde{m}_{n}(\hat{\theta}_{n},\hat{p}_{n})\Vert<\inf_{(\theta,p)\in(\Theta\times\mathcal{P})\backslash\mathcal{B}_{0}(\delta)}\Vert\tilde{m}_{n}(\theta,p)\Vert\right\vert X\right).\label{eq:consist.prob}
\end{equation}
By the triangle inequality and $\hat{\tilde{m}}_{n}(\hat{\theta}_{n},\hat{p}_{n})=o_{p}(1)$,
we obtain
\begin{eqnarray*}
\Vert\tilde{m}_{n}(\hat{\theta}_{n},\hat{p}_{n})\Vert & \leq & \Vert\hat{\tilde{m}}_{n}(\hat{\theta}_{n},\hat{p}_{n})-\tilde{m}_{n}(\hat{\theta}_{n},\hat{p}_{n})\Vert+\Vert\hat{\tilde{m}}_{n}(\hat{\theta}_{n},\hat{p}_{n})\Vert\\
 & \leq & \sup_{(\theta,p)\in\Theta\times\mathcal{P}}\Vert\hat{\tilde{m}}_{n}(\theta,p)-\tilde{m}_{n}(\theta,p)\Vert+o_{p}(1).
\end{eqnarray*}
The uniform LLN in Lemma \ref{lem:moment.lln} shows that $\sup_{(\theta,p)\in\Theta\times\mathcal{P}}\Vert\hat{m}_{n}(\theta,p)-m_{n}(\theta,p)\Vert=o_{p}(1)$.
Moreover, observe that for each $1\leq s,t\leq T$
\[
\hat{h}_{n,st}(p)-h_{n,st}(p)=\frac{1}{n(n-1)}\sum_{i}\sum_{j\neq i}1\{X_{i}=x_{s},X_{j}=x_{t}\}(G_{n,ij}-\mathbb{E}[G_{n,ij}|X]),
\]
which does not depend on $p$. Because given $X$, $G_{n,i}=(G_{n,ij},j\neq i)$
are independent across $i$, $\mathbb{E}[(\hat{h}_{n,st}(p)-h_{n,st}(p))^{2}|X]$
is given by $(n(n-1))^{-2}$ times 
\begin{eqnarray*}
 &  & \sum_{i}\sum_{j\neq i}1\{X_{i}=x_{s},X_{j}=x_{t}\}\mathbb{E}[(G_{n,ij}-\mathbb{E}[G_{n,ij}|X])^{2}|X]\\
 &  & +\sum_{i}\sum_{j\neq i}\sum_{k\neq i,j}1\{X_{i}=x_{s},X_{j}=X_{k}=x_{t}\}\mathbb{E}[(G_{n,ij}-\mathbb{E}[G_{n,ij}|X])(G_{n,ik}-\mathbb{E}[G_{n,ik}|X])|X].
\end{eqnarray*}
Because both $\mathbb{E}[(G_{n,ij}-\mathbb{E}[G_{n,ij}|X])^{2}|X]$
and $\mathbb{E}[(G_{n,ij}-\mathbb{E}[G_{n,ij}|X])(G_{n,ik}-\mathbb{E}[G_{n,ik}|X])|X]$
are bounded by $1$, we can bound $\mathbb{E}[(\hat{h}_{n,st}(p)-h_{n,st}(p))^{2}|X]$
by $(n(n-1))^{-2}(n(n-1)+n(n-1)(n-2))=o(1)$. Therefore, $\mathbb{E}[\|\hat{h}_{n}(p)-h_{n}(p)\|^{2}|X]=\sum_{s=1}^{T}\sum_{t=1}^{T}\mathbb{E}[(\hat{h}_{n,st}(p)-h_{n,st}(p))^{2}|X]=o(1)$.
By Markov's inequality, we obtain $\sup_{p\in\mathcal{P}}\|\hat{h}_{n}(p)-h_{n}(p)\|=o_{p}(1)$.
Combining the results yields $\sup_{(\theta,p)\in\Theta\times\mathcal{P}}\Vert\hat{\tilde{m}}_{n}(\theta,p)-\tilde{m}_{n}(\theta,p)\Vert=o_{p}(1)$
and thus $\tilde{m}_{n}(\hat{\theta}_{n},\hat{p}_{n})=o_{p}(1)$.

By Assumption \ref{ass:theta.consist}(ii) and (iv) and the definition
of $h_{n}(p)$, there is $\xi>0$ such that for $n$ sufficiently
large, $\Vert\tilde{m}_{n}(\theta,p)\Vert\geq\xi$ for all $\|(\theta,p)-(\theta_{0},p_{n})\|\geq\delta$.\footnote{Note that $\|(\theta,p)-(\theta_{0},p_{n})\|\geq\delta$ implies that
$\|\theta-\theta_{0}\|\geq\delta/\sqrt{2}$ or $\|p-p_{n}\|\geq\delta/\sqrt{2}$
because $\|(\theta,p)-(\theta_{0},p_{n})\|^{2}=\|\theta-\theta_{0}\|^{2}+\|p-p_{n}\|^{2}$.
If $\|\theta-\theta_{0}\|\geq\delta/\sqrt{2}$, then by Assumption
\ref{ass:theta.consist}(ii), there is $\xi>0$ such that for $n$
sufficiently large, $\Vert\tilde{m}_{n}(\theta,p)\Vert\geq\Vert m_{n}(\theta,p)\Vert\geq\|m_{n}(\theta,p_{n})\|+\|m_{n}(\theta,p)-m_{n}(\theta,p_{n})\|\geq\xi$.
If $\|p-p_{n}\|\geq\delta/\sqrt{2}$, then $\Vert\tilde{m}_{n}(\theta,p)\Vert\geq\|h_{n}(p)\|\geq\min_{s,t}f_{n,st}\cdot\|p-p_{n}\|$,
where $f_{n,st}=\frac{1}{n(n-1)}\sum_{i}\sum_{j\neq i}1\{X_{i}=x_{s},X_{j}=x_{t}\}$.
By Assumption \ref{ass:theta.consist}(iv), there is $\xi>0$ such
that $\Vert\tilde{m}_{n}(\theta,p)\Vert\geq\min_{s,t}f_{n,st}\cdot\|p-p_{n}\|\geq\xi$
for $n$ sufficiently large.} This, together with the compactness of $(\Theta\times\mathcal{P})\backslash\mathcal{B}_{0}(\delta)$
(Assumption \ref{ass:theta.consist}(i)) and the continuity of $\tilde{m}_{n}(\theta,p)$
in $\theta$ and $p$ (Lemma \ref{lem:ccp.cont.esti}), implies that
for $n$ sufficiently large, $\inf_{(\theta,p)\in(\Theta\times\mathcal{P})\backslash\mathcal{B}_{0}(\delta)}\Vert\tilde{m}_{n}(\theta,p)\Vert=\|\tilde{m}_{n}(\bar{\theta},\bar{p})\|\geq\xi$
for some $(\bar{\theta},\bar{p})\in(\Theta\times\mathcal{P})\backslash\mathcal{B}_{0}(\delta)$.
Combining the results we can see that the right-hand side of equation
(\ref{eq:consist.prob}) goes to $1$ and the consistency is proved.
\end{proof}

\begin{proof}[Proof of Theorem \ref{thm:theta.asymdist}]

By the consistency of $\hat{\theta}_{n}$ and $\hat{p}_{n}$, for
$n$ sufficiently large $(\hat{\theta}_{n},\hat{p}_{n})$ lies in
a neighborhood of $(\theta_{0},p_{n})$ where $P_{n,ij}(\theta,p)$
is continuously differentiable (Assumption \ref{ass:theta.asymdist}(i)).
The Taylor expansion gives $P_{n,ij}(\hat{\theta}_{n},\hat{p}_{n})=P_{n,ij}(\theta_{0},p_{n})+\nabla_{\theta^{\prime}}P_{n,ij}(\theta_{0},p_{n})(\hat{\theta}_{n}-\theta_{0})+\nabla_{p'_{n}}P_{n,ij}(\theta_{0},p_{n})(\hat{p}_{n}-p_{n})+o_{p}(\Vert(\hat{\theta}_{n},\hat{p}_{n})-(\theta_{0},p_{n})\Vert)$.
Because $\max_{1\leq i,j\leq n}\Vert\hat{q}_{n,ij}-q_{n,ij}\Vert=o_{p}(1)$
(Assumption \ref{ass:theta.consist}(iii)) and for $n$ sufficiently
large, $\max_{1\leq i,j\leq n}\|\nabla_{\theta^{\prime}}P_{n,ij}(\theta_{0},p_{n})\|<\infty$
and $\max_{1\leq i,j\leq n}\|\nabla_{p'_{n}}P_{n,ij}(\theta_{0},p_{n})\|<\infty$
(Assumption \ref{ass:theta.asymdist}(i)), upon rearranging the terms
we derive
\begin{eqnarray}
 &  & \frac{1}{n(n-1)}\sum_{i}\sum_{j\neq i}q_{n,ij}\nabla_{\theta^{\prime}}P_{n,ij}(\theta_{0},p_{n})(\hat{\theta}_{n}-\theta_{0})+o_{p}(1)(\hat{\theta}_{n}-\theta_{0})\nonumber \\
 & = & \frac{1}{n(n-1)}\sum_{i}\sum_{j\neq i}q_{n,ij}(G_{n,ij}-P_{n,ij}(\theta_{0},p_{n})-\nabla_{p'_{n}}P_{n,ij}(\theta_{0},p_{n})(\hat{p}_{n}-p_{n}))\nonumber \\
 &  & +\frac{1}{n(n-1)}\sum_{i}\sum_{j\neq i}(\hat{q}_{n,ij}-q_{n,ij})(G_{n,ij}-P_{n,ij}(\theta_{0},p_{n}))-o_{p}(1)(\hat{p}_{n}-p_{n})\nonumber \\
 &  & -\frac{1}{n(n-1)}\sum_{i}\sum_{j\neq i}\hat{q}_{n,ij}o_{p}(\Vert(\hat{\theta}_{n},\hat{p}_{n})-(\theta_{0},p_{n})\Vert).\label{eq:theta.dcmp}
\end{eqnarray}

The first-step estimator $\hat{p}_{n}=(\hat{p}_{n,(st)},1\leq s,t\leq T)'$
satisfies
\[
\hat{p}_{n}-p_{n}=\frac{1}{n(n-1)}\sum_{i}\sum_{j\neq i}w_{n,ij}(G_{n,ij}-P_{n,ij}(\theta_{0},p_{n})),
\]
where $w_{n,ij}=(w_{n,ij,(11)},\ldots,w_{n,ij,(TT)})^{\prime}\in\mathbb{R}^{T^{2}}$
is a $T^{2}\times1$ vector and $w_{n,ij,(st)}=1\{X_{i}=x_{s},X_{j}=x_{t}\}/\frac{1}{n(n-1)}\sum_{\tilde{i}}\sum_{\tilde{j}\neq\tilde{i}}1\{X_{\tilde{i}}=x_{s},X_{\tilde{j}}=x_{t}\}$
for $1\leq s,t\leq T$. Hence, we can write the first term on the
right-hand side of (\ref{eq:theta.dcmp}) in a sample average form
\[
\frac{1}{n(n-1)}\sum_{i}\sum_{j\neq i}\tilde{q}_{n,ij}(G_{n,ij}-P_{n,ij}(\theta_{0},p_{n})),
\]
where 
\begin{equation}
\tilde{q}_{n,ij}\equiv q_{n,ij}-(\frac{1}{n(n-1)}\sum_{k}\sum_{l\neq k}q_{n,kl}\nabla_{p'_{n}}P_{n,kl}(\theta_{0},p_{n}))w_{n,ij}\label{eq:q_tilde}
\end{equation}
 is the augmented instrument that incorporates the weight $w_{n,ij}$
in the first step. Because $\max_{1\leq i,j\leq n}\|q_{n,ij}\|<\infty$
(Assumption \ref{ass:theta.consist}(iii)) and for $n$ sufficiently
large, $\max_{1\leq i,j\leq n}\|w_{n,ij}\|<\infty$ (Assumption \ref{ass:theta.consist}(iv))
and $\max_{1\leq i,j\leq n}\|\nabla_{p'_{n}}P_{n,ij}(\theta_{0},p_{n})\|<\infty$
(Assumption \ref{ass:theta.asymdist}(i)), we have $\max_{1\leq i,j\leq n}\|\tilde{q}_{n,ij}\|<\infty$.
Applying Lemma \ref{lem:m.clt} for $\tilde{q}_{n,ij}$, we obtain
$\frac{1}{n(n-1)}\sum_{i}\sum_{j\neq i}\tilde{q}_{n,ij}(G_{n,ij}-P_{n,ij}(\theta_{0},p_{n}))=O_{p}(n^{-1})$. 

Observe that $\hat{p}_{n}-p_{n}$ has a similar form with $w_{n,ij}$
in place of $\tilde{q}_{n,ij}$. Similarly as in Lemma \ref{lem:m.clt},
we can prove that $\hat{p}_{n}-p_{n}=O_{p}(n^{-1})$. The second term
on the right-hand side of (\ref{eq:theta.dcmp}) also has a similar
form with weight $\hat{q}_{n,ij}-q_{n,ij}$. It is $o_{p}(n^{-1})$
following Lemma \ref{lem:m.clt} with slight modifications.\footnote{The proof is available upon request.}
Moreover, by $\Vert(\hat{\theta}_{n},\hat{p}_{n})-(\theta_{0},p_{n})\Vert\leq\Vert\hat{\theta}_{n}-\theta_{0}\Vert+\Vert\hat{p}_{n}-p_{n}\Vert$
and Assumption \ref{ass:theta.consist}(iii), we can bound the last
term on the right-hand side of (\ref{eq:theta.dcmp}) by $o_{p}(\Vert\hat{\theta}_{n}-\theta_{0}\Vert+\Vert\hat{p}_{n}-p_{n}\Vert)=o_{p}(\Vert\hat{\theta}_{n}-\theta_{0}\Vert)+o_{p}(\Vert\hat{p}_{n}-p_{n}\Vert)=o_{p}(\Vert\hat{\theta}_{n}-\theta_{0}\Vert)+o_{p}(n^{-1})$.
Note that the left-hand side of (\ref{eq:theta.dcmp}) is given by
$(J_{n}+o_{p}(1))(\hat{\theta}_{n}-\theta_{0})$. By Assumption \ref{ass:theta.asymdist}(ii),
we have the bound $\Vert J_{n}(\theta-\theta_{0})\Vert^{2}\geq c^{2}\Vert\theta-\theta_{0}\Vert^{2}$,
where $c^{2}=\lambda_{\min}(J'_{n}J_{n})>0$ for $n$ sufficiently
large.\footnote{If $J_{n}$ is nonsingular, then $J'_{n}J_{n}$ is positive definite
and $\lambda_{\min}(J'_{n}J_{n})>0$.} Combining these results we derive from equation (\ref{eq:theta.dcmp})
that $\Vert\hat{\theta}_{n}-\theta_{0}\Vert(c+o_{p}(1))\leq O_{p}(n^{-1})+o_{p}(n^{-1})$.
This implies that $\hat{\theta}_{n}-\theta_{0}=O_{p}(n^{-1})$, that
is, $\hat{\theta}_{n}$ is $n$-consistent for $\theta_{0}$.

To derive the asymptotic distribution of $\hat{\theta}_{n}$, write
equation (\ref{eq:theta.dcmp}) as
\[
\sqrt{n(n-1)}J_{n}(\hat{\theta}_{n}-\theta_{0})=\frac{1}{\sqrt{n(n-1)}}\sum_{i}\sum_{j\neq i}\tilde{q}_{n,ij}(G_{n,ij}-P_{n,ij}(\theta_{0},p_{n}))+o_{p}(1),
\]
and apply Lemma \ref{lem:m.clt} to the leading term on the right-hand
side. Define the $d_{\theta}\times1$ vector 
\begin{equation}
\phi_{n,ij}^{\theta}\equiv\tilde{q}_{n,ij}(g_{n,ij}(\omega_{ni}^{\ast},\epsilon_{ij})-P_{n,ij}^{\ast}(\omega_{ni}^{\ast}))+J_{ni}^{\omega}(\omega_{ni}^{\ast},\tilde{q}_{ni})\phi_{n,ij}^{\omega}(\omega_{ni}^{\ast},\epsilon_{ij}),\label{eq:theta.influence}
\end{equation}
where $g_{n,ij}(\omega_{ni}^{\ast},\epsilon_{ij})=1\{U_{n,ij}+\frac{n-1}{n-2}Z_{j}^{\prime}\Phi_{ni}\Lambda_{ni}\omega_{ni}^{\ast}\geq\epsilon_{ij}\}$,
$P_{n,ij}^{\ast}(\omega_{ni}^{\ast})=F_{\epsilon}(U_{n,ij}+\frac{n-1}{n-2}Z_{j}^{\prime}\Phi_{ni}\Lambda_{ni}\omega_{ni}^{\ast})$,
$J_{ni}^{\omega}(\omega_{ni}^{\ast},\tilde{q}_{ni})=\frac{1}{n-1}\sum_{j\neq i}\tilde{q}_{n,ij}\nabla_{\omega^{\lambda\prime}}P_{n,ij}^{\ast}(\omega_{ni}^{\ast})$,
$\phi_{n,ij}^{\omega}(\omega_{ni}^{\ast},\epsilon_{ij})\in\mathbb{R}^{T}$
is the influence function defined in Lemma \ref{lem:w.asymlin}, and
$\omega_{ni}^{\ast}$ is a maximin solution of (\ref{eq:Pai*}).
Define the $d_{\theta}\times d_{\theta}$ variance matrix $\Sigma_{n}=\frac{1}{n(n-1)}\sum_{i}\sum_{j\neq i}\mathbb{E}[\phi_{n,ij}^{\theta}\phi_{n,ij}^{\theta\prime}|X]$.
By Lemma \ref{lem:m.clt}, we obtain $\sqrt{n(n-1)}\Sigma_{n}^{-1/2}J_{n}(\hat{\theta}_{n}-\theta_{0})\overset{d}{\rightarrow}N(0,I_{d_{\theta}})$.
\end{proof}

\subsection{\label{app:proof.extension}Proofs in Section \ref{sec:extension}}

\begin{proof}[Proof of Proposition \ref{prop:CCP.limit.consist}]
Denote $U_{ij}^{*}(p)=U^{*}(X_{i},X_{j},p)$, $V_{i}^{*}(p)=V^{*}(X_{i},p)$
and $\omega_{i}^{*}(p)=\omega^{*}(X_{i},p)$. Suppress the argument
$(X,p)$ or $p$ whenever possible. By definition
\begin{eqnarray}
 &  & P_{n,ij}(X,p)-P^{*}(X_{i},X_{j},p)\nonumber \\
 & = & \mathbb{E}[1\{U_{n,ij}+\frac{n-1}{n-2}Z_{j}^{\prime}\Phi_{ni}\Lambda_{ni}\omega_{ni}(\epsilon_{i})\geq\epsilon_{ij}\}-1\{U_{ij}^{*}+Z_{j}^{\prime}\Phi_{i}^{*}\Lambda_{i}^{*}\omega_{i}^{*}\geq\epsilon_{ij}\}|X].\label{eq:Pnij.diff}
\end{eqnarray}
Denote $\omega_{ni}^{\phi\lambda}(\epsilon_{i})\equiv\Phi_{ni}\Lambda_{ni}\omega_{ni}(\epsilon_{i})\in\Omega$
and $\omega_{i}^{\phi\lambda*}\equiv\Phi_{i}^{*}\Lambda_{i}^{*}\omega_{i}^{*}\in\Omega$.\footnote{Because $\|V_{ni}\|,\|V_{i}^{*}\|<\infty$, without loss of generality
we use $\Omega$ in Assumption \ref{ass:lim}(i) to denote the spaces
of $\omega_{ni}^{\phi\lambda}(\epsilon_{i})$ and $\omega_{i}^{\phi\lambda*}$.} The right-hand side of (\ref{eq:Pnij.diff}) is bounded by the probability
that $\epsilon_{ij}$ lies between $U_{n,ij}+\frac{n-1}{n-2}Z_{j}^{\prime}\omega_{ni}^{\phi\lambda}(\epsilon_{i})$
and $U_{ij}^{*}+Z_{j}^{\prime}\omega_{i}^{\phi\lambda*}$. Define
$\Delta_{n,ij}(\epsilon_{i})\equiv(U_{n,ij}-U_{ij}^{*})+Z_{j}^{\prime}(\omega_{ni}^{\phi\lambda}(\epsilon_{i})-\omega_{i}^{\phi\lambda*})+\frac{1}{n-2}Z_{j}^{\prime}\omega_{ni}^{\phi\lambda}(\epsilon_{i})$.
For any $\delta_{n}>0$, if $\epsilon_{ij}$ lies between $U_{n,ij}+\frac{n-1}{n-2}Z_{j}^{\prime}\omega_{ni}^{\phi\lambda}(\epsilon_{i})$
and $U_{ij}^{*}+Z_{j}^{\prime}\omega_{i}^{\phi\lambda*}$, and if
their difference $|\Delta_{n,ij}(\epsilon_{i})|$ is at most $\delta_{n}$,
then $\epsilon_{ij}$ must lie between $U_{ij}^{*}+Z_{j}^{\prime}\omega_{i}^{\phi\lambda*}-\delta_{n}$
and $U_{ij}^{*}+Z_{j}^{\prime}\omega_{i}^{\phi\lambda*}+\delta_{n}$.
Therefore, we can further bound the right-hand side of (\ref{eq:Pnij.diff})
by
\begin{equation}
\Pr(|\Delta_{n,ij}(\epsilon_{i})|>\delta_{n}|X)+\Pr(U_{ij}^{*}+Z_{j}^{\prime}\omega_{i}^{\phi\lambda*}-\delta_{n}\leq\epsilon_{ij}\leq U_{ij}^{*}+Z_{j}^{\prime}\omega_{i}^{\phi\lambda*}+\delta_{n}|X).\label{eq:Pnij.diff.bdd}
\end{equation}
Consider the first term in (\ref{eq:Pnij.diff.bdd}). By Lemma \ref{lem:wlim.consist},
$\omega_{ni}^{\phi\lambda}(\epsilon_{i})-\omega_{i}^{\phi\lambda*}=o_{p}(1)$.\footnote{Lemma \ref{lem:wlim.consist} fixes $X_{i}$ and treats $(X_{k},k\neq i)$
as random. The result holds if $X_{j}$ is fixed in addition to $X_{i}$.} Moreover, $U_{n,ij}-U_{ij}^{*}=o_{p}(1)$ by Assumption \ref{ass:lim}(iv)
and $\omega_{ni}^{\phi\lambda}(\epsilon_{i})$ is bounded. Hence,
for any $\delta_{n}>0$ we have $\Pr(|\Delta_{n,ij}(\epsilon_{i})|>\delta_{n}|X_{i},X_{j})\rightarrow0$
as $n\rightarrow\infty$. By iterated expectations $\Pr(|\Delta_{n,ij}(\epsilon_{i})|>\delta_{n}|X_{i},X_{j})=\mathbb{E}[\Pr(|\Delta_{n,ij}(\epsilon_{i})|>\delta_{n}|X)|X_{i},X_{j}]$
and Markov's inequality, given $X_{i}$ and $X_{j}$ we must have
$\Pr(|\Delta_{n,ij}(\epsilon_{i})|>\delta_{n}|X)=o_{p}(1)$.\footnote{Denote $W_{n}=\Pr(|\Delta_{n,ij}(\epsilon_{i})|>\delta_{n}|X)$. For
any $k>0$, $\Pr(W_{n}>k|X_{i},X_{j})\leq k^{-1}\mathbb{E}[W_{n}|X_{i},X_{j}]\rightarrow0$.
Hence, $W_{n}=o_{p}(1)$.}

For the second term in (\ref{eq:Pnij.diff.bdd}), by the mean-value
theorem we derive $\Pr(U_{ij}^{*}+Z_{j}^{\prime}\omega_{i}^{\phi\lambda*}-\delta_{n}\leq\epsilon_{ij}\leq U_{ij}^{*}+Z_{j}^{\prime}\omega_{i}^{\phi\lambda*}+\delta_{n}|X)=2f_{\epsilon}(U_{ij}^{*}+Z_{j}^{\prime}\omega_{i}^{\phi\lambda*}+t_{n,ij}\delta_{n})\delta_{n}$,
for some $-1\leq t_{n,ij}\leq1$. Because $f_{\epsilon}(U_{ij}^{*}+Z_{j}^{\prime}\omega_{i}^{\phi\lambda*}+t_{n,ij}\delta_{n})$
is bounded, by choosing $\delta_{n}>0$ with $\delta_{n}\downarrow0$
as $n\rightarrow\infty$, we derive that the second term in (\ref{eq:Pnij.diff.bdd})
is $o(1)$. The proposition is proved.
\end{proof}
\begin{spacing}{1.0} \bibliographystyle{ecta}
\bibliography{network}
\end{spacing} 

\newpage{}

\appendix
\begin{center}
{\LARGE Online Appendix to}\\[1cm]{\LARGE Two-Step Estimation of A
Strategic Network Formation }\\[0.25cm] {\LARGE Model with Clustering}\\[1cm]
{\large Geert Ridder \hspace{1cm} Shuyang Sheng}\\[1cm] 
\par\end{center}

\pagestyle{appendix}
\setcounter{page}{1}
\renewcommand*{\thesection}{O.\Alph{section}}
\numberwithin{equation}{section} \numberwithin{assumption}{section}
\numberwithin{figure}{section} \numberwithin{table}{section}

\section{\label{online:dynamic}Dynamic Processes of Network Formation}

\subsection{Sequential Updating}

In this section, we follow \citet{Myatt_Wallace_2004} and construct
a dynamic process of network formation, where links in a network are
formed and updated in a random sequence. Assume that the network is
directed. We consider a simplified version of the utility in (\ref{eq:U})--(\ref{eq:U.nonsep}),
where there exists no $X$, nor the effect of friends in common ($v_{i,jk}(G_{-i})=0$).
Hence, individual $i$'s utility is given by $U_{i}(G,\epsilon_{i})=\frac{1}{n-1}\sum_{j\neq i}G_{ij}(u_{ij}(G_{-i})-\epsilon_{ij})$,
where $u_{ij}(G_{-i})=\beta_{1}+G_{ji}\beta_{4}+\frac{1}{n-2}\sum_{k\neq i,j}G_{kj}\beta_{5}+\frac{1}{n-2}\sum_{k\neq i,j}G_{jk}\beta_{6}.$ 

We construct the dynamic process as follows. In an initial period,
individuals simultaneously form links, based on their arbitrary beliefs
about the links formed by other individuals $G_{-i}$. After the initial
period, individuals update links sequentially. Specifically, in each
period, a link $G_{ij}$ is selected at random; individual $i$, who
determines the link, receives a new draw of $\epsilon_{ij}$. The
active individual observes $\epsilon_{ij}$ as well as the distribution
of the links formed in the past, and updates the link accordingly.
Assume individuals make myopic decisions. 

Let $Z\in\mathbb{Z}=\{0,1,\ldots,n(n-1)\}$ denote the total number
of links formed in the network. Because individuals are of the same
type (no $X$), $Z$ captures the state of play. Given state $z$,
an active individual $i$ forms a belief about link $G_{jk}$ by the
fraction of links that are formed in the network $\mathbb{E}[G_{jk}|Z=z]=\frac{z}{n(n-1)}$
for all $j\neq i$ and $k\neq j$. The optimal choice for link $G_{ij}$
is $G_{ij}=1\{\beta_{1}+\frac{z}{n(n-1)}(\beta_{4}+\beta_{5}+\beta_{6})\geq\epsilon_{ij}\}$.
The probability that individual $i$ forms link $G_{ij}$ is $F_{\epsilon}(\beta_{1}+\frac{z}{n(n-1)}(\beta_{4}+\beta_{5}+\beta_{6}))=\Psi(\frac{z}{n(n-1)})$,
where we define $\Psi(p)=F_{\epsilon}(\beta_{1}+p(\beta_{4}+\beta_{5}+\beta_{6}))$. 

To analyze this dynamic process, we calculate the transition probabilities
between states. Let $p_{z,z'}=\Pr(Z_{t+1}=z'|Z_{t}=z)$ denote the
probability of a transition from state $z$ to state $z'$. Given
state $z$, the state in the next period can be $z+1$, $z$, or $z-1$.
The state moves from $z$ to $z+1$ if an unformed link $G_{ij}=0$
is selected and individual $i$ decides to form the link. The probability
from $z$ to $z+1$ is $p_{z,z+1}=(1-\frac{z}{n(n-1)})\Psi(\frac{z}{n(n-1)})$.
Similarly, the state moves from $z$ to $z-1$ if a formed link $G_{ij}=1$
is selected and individual $i$ decides to sever the link. The probability
from $z$ to $z-1$ is $p_{z,z-1}=\frac{z}{n(n-1)}(1-\Psi(\frac{z}{n(n-1)}))$.
The state remains at $z$ if either a formed link is selected and
remains formed or an unformed link is selected and remains unformed.
The probability of remaining at $z$ is $p_{z,z}=\frac{z}{n(n-1)}\Psi(\frac{z}{n(n-1)})+(1-\frac{z}{n(n-1)})(1-\Psi(\frac{z}{n(n-1)}))$.

The dynamic process defines a Markov chain over the state space $\mathbb{Z}$.
It is irreducible (because there is a positive probability of moving
between any two states in a finite number of steps) and aperiodic
(because there is a positive probability of remaining in a state).
Therefore, the Markov chain is ergodic, and hence by the Ergodic Theorem
it has a unique stationary distribution. Let $\pi=(\pi_{z},z\in\mathbb{Z})$
denote the stationary distribution. Following \citet{Myatt_Wallace_2004},
we can show that for any state $0\leq z<n(n-1)$,
\[
\frac{\pi_{z}}{\pi_{z+1}}=\frac{p_{z+1,z}}{p_{z,z+1}}=\frac{\frac{z+1}{n(n-1)}\left(1-\Psi\left(\frac{z+1}{n(n-1)}\right)\right)}{\left(1-\frac{z}{n(n-1)}\right)\Psi\left(\frac{z}{n(n-1)}\right)}.
\]

Consider Bayesian Nash equilibria (BNE) in the static model. Let $p=\Pr(G_{ij}=1)$
denote the probability of forming a link. Because links are independent
(because of the absence of $X$ and separability of utility), a BNE
can be represented by the link choice probability $p$. In particular,
an equilibrium $p$ solves the fixed point equation $p=\Psi(p)$.\footnote{This equation does not depend on $n$, nor does an equilibrium $p$. }
We show that the BNE in the static model can be related to the stationary
distribution in the dynamic model. In particular, specific, we follow
\citet[Proposition 1]{Myatt_Wallace_2004} and show that the local
maxima in the stationary distribution coincide with the stable Bayesian
Nash equilibria in the static model. A BNE $p^{*}$ such that $p^{*}=\Psi(p^{*})$
is stable if $p<\Psi(p)$ for $p<p^{*}$ and $p>\Psi(p)$ for $p>p^{*}$.

\begin{prop}
The local maxima (modes) of the stationary distribution $\pi$ coincide
with the stable Bayesian Nash equilibria of the static model. The
local minima of the stationary distribution $\pi$ coincide with the
unstable Bayesian Nash equilibria of the static model. Formally, let
$\left\lfloor x\right\rfloor $ denote the largest integer below $x$,
and $\left\lceil x\right\rceil $ denote the smallest integer above
$x$. For sufficiently large $n$, we have $\pi_{\left\lfloor pn(n-1)\right\rfloor }<\pi_{\left\lceil pn(n-1)\right\rceil }$
for $p<\Psi(p)$ and $\pi_{\left\lfloor pn(n-1)\right\rfloor }>\pi_{\left\lceil pn(n-1)\right\rceil }$
for $p>\Psi(p)$.
\end{prop}
\begin{proof}
For any $p$, 
\begin{equation}
\frac{\pi_{\left\lfloor pn(n-1)\right\rfloor }}{\pi_{\left\lceil pn(n-1)\right\rceil }}=\frac{p_{\left\lceil pn(n-1)\right\rceil ,\left\lfloor pn(n-1)\right\rfloor }}{p_{\left\lfloor pn(n-1)\right\rfloor ,\left\lceil pn(n-1)\right\rceil }}=\frac{\frac{\left\lceil pn(n-1)\right\rceil }{n(n-1)}\left(1-\Psi\left(\frac{\left\lceil pn(n-1)\right\rceil }{n(n-1)}\right)\right)}{\left(1-\frac{\left\lfloor pn(n-1)\right\rfloor }{n(n-1)}\right)\Psi\left(\frac{\left\lfloor pn(n-1)\right\rfloor }{n(n-1)}\right)}.\label{eq:dynamic1.ratio}
\end{equation}
Note that $\frac{\left\lceil pn(n-1)\right\rceil }{n(n-1)}$ and $\frac{\left\lfloor pn(n-1)\right\rfloor }{n(n-1)}\rightarrow p$
as $n\rightarrow\infty$. Because $\Psi(p)$ is continuous in $p$,
for sufficiently large $n$, the ratio in (\ref{eq:dynamic1.ratio})
can be made arbitrarily close to $\frac{p(1-\Psi(p))}{(1-p)\Psi(p)}$.
Since $\frac{p(1-\Psi(p))}{(1-p)\Psi(p)}<1$ if and only if $p<\Psi(p)$,
we derive that for sufficiently large $n$, $\frac{\pi_{\left\lfloor pn(n-1)\right\rfloor }}{\pi_{\left\lceil pn(n-1)\right\rceil }}<1$
for $p<\Psi(p)$ and $\frac{\pi_{\left\lfloor pn(n-1)\right\rfloor }}{\pi_{\left\lceil pn(n-1)\right\rceil }}>1$
for $p>\Psi(p)$.
\end{proof}

\subsection{Simultaneous Updating}

We follow \citet{Myatt_Wallace_2003} to construct the second dynamic
process of network formation, where all the links in a network are
updated in each period. We follow the setting in the first process
and assume that there is no $X$ nor the effect from friends in common. 

The initial period is the same as that in the first process: individuals
simultaneously form links based on their arbitrary beliefs. After
the initial period, individuals simultaneously update their links
in each period. Specifically, in period $t$, each individual $i$
receives a new draw of $\epsilon_{i,t}=(\epsilon_{ij,t},j\neq i)$
and observes the distribution of links formed in period $t-1$. Based
on the information, each $i$ updates links $G_{i,t}=(G_{ij,t},j\neq i)$
accordingly.

Let $Z_{t-1}\in\mathbb{Z}=\{0,1,\ldots,n(n-1)\}$ denote the total
number of links formed in period $t-1$. In period $t$, individual
$i$ forms a belief about link $G_{jk,t}$ by the fraction of links
formed in the previous period $\mathbb{E}[G_{jk,t}|Z_{t-1}]=\frac{Z_{t-1}}{n(n-1)}$
for all $j\neq i$ and $k\neq j$. The optimal choice for link $G_{ij,t}$
is $G_{ij,t}=1\{\beta_{1}+\frac{Z_{t-1}}{n(n-1)}(\beta_{4}+\beta_{5}+\beta_{6})-\epsilon_{ij,t}\geq0\}$.
The probability that individual $i$ forms link $G_{ij,t}$ is $\Pr(G_{ij,t}=1|Z_{t-1})=F_{\epsilon}(\beta_{1}+\frac{Z_{t-1}}{n(n-1)}(\beta_{4}+\beta_{5}+\beta_{6}))=\Psi(\frac{Z_{t-1}}{n(n-1)})$,
where we define $\Psi(p)=F_{\epsilon}(\beta_{1}+p(\beta_{4}+\beta_{5}+\beta_{6}))$.

Let $p_{t}=\frac{Z_{t}}{n(n-1)}=\frac{1}{n(n-1)}\sum_{i}\sum_{j\neq i}G_{ij,t}$
denote the fraction of links formed in period $t$. Proposition \ref{prop:dynamic2}
shows that the link fraction $p_{t}$ converges to a Bayesian Nash
equilibrium (BNE) in the static model. Recall that a BNE $p^{\ast}$
solves the equation $p^{\ast}=\Psi(p^{\ast})$.
\begin{prop}
\label{prop:dynamic2}Let $\bar{p}_{t}$, $t\geq0$, be an iterative
process defined by $\bar{p}_{t}=\Psi(\bar{p}_{t-1})$ with $\bar{p}_{0}=p_{0}$.
If there exists an equilibrium $p^{\ast}$ such that $p^{\ast}=\Psi(p^{\ast})$
and $\bar{p}_{t}\rightarrow p^{\ast}$ as $t\rightarrow\infty$, then
$p_{t}-p^{\ast}=o_{p}(1)$ as $n,t\rightarrow\infty$.
\end{prop}
\begin{proof}
Conditional on $p_{t-1}$, the link fraction $p_{t}$ has expectation
$\mathbb{E}[p_{t}|p_{t-1}]=\Psi(p_{t-1})$. By the law of large numbers
$p_{t}=\Psi(p_{t-1})+o_{p}(1)$ as $n\rightarrow\infty$. Because
$\Psi(p)$ is continuous in $p$, applying the continuous mapping
theorem iteratively we derive $p_{t}=\bar{p}_{t}+o_{p}(1)$, $t=1,2,\ldots$.
That is, each link fraction $p_{t}$ is asymptotically close to its
deterministic counterpart $\bar{p}_{t}$ as $n\rightarrow\infty$.
By the triangle inequality $|p_{t}-p^{\ast}|\leq|p_{t}-\bar{p}_{t}|+|\bar{p}_{t}-p^{\ast}|$,
where $|p_{t}-\bar{p}_{t}|=o_{p}(1)$ for all $t$ as $n\rightarrow\infty$,
and $|\bar{p}_{t}-p^{*}|=o(1)$ as $t\rightarrow\infty$. Therefore,
as $n,t\rightarrow\infty$, we have $p_{t}-p^{\ast}=o_{p}(1)$. 
\end{proof}

\section{\label{online:partition}Partition Representation of Optimal Decisions}

In this section, we establish a one-to-one mapping between the optimal
decision of an individual and a partition of the $\epsilon_{i}$ space
$\mathbb{R}^{n-1}$. The partition representation of optimal decisions
is useful for analyzing the properties of link choice probabilities.

Recall that for the expected utility in (\ref{eq:EU}), the expected
marginal utility of individual $i$ from forming a link to individual
$j$ is given by $\frac{1}{n-1}(\mathbb{E}[u_{ij}|X]+\frac{1}{n-2}\sum_{k\neq i,j}G_{ik}\mathbb{E}[v_{i,jk}|X]-\epsilon_{ij})$,
where we have used the symmetry of $v_{i,jk}$ in $j$ and $k$ ($v_{i,jk}=v_{i,kj}$).
If $G_{i}\in\mathcal{G}_{i}$ is an optimal decision, it must satisfy
that for each $j\neq i$, $G_{ij}=1$ if and only if the expected
marginal utility from the link is nonnegative. This yields the system
of equations
\begin{equation}
G_{ij}=1\left\{ \mathbb{E}[u_{ij}|X]+\frac{1}{n-2}\sum_{k\neq i,j}G_{ik}\mathbb{E}[v_{i,jk}|X]\geq\epsilon_{ij}\right\} ,\text{ }\forall j\neq i.\label{eq:optG.local}
\end{equation}
There may be multiple solutions to (\ref{eq:optG.local}). For example,
assume that $\mathbb{E}[v_{i,jk}|X]>0$, so link choices are strategic
complements. For $\epsilon_{i}\in\mathbb{R}^{n-1}$ such that $\mathbb{E}[u_{ij}|X]<\epsilon_{ij}\leq\mathbb{E}[u_{ij}|X]+\frac{1}{n-2}\sum_{k\neq i,j}\mathbb{E}[v_{i,jk}|X]$,
$\forall j\neq i$, we find that both $G_{i}^{1}=(G_{ij}=1,\forall j\neq i)$
and $G_{i}^{0}=(G_{ij}=0,\forall j\neq i)$ are solutions to (\ref{eq:optG.local}).
We refer to (\ref{eq:optG.local}) as the local optimal condition.

Among the local solutions to (\ref{eq:optG.local}), an optimal decision
$G_{i}$ achieves the highest expected utility, that is,
\begin{eqnarray}
 &  & \frac{1}{n-1}\sum_{j\neq i}G_{ij}\left(\mathbb{E}[u_{ij}|X]+\frac{1}{2(n-2)}\sum_{k\neq i,j}G_{ik}\mathbb{E}[v_{i,jk}|X]-\epsilon_{ij}\right)\nonumber \\
 & \geq & \max_{\substack{\tilde{g}_{i}\in\mathcal{G}_{i}\\
\tilde{g}_{i}\text{ satisfies (\ref{eq:optG.local})}
}
}\frac{1}{n-1}\sum_{j\neq i}\tilde{g}_{ij}\left(\mathbb{E}[u_{ij}|X]+\frac{1}{2(n-2)}\sum_{k\neq i,j}\tilde{g}_{ik}\mathbb{E}[v_{i,jk}|X]-\epsilon_{ij}\right).\label{eq:optG.global}
\end{eqnarray}
Because $\epsilon_{i}$ follows a continuous distribution (Assumption
\ref{ass:e=000026x}(ii)), two decisions achieve the same expected
utility with probability zero. Therefore, there is a unique solution
to (\ref{eq:optG.global}) with probability one. We refer to (\ref{eq:optG.global})
as the global optimal condition.

For each $g_{i}\in\mathcal{G}_{i}$, define the set 
\begin{equation}
\mathcal{E}_{i}(g_{i},X)\equiv\{\epsilon_{i}\in\mathbb{R}^{n-1}:g_{i}\text{ satisfies conditions (\ref{eq:optG.local}) and (\ref{eq:optG.global})}\}.\label{eq:optG.set}
\end{equation}
It represents the collection of $\epsilon_{i}\in\mathbb{R}^{n-1}$
such that $g_{i}$ is the optimal solution. Note that since $\epsilon_{i}$
has an unbounded support $\mathbb{R}^{n-1}$, the set $\mathcal{E}_{i}(g_{i},X)$
is nonempty for all $g_{i}\in\mathcal{G}_{i}$. Because there is a
unique optimal decision for almost all $\epsilon_{i}\in\mathbb{R}^{n-1}$,
the sets $\{\mathcal{E}_{i}(g_{i},X),g_{i}\in\mathcal{G}_{i}\}$ form
a partition of $\mathbb{R}^{n-1}$ with probability one. The results
are summarized in Lemma \ref{lem:partE}.
\begin{lem}
\label{lem:partE}Suppose that Assumption \ref{ass:e=000026x} is
satisfied. For each $i$, an optimal decision $G_{i}\in\mathcal{G}_{i}$
satisfies\textup{ conditions $(\ref{eq:optG.local})$ and $(\ref{eq:optG.global})$}.
Moreover, the sets\textup{ $\{\mathcal{E}_{i}(g_{i},X),g_{i}\in\mathcal{G}_{i}\}$}
form a partition of $\mathbb{R}^{n-1}$ with probability one.
\end{lem}
The local optimal condition $(\ref{eq:optG.local})$ resembles pure-strategy
Nash equilibria in static games of complete information \citep{Ciliberto_Tamer_2009}.
Strategic interactions between link choices arise due to the effect
of friends in common ($\mathbb{E}[v_{i,jk}|X]\neq0$). The multiplicity
of solutions to $(\ref{eq:optG.local})$ mirrors the multiplicity
of equilibria, while the global optimal condition $(\ref{eq:optG.global})$
serves as an equilibrium selection mechanism \citep{Ciliberto_Tamer_2009}.

\section{\label{online:implementation}Implementing the Estimation and Inference}

Now we discuss how to implement the estimation and inference procedure
in Section \ref{sec:estimation}. Let $\hat{p}_{n}=(\hat{p}_{n,(st)},1\leq s,t\leq T)'$
denote the first-step estimator. In the second step, we estimate the
parameter $\theta$ by GMM based on the sample moment (\ref{eq:m_hat}).
The main challenge lies in computing $P_{n,ij}(\theta,\hat{p}_{n})$.
As mentioned in Remark \ref{rem:P_sim}, this finite-$n$ link choice
probability must be computed through simulation. Algorithm \ref{alg:P_sim}
outlines the specific steps to simulate the type-specific link choice
probability $P_{n,(st)}(\theta,\hat{p}_{n})$, $1\leq s,t\le T$,
for each $\theta$ to be evaluated.
\begin{lyxalgorithm}
\label{alg:P_sim}For each $\theta$ to be evaluated, we simulate
$P_{n,(st)}(\theta,\hat{p}_{n})$, $1\leq s,t\le T$, in the following
steps:

1. Draw i.i.d. $\epsilon_{ij,r}$ from $F_{\epsilon}(\cdot;\theta_{\epsilon})$,
$1\le i\neq j\leq n$, $1\leq r\leq R$. Let $\epsilon_{i,r}=(\epsilon_{ij,r},j\neq i)$.

2. For each $r$ and each $i$, solve for $\omega_{ni}(\epsilon_{i,r},\theta,\hat{p}_{n})$
from the first-order condition (\ref{eq:foc.maxmin}).\footnote{\label{fn:foc_wn}To simplify computation, we multiply equation (\ref{eq:foc.maxmin})
by $\Phi_{ni}$ and solve for $\Phi_{ni}\Lambda_{ni}\omega_{ni}(\epsilon_{i,r})$
directly from the equation $\frac{1}{n-1}\sum_{j\neq i}1\{U_{n,ij}+\frac{n-1}{n-2}Z'_{j}\Phi_{ni}\Lambda_{ni}\omega\geq\epsilon_{ij,r}\}V_{ni}Z_{j}=\Phi_{ni}\Lambda_{ni}\omega$.}

3. For each $r$, each $i$, and each $j\neq i$, compute $G_{n,ij}(\epsilon_{i,r},\theta,\hat{p}_{n})$
using equation (\ref{eq:gij}).

4. For $1\leq s,t\leq T$, compute $P_{n,(st)}(\theta,\hat{p}_{n})$
by $\sum_{r=1}^{R}\sum_{i=n}^{n}\sum_{j\neq i}G_{n,ij}(\epsilon_{i,r},\theta,\hat{p}_{n})1\{X_{i}=x_{s},X_{j}=x_{t}\}/(R\sum_{i=1}^{n}\sum_{j\neq i}1\{X_{i}=x_{s},X_{j}=x_{t}\}).$
\end{lyxalgorithm}

\paragraph*{Standard errors.}

The asymptotic variance of the estimator $\hat{\theta}_{n}$ is $\frac{1}{n(n-1)}J_{n}^{-1}\Sigma_{n}(J'_{n})^{-1}$.
This asymptotic variance can be estimated using a plug-in estimator.
Specifically, recall that $J_{n}=\frac{1}{n(n-1)}\sum_{i}\sum_{j\neq i}q_{n,ij}\nabla_{\theta^{\prime}}P_{n,ij}(\theta_{0},p_{n})$
(Assumption \ref{ass:theta.asymdist}). We estimate $J_{n}$ by $\hat{J}_{n}=\frac{1}{n(n-1)}\sum_{i}\sum_{j\neq i}\hat{q}_{n,ij}\nabla_{\theta^{\prime}}P_{n,ij}(\hat{\theta}_{n},\hat{p}_{n})$,
where the derivative $\nabla_{\theta^{\prime}}P_{n,ij}$ is computed
numerically. Moreover, $\Sigma_{n}=\frac{1}{n(n-1)}\sum_{i}\sum_{j\neq i}\mathbb{E}[\phi_{n,ij}^{\theta}\phi_{n,ij}^{\theta\prime}|X]$
(Theorem \ref{thm:theta.asymdist}), where $\phi_{n,ij}^{\theta}=\tilde{q}_{n,ij}(g_{n,ij}(\omega_{ni}^{\ast},\epsilon_{ij})-P_{n,ij}^{\ast}(\omega_{ni}^{\ast}))+J_{ni}^{\omega}(\omega_{ni}^{\ast},\tilde{q}_{ni})\phi_{n,ij}^{\omega}(\omega_{ni}^{\ast},\epsilon_{ij})$
(equation (\ref{eq:theta.influence})). Hence, for each $i\neq j$,
\begin{eqnarray}
\mathbb{E}[\phi_{n,ij}^{\theta}\phi_{n,ij}^{\theta\prime}|X] & = & \tilde{q}_{n,ij}\tilde{q}'_{n,ij}P_{n,ij}^{\ast}(\omega_{ni}^{\ast})(1-P_{n,ij}^{\ast}(\omega_{ni}^{\ast}))\nonumber \\
 &  & +J_{ni}^{\omega}(\omega_{ni}^{\ast},\tilde{q}_{ni})\mathbb{E}[\phi_{n,ij}^{\omega}(\omega_{ni}^{\ast},\epsilon_{ij})\phi_{n,ij}^{\omega\prime}(\omega_{ni}^{\ast},\epsilon_{ij})|X]J_{ni}^{\omega\prime}(\omega_{ni}^{\ast},\tilde{q}_{ni})\nonumber \\
 &  & +\tilde{q}_{n,ij}\mathbb{E}[(g_{n,ij}(\omega_{ni}^{\ast},\epsilon_{ij})-P_{n,ij}^{\ast}(\omega_{ni}^{\ast}))\phi_{n,ij}^{\omega\prime}(\omega_{ni}^{\ast},\epsilon_{ij})|X]J_{ni}^{\omega\prime}(\omega_{ni}^{\ast},\tilde{q}_{ni})\nonumber \\
 &  & +J_{ni}^{\omega}(\omega_{ni}^{\ast},\tilde{q}_{ni})\mathbb{E}[\phi_{n,ij}^{\omega}(\omega_{ni}^{\ast},\epsilon_{ij})(g_{n,ij}(\omega_{ni}^{\ast},\epsilon_{ij})-P_{n,ij}^{\ast}(\omega_{ni}^{\ast}))|X]\tilde{q}'_{n,ij}.\label{eq:phi_theta_var}
\end{eqnarray}
To estimate $\mathbb{E}[\phi_{n,ij}^{\theta}\phi_{n,ij}^{\theta\prime}|X]$,
we estimate $\tilde{q}_{n,ij}$ (equation (\ref{eq:q_tilde})) by
a plug-in estimator. The auxiliary variable $\omega_{ni}^{\ast}$
is a maximin solution of (\ref{eq:Pai*}), which can be solved from
the first-order condition (\ref{eq:foc.w}).\footnote{Similarly to footnote \ref{fn:foc_wn}, we multiply equation (\ref{eq:foc.w})
by $\Phi_{ni}$ and solve for $\Phi_{ni}\Lambda_{ni}\omega_{ni}^{*}$
directly from the equation $\frac{1}{n-1}\sum_{j\neq i}F_{\epsilon}(U_{n,ij}+\frac{n-1}{n-2}Z'_{j}\Phi_{ni}\Lambda_{ni}\omega)V_{ni}Z_{j}=\Phi_{ni}\Lambda_{ni}\omega$.
Note that all the terms in (\ref{eq:phi_theta_var}) involving $\omega_{ni}^{*}$
can be expressed as a function of $\Phi_{ni}\Lambda_{ni}\omega_{ni}^{*}$.} Both $P_{n,ij}^{\ast}(\omega_{ni}^{\ast})$ and $J_{ni}^{\omega}(\omega_{ni}^{\ast},\tilde{q}_{ni})$
(Lemma \ref{lem:m.clt}) are thus estimated by plug-in estimators,
where the derivatives in $J_{ni}^{\omega}(\omega_{ni}^{\ast},\tilde{q}_{ni})$
are computed numerically. Note that $\phi_{n,ij}^{\omega}(\omega_{ni}^{\ast},\epsilon_{ij})$
(Lemma \ref{lem:w.asymlin}) depends on $\epsilon_{ij}$ only through
$g_{n,ij}(\omega_{ni}^{\ast},\epsilon_{ij})$. Therefore, the three
expectation terms in (\ref{eq:phi_theta_var}) have analytical forms
and are estimated by plug-in estimators.

\section{\label{online:undi.estimation}Estimation in Undirected Networks}

In this section, we extend the estimation procedure in Section \ref{sec:estimation}
to undirected networks. Because proposals depend on the network size
$n$, we add the subscript $n$ hereafter. 

Recall that the expected utility in (\ref{eq:EU.undi}) involves $\mathbb{E}[D_{n,ij}|X]$
and $\mathbb{E}[D_{n,ij}D_{n,ik}|X]$. For $1\leq r,s,t\leq T$, we
define the type-specific counterparts
\begin{eqnarray*}
p_{n,(st)} & = & \mathbb{E}[D_{n,ij}|X_{i}=x_{s},X_{j}=x_{t},X],\\
q_{n,(rst)} & = & \mathbb{E}[D_{n,ij}D_{n,ik}|X_{i}=x_{r},X_{j}=x_{s},X_{k}=x_{t},X].
\end{eqnarray*}
These conditional probabilities are not directly estimable because
the proposals are not observed. Nevertheless, we can treat them as
parameters and estimate them jointly with the model parameters in
$\theta$ using the observed links. Define the $T\times T$ matrix
$p_{n}=(p_{n,(st)},1\leq s,t\leq T)$ and the $T\times T\times T$
array $q_{n}=(q_{n,(rst)},1\leq r,s,t\leq T)$. Note that $\mathbb{E}[D_{n,ij}D_{n,ik}|X]$
is symmetric in $j$ and $k$, and thus $q_{n,(rst)}$ is symmetric
in $s$ and $t$. In total, we have $T^{2}+\frac{1}{2}T^{2}(T+1)$
parameters in $p_{n}$ and $q_{n}$, in addition to the $d_{\theta}$
parameters in $\theta$.

Next we relate $p_{n}$ and $q_{n}$ to the observed links. For $1\leq r,s,t\leq T$,
we define 
\begin{eqnarray*}
\pi_{n,(st)} & = & \mathbb{E}[G_{n,ij}|X_{i}=x_{s},X_{j}=x_{t},X],\\
\tau_{n,(rst)} & = & \mathbb{E}[G_{n,ij}G_{n,ik}|X_{i}=x_{r},X_{j}=x_{s},X_{k}=x_{t},X].
\end{eqnarray*}
These conditional probabilities can be estimated from the observed
links using relative frequency estimators. Because $G_{ij}=D_{ij}D_{ji}$,
and $D_{i}$ and $D_{j}$ are independent conditional on $X$, it
follows that
\begin{eqnarray}
\pi_{n,(st)} & = & p_{n,(st)}p_{n,(ts)},\nonumber \\
\tau_{n,(rst)} & = & q_{n,(rst)}p_{n,(sr)}p_{n,(tr)}.\label{eq:mom1}
\end{eqnarray}
Because both $\pi_{n,(st)}$ and $\tau_{n,(rst)}$ are symmetric in
$s$ and $t$, we have a total of $\frac{1}{2}T(T+1)+\frac{1}{2}T^{2}(T+1)$
equations in (\ref{eq:mom1}). For $T\geq2$, additional restrictions
are necessary for identification because $T^{2}>\frac{1}{2}T(T+1)$.

These additional restrictions can be derived from model-implied probabilities.
From Corollary \ref{cor:optD}, an optimal proposal can be represented
as $D_{n,ij}(\epsilon_{i},\theta,p_{n},q_{n})$. For $1\leq r,s,t\leq T$,
we define the type-specific model-implied probabilities
\begin{eqnarray*}
P_{n,(st)}(\theta,p_{n},q_{n}) & = & \mathbb{E}[D_{n,ij}(\epsilon_{i},\theta,p_{n},q_{n})|X_{i}=x_{s},X_{j}=x_{t},X],\\
Q_{n,(rst)}(\theta,p_{n},q_{n}) & = & \mathbb{E}[D_{n,ij}(\epsilon_{i},\theta,p_{n},q_{n})D_{n,ik}(\epsilon_{i},\theta,p_{n},q_{n})|X_{i}=x_{r},X_{j}=x_{s},X_{k}=x_{t},X].
\end{eqnarray*}
Relating these model-implied probabilities to the distribution of
observed links yields
\begin{eqnarray}
\pi_{n,(st)} & = & P_{n,(st)}(\theta,p_{n},q_{n})P_{n,(ts)}(\theta,p_{n},q_{n}),\nonumber \\
\tau_{n,(rst)} & = & Q_{n,(rst)}(\theta,p_{n},q_{n})P_{n,(sr)}(\theta,p_{n},q_{n})P_{n,(tr)}(\theta,p_{n},q_{n}).\label{eq:mom2}
\end{eqnarray}
The equations in (\ref{eq:mom1}) and (\ref{eq:mom2}) resemble the
moment conditions in the first and second steps in the directed setting,
respectively. By combining (\ref{eq:mom1}) and (\ref{eq:mom2}),
we obtain a total of $T(T+1)+T^{2}(T+1)$ equations. Therefore, it
is possible to identify the parameters if the number of types $T$
is sufficiently large (i.e., $T+\frac{1}{2}T^{2}(T+1)>d_{\theta}$).\footnote{For instance, when $T=2$, we have $18$ equations and $10+d_{\theta}$
parameters. When $T=3$, we have $48$ equations and $27+d_{\theta}$
parameters.} An in-depth analysis of this procedure is left for future research.

\section{\label{online:simulation}Monte Carlo Simulation}

In this section, we evaluate our approach in a simulation study. We
consider the specification
\begin{eqnarray*}
U_{i}(G,X,\epsilon_{i}) & = & \frac{1}{n-1}\sum_{j\neq i}G_{ij}\left(\beta_{1}+X_{i}\beta_{2}+|X_{i}-X_{j}|\beta_{3}+\frac{1}{n-2}\sum_{k\neq i,j}G_{jk}\beta_{4}\right.\\
 &  & \left.+\frac{1}{2(n-2)}\sum_{k\neq i,j}G_{ik}(G_{jk}+G_{kj})\gamma-\epsilon_{ij}\right),
\end{eqnarray*}
where $X_{i}$ is an i.i.d. binary variable with equal probability
of being $0$ or $1$, and $\epsilon_{ij}$ is i.i.d. following $N(\ensuremath{0,1)}$.
The true parameter values are given by $(\beta_{1},\beta_{2},\beta_{3},\beta_{4},\gamma)=(-1,1,-2,1,1)$,
where $\beta_{3}$ represents the homophily effect, $\beta_{4}$ represents
the effect of friends of friends, and $\gamma$ represents the effect
of friends in common. We consider a variety of network sizes $n=10,25,50,100,250,500$.

For each $n$, we generate a single directed network as follows. First,
we generate a characteristic profile $X$ and compute a Bayesian Nash
equilibrium $\sigma$.\footnote{We compute an equilibrium by iterating equation (\ref{eq:BNE}) from
an initial value. The initial value we use is an equilibrium in the
limiting game, which is computed by solving for the limiting version
of equation (\ref{eq:BNE}), where we replace the finite-$n$ choice
probability on the right-hand side of (\ref{eq:BNE}) by its limiting
counterpart.} Second, we use the equilibrium $\sigma$ to compute $U_{n,ij}(X,\sigma)$
and $V_{ni}(X,\sigma)$. Third, we compute $\omega_{ni}(\epsilon_{i})$
for a simulated $\epsilon_{i}$ and obtain the optimal link choices
from equation (\ref{eq:gij}).\footnote{For small networks, it could be computationally competitive to maximize
expected utility (\ref{eq:EU}) directly by quadratic integer programming
(QIP). In our simulation study, we compute the optimal link choices
using QIP for $n\leq100$ and equation (\ref{eq:gij}) for $n>100$.
We solve QIP using the solver \textit{cplexmiqp} provided in CPLEX.
QIP is also a check on whether the link choices in (\ref{eq:gij})
maximize the expected utility. We find that the simulated link choice
probabilities based on QIP and those based on (\ref{eq:gij}) are
indeed the same.} Each experiment is repeated 100 times.

We estimate the parameters by two-step GMM, where we first estimate
the link choice probabilities by the frequency estimator and then
estimate the parameters by GMM. We use the moment function in (\ref{eq:m_hat})
and the instrument in (\ref{eq:instrument}) and consider three varieties
of the second step. (i) Both the moment function and instrument are
constructed using the finite-$n$ link probabilities. (ii) The moment
function is constructed using the finite-$n$ link probabilities,
but the instrument is constructed using the limiting link probabilities.
(iii) Both the moment function and instrument are constructed using
the limiting link probabilities. Table \ref{tab:simulation} reports
the average biases and root MSEs of the estimates in the three cases.
\begin{table}
\centering  \begin{threeparttable}

\caption{\label{tab:simulation}Two-Step GMM Estimation in Simulated Data}
\begin{tabular}{llr@{\extracolsep{0pt}.}lr@{\extracolsep{0pt}.}lr@{\extracolsep{0pt}.}lr@{\extracolsep{0pt}.}lr@{\extracolsep{0pt}.}lr@{\extracolsep{0pt}.}l}
\toprule 
 &  & \multicolumn{4}{c}{Case (i)} & \multicolumn{4}{c}{Case (ii)} & \multicolumn{4}{c}{Case (iii)}\tabularnewline
\cmidrule{3-14}
 & Para. & \multicolumn{2}{c}{Bias} & \multicolumn{2}{c}{rMSE} & \multicolumn{2}{c}{Bias} & \multicolumn{2}{c}{rMSE} & \multicolumn{2}{c}{Bias} & \multicolumn{2}{c}{rMSE}\tabularnewline
\midrule 
$n=10$ & $\beta_{1}$ & 0&005 & 0&206 & -0&008 & 0&438 & -0&152 & 2&284\tabularnewline
 & $\beta_{2}$ & -0&042 & 0&191 & 0&273 & 1&010 & 1&806 & 3&004\tabularnewline
 & $\beta_{3}$ & 0&063 & 0&402 & -0&940 & 3&338 & -4&469 & 3&649\tabularnewline
 & $\beta_{4}$ & -0&019 & 0&185 & -0&166 & 1&900 & -3&626 & 8&890\tabularnewline
 & $\gamma$ & -0&021 & 0&182 & -0&057 & 1&004 & -1&194 & 6&438\tabularnewline
\midrule
$n=25$ & $\beta_{1}$ & -0&010 & 0&066 & -0&017 & 0&097 & 0&281 & 0&447\tabularnewline
 & $\beta_{2}$ & 0&006 & 0&109 & 0&012 & 0&185 & 1&639 & 2&029\tabularnewline
 & $\beta_{3}$ & -0&038 & 0&194 & -0&065 & 0&268 & -1&899 & 2&152\tabularnewline
 & $\beta_{4}$ & 0&003 & 0&092 & 0&016 & 0&232 & -2&835 & 3&710\tabularnewline
 & $\gamma$ & -0&004 & 0&098 & -0&014 & 0&146 & -1&887 & 3&948\tabularnewline
\midrule
$n=50$ & $\beta_{1}$ & -0&003 & 0&042 & -0&010 & 0&052 & 0&014 & 0&126\tabularnewline
 & $\beta_{2}$ & -0&001 & 0&065 & -0&005 & 0&070 & 0&058 & 0&499\tabularnewline
 & $\beta_{3}$ & -0&001 & 0&097 & 0&005 & 0&101 & -0&064 & 0&499\tabularnewline
 & $\beta_{4}$ & 0&020 & 0&083 & 0&050 & 0&110 & -0&142 & 0&921\tabularnewline
 & $\gamma$ & -0&012 & 0&072 & -0&016 & 0&094 & -0&091 & 0&551\tabularnewline
\midrule
$n=100$ & $\beta_{1}$ & 0&004 & 0&023 & 0&005 & 0&023 & 0&005 & 0&034\tabularnewline
 & $\beta_{2}$ & -0&007 & 0&036 & -0&009 & 0&040 & 0&008 & 0&084\tabularnewline
 & $\beta_{3}$ & -0&010 & 0&052 & -0&010 & 0&050 & -0&007 & 0&084\tabularnewline
 & $\beta_{4}$ & 0&031 & 0&064 & 0&034 & 0&073 & -0&015 & 0&165\tabularnewline
 & $\gamma$ & -0&019 & 0&055 & -0&021 & 0&062 & -0&041 & 0&208\tabularnewline
\midrule
$n=250$ & $\beta_{1}$ & 0&002 & 0&008 & 0&002 & 0&008 & -0&001 & 0&014\tabularnewline
 & $\beta_{2}$ & -0&001 & 0&017 & 0&000 & 0&018 & 0&004 & 0&039\tabularnewline
 & $\beta_{3}$ & 0&000 & 0&020 & -0&001 & 0&021 & -0&003 & 0&037\tabularnewline
 & $\beta_{4}$ & 0&027 & 0&035 & 0&031 & 0&038 & 0&009 & 0&075\tabularnewline
 & $\gamma$ & -0&013 & 0&033 & -0&017 & 0&036 & -0&031 & 0&173\tabularnewline
\midrule
$n=500$ & $\beta_{1}$ & -0&001 & 0&006 & -0&001 & 0&005 & -0&001 & 0&010\tabularnewline
 & $\beta_{2}$ & 0&007 & 0&011 & 0&010 & 0&013 & 0&001 & 0&022\tabularnewline
 & $\beta_{3}$ & 0&003 & 0&011 & 0&000 & 0&012 & 0&000 & 0&022\tabularnewline
 & $\beta_{4}$ & -0&002 & 0&028 & -0&001 & 0&034 & 0&006 & 0&047\tabularnewline
 & $\gamma$ & -0&005 & 0&019 & -0&011 & 0&025 & -0&014 & 0&103\tabularnewline
\bottomrule
\end{tabular}

\begin{tablenotes}[flushleft]\footnotesize \item
Note: Average biases and root MSEs  from 100 repeated samples. Case (i) uses the moment function and instrument based on the finite-$n$ link probabilities. Case (ii) uses the moment function based on the finite-$n$ link probabilities and the instrument based on the limiting link probabilities. Case (iii) uses the moment function and instrument based on the limiting link probabilities. The finite-$n$ link probabilities are computed from simulations by either solving quadratic integer programming (for $n\leq100$)  or applying  equation (\ref{eq:gij}) (for $n>100$).
\end{tablenotes}\end{threeparttable}
\end{table}

Columns 1 and 2 present the results in case (i). We compute the finite-$n$
link choice probabilities by simulation. Specifically, we draw a random
sample of $\epsilon_{i}$; for each simulated $\epsilon_{i}$, we
compute $\omega_{ni}(\epsilon_{i})$ and derive the optimal link choices
from equation (\ref{eq:gij}). The fraction of draws that result in
a link gives a simulated link choice probability. We calculate the
GMM estimator by continuous updating, where the instrument is computed
by simulation\footnote{We simulate the instrument using $\epsilon_{i}$ that are drawn independently
of those drawn to simulate the link choice probabilities in the moment
function.} and the derivative in the numerator in (\ref{eq:instrument}) is
approximated by a numerical derivative.\footnote{Because the sample moment is not everywhere differentiable, we use
the derivative-free optimization solver \textit{fminsearch} provided
in MATLAB when searching for the estimate of $\theta$.} The results show that the two-step GMM based on the finite-$n$ link
probabilities performs well. The estimates are close to the true values
even for network sizes as small as $n=25$. The root MSEs also decrease
as the network size increases, as expected.

Columns 3 and 4 present the results in case (ii). We simulate the
finite-$n$ link probabilities in the moment as described in case
(i), but the limiting link probabilities and their derivatives in
the instrument are computed without simulation. We find that for small
networks (e.g. $n=10$), the estimates are biased and the root MSEs
are larger than those in case (i). But for larger networks, the biases
and root MSEs become close to those in case (i). These results suggest
that in large networks, the computationally convenient limiting link
probabilities can be used to approximate the instrument without sacrificing
the estimation precision.

Columns 5 and 6 present the results in case (iii). Because both the
moment function and instrument are constructed using the limiting
link probabilities, which can be computed without simulation, this
case is the most computationally convenient among the three.\footnote{The moment condition in this case is equal to the first-order condition
from QMLE based on the limiting link probabilities, so we estimate
$\theta$ equivalently by QMLE based on the limiting link probabilities.} The results show that the estimates present large biases in small
networks, but these biases vanishes as the network size grows, suggesting
that the estimator based on the limiting link probabilities is consistent.
The root MSEs are generally larger than those in cases (i) and (ii),
but become similar once the networks are sufficiently large (e.g.,
$n=500$).

In sum, the two-step estimation procedure based on the finite-$n$
link probabilities performs well even in relatively small networks.
In sufficiently large networks, the estimates based on the limiting
link probabilities can perform as well as those based on the finite-$n$
ones, regardless of whether we use the limiting link probabilities
for the instrument or the moment function. This parity suggests that
the limiting link probabilities can provide a useful approximation
for reducing the computational burden in large networks.

\section{\label{online:lemma}Additional Lemmas for Sections \ref{sec:model}--\ref{sec:extension}}

\paragraph{Notation}

For any random variable $Z\in\mathbb{R}^{n}$, $\Vert Z\Vert_{\psi|X}$
denotes the conditional Orlicz norm of $Z$ given $X$ for a non-decreasing,
convex function $\psi$ with $\psi(0)=0$; that is, $\Vert Z\Vert_{\psi|X}=\inf\{C>0:\mathbb{E}[\psi(\Vert Z\Vert/C)|X]\leq1\}$.
Conditional Orlicz norms satisfy the triangle inequality. For conditional
Orlicz norms $\Vert\cdot\Vert_{\psi_{p}|X}$ with the functions $\psi_{p}(z)=e^{z^{p}}-1$
for $p\geq1$, the bound $z^{p}\leq\psi_{p}(z)$ for all $z\geq0$
implies that $(\mathbb{E}[\Vert Z\Vert^{p}|X])^{1/p}\leq\Vert Z\Vert_{\psi_{p}|X}$
for all $p\geq1$.\footnote{This is true because $\mathbb{E}[\psi_{p}(\Vert Z\Vert/\Vert Z\Vert_{\psi_{p}|X})|X]\leq1\leq\mathbb{E}[\psi_{p}(\Vert Z\Vert/(\mathbb{E}[\Vert Z\Vert^{p}|X])^{1/p})|X]$,
where the second inequality follows from $z^{p}\leq\psi_{p}(z)$.} Moreover, $\Vert Z\Vert_{\psi_{p}|X}\leq\Vert Z\Vert_{\psi_{q}|X}(\ln2)^{1/q-1/p}$
for $p\leq q$ and $(\mathbb{E}[\Vert Z\Vert^{p}|X])^{1/p}\leq p!\Vert Z\Vert_{\psi_{1}|X}$
for $p\geq1$.

\subsection{Lemmas for Section \ref{sec:model}}
\begin{lem}[Continuity of CCP]
\label{lem:ccp.cont}Suppose that Assumption \ref{ass:e=000026x}
is satisfied. For any $g_{i}\in\mathcal{G}_{i}$, $P_{i}(g_{i},X,\sigma)=\Pr(G_{i}(\epsilon_{i},X,\sigma)=g_{i}|X)$
is continuous in $\sigma$.
\end{lem}
\begin{proof}
By Lemma \ref{lem:partE}, the optimal decision $G_{i}(\epsilon_{i},X,\sigma)$
takes the value $g_{i}\in\mathcal{G}_{i}$ if and only if $\epsilon_{i}\in\mathcal{E}_{i}(g_{i},X,\sigma)$,
where the set $\mathcal{E}_{i}(g_{i},X,\sigma)$ is defined in equation
(\ref{eq:optG.set}). We can write $P_{i}(g_{i},X,\sigma)=\int1\{\epsilon_{i}\in\mathcal{E}_{i}(g_{i},X,\sigma)\}f_{\epsilon_{i}}(\epsilon_{i};\theta_{\epsilon})d\epsilon_{i}$. 

The equations in (\ref{eq:optG.local}) define an orthant $\mathcal{E}_{i}^{(1)}(g_{i},X,\sigma)$
in $\mathbb{R}^{n-1}$ given by
\begin{equation}
\epsilon_{ij}\begin{cases}
\leq U_{ij}^{0}(X,\sigma)+\frac{1}{n-2}\sum_{k\neq i,j}g_{ik}V_{i,jk}^{0}(X,\sigma), & \text{if }g_{ij}=1,\\
>U_{ij}^{0}(X,\sigma)+\frac{1}{n-2}\sum_{k\neq i,j}g_{ik}V_{i,jk}^{0}(X,\sigma), & \text{if }g_{ij}=0,
\end{cases}\text{ }\label{eq:e.ineq.local}
\end{equation}
for all $j\neq i$, where $U_{ij}^{0}(X,\sigma)\equiv\mathbb{E}[u_{ij}|X]$,
and $V_{i,jk}^{0}(X,\sigma)\equiv\mathbb{E}[v_{i,jk}|X]$. Because
both $U_{ij}^{0}(X,\sigma)$ and $V_{i,jk}^{0}(X,\sigma)$ are continuous
in $\sigma$, the indicator function $1\{\epsilon_{i}\in\mathcal{E}_{i}^{(1)}(g_{i},X,\sigma)\}$
is continuous in $\sigma$, except on a null set of $\epsilon_{i}$.

Moreover, the inequality in (\ref{eq:optG.global}) defines a half-space
$\mathcal{E}_{i}^{(2)}(g_{i},X,\sigma)$ in $\mathbb{R}^{n-1}$ given
by
\begin{align}
 & \frac{1}{n-1}\sum_{j\neq i}(g_{ij}-\tilde{g}_{ij})\epsilon_{ij}\nonumber \\
\leq & \max_{\substack{\tilde{g}_{i}\in\mathcal{G}_{i}\\
\tilde{g}_{i}\text{ satisfies (\ref{eq:e.ineq.local}) }
}
}\frac{1}{n-1}\sum_{j\neq i}\left((g_{ij}-\tilde{g}_{ij})U_{ij}^{0}(X,\sigma)+\frac{1}{2(n-2)}\sum_{k\neq i,j}(g_{ij}g_{ik}-\tilde{g}_{ij}\tilde{g}_{ik})V_{i,jk}^{0}(X,\sigma)\right).\label{eq:e.ineq.global}
\end{align}
Each function inside the maximization on the right-hand side of (\ref{eq:e.ineq.global})
is continuous in $\sigma$. While the set of solutions to (\ref{eq:e.ineq.local})
for a given $\epsilon_{i}$ can be discontinuous in $\sigma$ (e.g.,
some link choices in an optimal $g_{i}$ may switch from $0$ to $1$
or the opposite as $\sigma$ changes), this event occurs on a null
set of $\epsilon_{i}$. Because max is a continuous operation, the
right-hand side of (\ref{eq:e.ineq.global}) is continuous in $\sigma$,
except on a null set. Hence, the indicator function $1\{\epsilon_{i}\in\mathcal{E}_{i}^{(2)}(g_{i},X,\sigma)\}$
is continuous in $\sigma$ almost surely in $\epsilon_{i}$. 

For any $g_{i}\in\mathcal{G}_{i}$, the set $\mathcal{E}_{i}(g_{i},X,\sigma)$
is the intersection of the orthant in (\ref{eq:e.ineq.local}) and
the half-space defined by (\ref{eq:e.ineq.global}), and thus $1\{\epsilon_{i}\in\mathcal{E}_{i}(g_{i},X,\sigma)\}=1\{\epsilon_{i}\in\mathcal{E}_{i}^{(1)}(g_{i},X,\sigma)\}\cdot1\{\epsilon_{i}\in\mathcal{E}_{i}^{(2)}(g_{i},X,\sigma)\}$.
This indicator function is continuous in $\sigma$ almost surely in
$\epsilon_{i}$. Moreover, $f_{\epsilon_{i}}(\epsilon_{i};\theta_{\epsilon})$
is bounded and continuous in $\theta_{\epsilon}$ by Assumption \ref{ass:e=000026x}(i)-(ii).
Therefore, by dominated convergence we conclude that $P_{i}(g_{i},X,\sigma)$
is continuous in $\sigma$.

\end{proof}

\subsection{Lemmas for Section \ref{sec:legendre}}
\begin{lem}[FOC for $\omega$]
\label{lem:foc.w}Suppose that Assumptions \ref{ass:e=000026x}--\ref{ass:discX}
are satisfied. An optimal $\omega_{i}(\epsilon_{i},X,\sigma)$ that
solves problem (\ref{eq:maxmin}) satisfies the first-order condition
\begin{eqnarray}
 &  & \frac{1}{n-1}\sum_{j\neq i}1\left\{ U_{ij}(X,\sigma)+\frac{n-1}{n-2}Z_{j}^{\prime}\Phi_{i}(X,\sigma)\Lambda_{i}(X,\sigma)\omega\geq\epsilon_{ij}\right\} \Lambda_{i}(X,\sigma)\Phi_{i}^{\prime}(X,\sigma)Z_{j}\nonumber \\
 & = & \Lambda_{i}(X,\sigma)\omega,\text{ a.s.}\label{eq:foc.maxmin}
\end{eqnarray}
\end{lem}
\begin{proof}
Observe that the objective function $\Pi_{i}(\omega,\epsilon_{i},X,\sigma)$
in problem (\ref{eq:maxmin}) is sub-differentiable in $\omega$.\footnote{Note that the function $\max\{x,0\}$ is differentiable at $x\neq0$
and sub-differentiable at $x=0$ with subderivatives in $[0,1]$.} By the optimality of $\omega_{i}(\epsilon_{i},X,\sigma)$, $\Pi_{i}(\omega,\epsilon_{i},X,\sigma)$
has subgradient $0$ at $\omega_{i}(\epsilon_{i},X,\sigma)$, that
is, $\omega_{i}(\epsilon_{i},X,\sigma)$ satisfies the first-order
condition, omitting $X$ and $\sigma$ in the notation
\begin{eqnarray}
 &  & \frac{1}{n-1}\sum_{j\neq i}1\left\{ U_{ij}+\frac{n-1}{n-2}Z_{j}^{\prime}\Phi_{i}\Lambda_{i}\omega>\epsilon_{ij}\right\} \Lambda_{i}\Phi_{i}^{\prime}Z_{j}-\Lambda_{i}\omega\nonumber \\
 & = & -\frac{1}{n-1}\sum_{j\neq i}1\left\{ U_{ij}+\frac{n-1}{n-2}Z_{j}^{\prime}\Phi_{i}\Lambda_{i}\omega=\epsilon_{ij}\right\} \text{diag}(\tau)\Lambda_{i}\Phi_{i}^{\prime}Z_{j},\label{eq:foc.w.sub}
\end{eqnarray}
for some $\tau=(\tau_{1},\ldots,\tau_{T})\in[0,1]^{T}$. Define the
right-hand side of (\ref{eq:foc.w.sub}) as $\Delta_{n}(\omega,\epsilon_{i},X,\sigma)$.
For any $\omega$, 
\begin{eqnarray}
 &  & \Pr(\|\Delta_{n}(\omega,\epsilon_{i},X,\sigma)\|>0|X)\nonumber \\
 & \leq & \Pr\left(\left.\exists j\neq i,U_{ij}(X,\sigma)+\frac{n-1}{n-2}Z_{j}^{\prime}\Phi_{i}(X,\sigma)\Lambda_{i}(X,\sigma)\omega=\epsilon_{ij}\right|X\right)\nonumber \\
 & \leq & \sum_{j\neq i}\Pr\left(\left.U_{ij}(X,\sigma)+\frac{n-1}{n-2}Z_{j}^{\prime}\Phi_{i}(X,\sigma)\Lambda_{i}(X,\sigma)\omega=\epsilon_{ij}\right|X\right)=0,\label{eq:foc.w.prob0}
\end{eqnarray}
where the last equality follows because $\epsilon_{ij}$ has a continuous
distribution. Hence the first-order condition (\ref{eq:foc.w.sub})
holds with $\Delta_{n}(\omega,\epsilon_{i},X,\sigma)$ replaced by
$0$ with probability one. By (\ref{eq:foc.w.prob0}) again, we can
replace $>$ in the indicator on the left-hand side of (\ref{eq:foc.w.sub})
by $\geq$ with probability one and the lemma is proved.
\end{proof}

\subsection{Lemmas for Section \ref{sec:estimation}}

\subsubsection{\label{online:lemma.theta.consist}Consistency of $\hat{\theta}_{n}$
and $\hat{p}_{n}$}
\begin{lem}[Uniform LLN of sample moments]
\label{lem:moment.lln}Suppose that Assumptions \ref{ass:e=000026x}--\ref{ass:discX}
and \ref{ass:theta.consist}(iii) are satisfied. Conditional on $X$,
$\sup_{(\theta,p)\in\Theta\times\mathcal{P}}\Vert\hat{m}_{n}(\theta,p)-m_{n}(\theta,p)\Vert=o_{p}(1)$.
\end{lem}
\begin{proof}
By the definitions of $\hat{m}_{n}$ and $m_{n}$,
\begin{eqnarray}
 &  & \hat{m}_{n}(\theta,p)-m_{n}(\theta,p)\nonumber \\
 & = & \frac{1}{n(n-1)}\sum_{i}\sum_{j\neq i}(\hat{q}_{n,ij}(G_{n,ij}-P_{n,ij}(\theta,p))-q_{n,ij}(\mathbb{E}[G_{n,ij}|X]-P_{n,ij}(\theta,p)))\nonumber \\
 & = & \frac{1}{n(n-1)}\sum_{i}\sum_{j\neq i}(\hat{q}_{n,ij}-q_{n,ij})(G_{n,ij}-P_{n,ij}(\theta,p))\nonumber \\
 &  & +\frac{1}{n(n-1)}\sum_{i}\sum_{j\neq i}q_{n,ij}(G_{n,ij}-\mathbb{E}[G_{n,ij}|X]).\label{eq:m_hat.diff}
\end{eqnarray}
The second to last term in (\ref{eq:m_hat.diff}) is $o_{p}(1)$ uniformly
in $(\theta,p)\in\Theta\times\mathcal{P}$, because it is uniformly
bounded by $(n(n-1))^{-1}\sum_{i}\sum_{j\neq i}\sup_{(\theta,p)\in\Theta\times\mathcal{P}}\Vert(\hat{q}_{n,ij}-q_{n,ij})(G_{n,ij}-P_{n,ij}(\theta,p))\Vert\leq\max_{1\leq i,j\le n}\Vert\hat{q}_{n,ij}-q_{n,ij}\Vert=o_{p}(1)$
(Assumption \ref{ass:theta.consist}(iii)). Further, define
\[
Y_{ni}=\frac{1}{n-1}\sum_{j\neq i}q_{n,ij}(G_{n,ij}-\mathbb{E}[G_{n,ij}|X]),
\]
so the last term in (\ref{eq:m_hat.diff}) is given by $n^{-1}\sum_{i}Y_{ni}$.
This term does not depend on $\theta$ or $p$. We show that it is
$o_{p}(1)$ following a pointwise LLN. Given $X$, $Y_{ni}$, $i=1,\dots,n$,
are independent with mean $0$, so $\mathbb{E}[\Vert n^{-1}\sum_{i}Y_{ni}\Vert^{2}|X]=n^{-2}\sum_{i}\mathbb{E}[\Vert Y_{ni}\Vert^{2}|X]$.
For each $i$,
\begin{eqnarray*}
\mathbb{E}[\Vert Y_{ni}\Vert^{2}|X] & = & \frac{1}{(n-1)^{2}}\sum_{j\neq i}q_{n,ij}^{\prime}\mathbb{E}[(G_{n,ij}-\mathbb{E}[G_{n,ij}|X])^{2}|X]q_{n,ij}\\
 &  & +\frac{1}{(n-1)^{2}}\sum_{j\neq i}\sum_{k\neq i,j}q_{n,ij}^{\prime}\mathbb{E}[(G_{n,ij}-\mathbb{E}[G_{n,ij}|X])(G_{n,ik}-\mathbb{E}[G_{n,ik}|X])|X]q_{n,ik}.
\end{eqnarray*}
Because both $\mathbb{E}[(G_{n,ij}-\mathbb{E}[G_{n,ij}|X])^{2}|X]$
and $\mathbb{E}[(G_{n,ij}-\mathbb{E}[G_{n,ij}|X])(G_{n,ik}-\mathbb{E}[G_{n,ik}|X])|X]$
are bounded by $1$, we can bound $\mathbb{E}[\Vert n^{-1}\sum_{i}Y_{ni}\Vert^{2}|X]$
by $\frac{1}{n^{2}(n-1)^{2}}(n(n-1)\max_{1\leq i,j\le n}\Vert q_{n,ij}\Vert^{2}+n(n-1)(n-2)\max_{1\leq i,j,k\le n}\Vert q_{n,ij}\Vert\Vert q_{n,ik}\Vert)=o(1)$,
where the last equality holds by Assumption \ref{ass:theta.consist}(iii).
By Markov's inequality, we conclude that $n^{-1}\sum_{i}Y_{ni}=o_{p}(1)$
and hence $\sup_{(\theta,p)\in\Theta\times\mathcal{P}}\Vert\hat{m}_{n}(\theta,p)-m_{n}(\theta,p)\Vert=o_{p}(1)$.
\end{proof}
\begin{lem}[Continuity of $P_{n,ij}(\theta,p)$]
\label{lem:ccp.cont.esti}Suppose that Assumptions \ref{ass:e=000026x}-\ref{ass:discX}
are satisfied. Conditional on $X$, the conditional choice probability
$P_{n,ij}(\theta,p)$ is continuous in $\theta$ and $p$.
\end{lem}
\begin{proof}
Following Lemma \ref{lem:ccp.cont}, the joint probability $P_{ni}(g_{i},\theta,p)\equiv\Pr(G_{ni}(\epsilon_{i},\theta,p)=g_{i}|X)$
is continuous in $\theta$ and $p$ for all $g_{i}\in\mathcal{G}_{i}$.
Therefore, the marginal probability $P_{n,ij}(\theta,p)=\sum_{g_{i}\in\mathcal{G}_{i}:g_{ij}=1}P_{ni}(g_{i},\theta,p)$
is continuous in $\theta$ and $p$.
\end{proof}

\subsubsection{\label{online:lemma.w}Asymptotic Properties of $\omega_{ni}(\epsilon_{i})$}

In this section, we establish in a few lemmas the asymptotic properties
of $\omega_{ni}(\epsilon_{i})$ that are needed to prove Theorem \ref{thm:theta.asymdist}.
We show that $\Lambda_{ni}\omega_{ni}(\epsilon_{i})$ is consistent
for $\Lambda_{ni}\omega_{ni}^{\ast}$ (Lemma \ref{lem:w.consist}).
Moreover, $\Lambda_{ni}\omega_{ni}(\epsilon_{i})$ has an asymptotically
linear representation (Lemma \ref{lem:w.asymlin}) and satisfies certain
uniformity properties (Lemma \ref{lem:w.rem}). Additional results
that are needed to prove these lemmas are given in Lemmas \ref{lem:lln.pai}
and \ref{lem:w.emp}.

We make the following assumptions on the auxiliary variable $\omega$.
All the random quantities are evaluated at $(\theta_{0},p_{n})$,
which is suppressed for simplicity.
\begin{assumption}
\label{ass:w}(i) The auxiliary variable $\omega$ is in a compact
set $\Omega\subseteq\mathbb{R}^{T}$, which contains a compact neighborhood
of $0$. (ii) For any $\delta>0$, there is $\xi>0$ such that for
all $n$ sufficiently large, there exists $\omega_{ni}^{\ast}=(\omega_{ni,+}^{\ast},\omega_{ni,-}^{\ast},\omega_{ni,0}^{\ast})\in\Omega$
satisfying $\Pi_{ni}^{\ast}(\omega_{+},\omega_{ni,-}^{\ast})+\xi\leq\Pi_{ni}^{\ast}(\omega_{ni,+}^{\ast},\omega_{ni,-}^{\ast})\leq\Pi_{ni}^{\ast}(\omega_{ni,+}^{\ast},\omega_{-})-\xi$
for all $\omega=(\omega_{+},\omega_{-},\omega_{0})\in\Omega$ with
$\|\Lambda_{ni}(\omega-\omega_{ni}^{\ast})\|\geq\delta$, $1\leq i\leq n$.
(iii) The matrix \textup{$H_{ni}^{\omega}(\omega)\equiv\frac{1}{n-2}\sum_{j\neq i}f_{\epsilon}(U_{n,ij}+\frac{n-1}{n-2}Z_{j}^{\prime}\Phi_{ni}\Lambda_{ni}\omega)\Lambda_{ni}\Phi'_{ni}Z_{j}Z_{j}^{\prime}\Phi_{ni}-I_{T}$
}satisfies $\lim\inf_{n\rightarrow\infty}\min_{1\leq i\leq n}\lambda_{\min}(H_{ni}^{\omega}(\omega_{ni}^{*})'H_{ni}^{\omega}(\omega_{ni}^{*}))>0$.
\end{assumption}
By the first-order condition in Lemma \ref{lem:foc.w}, we obtain
that $\Vert\Lambda_{ni}\omega_{ni}(\epsilon_{i})\Vert\leq\Vert\Lambda_{ni}\Phi'_{ni}\Vert<\infty$,
so $\Lambda_{ni}\omega_{ni}(\epsilon_{i})$ is bounded almost surely.
Without loss of generality, we can assume that $\omega$ lies in a
compact set $\Omega\subseteq\mathbb{R}^{T}$ as in Assumption \ref{ass:w}(i).
Assumption \ref{ass:w}(iii) is a standard regularity condition. To
derive a sufficient condition for it, define the matrix $D_{ni}(\omega)\equiv\frac{1}{n-2}\sum_{j\neq i}f_{\epsilon}(U_{n,ij}+\frac{n-1}{n-2}Z_{j}^{\prime}\Phi_{ni}\Lambda_{ni}\omega)Z_{j}Z_{j}^{\prime}$.
This is a $T\times T$ diagonal matrix, with the $t$th diagonal element
given by $\frac{1}{n-2}\sum_{j\neq i}1\{X_{j}=x_{t}\}f_{\text{\ensuremath{\epsilon}},it}(\omega)>0$,
where $f_{\epsilon,it}(\omega)=f_{\epsilon}(U_{n,ij}+\frac{n-1}{n-2}Z_{j}^{\prime}\Phi_{ni}\Lambda_{ni}\omega)$
for $X_{j}=x_{t}$. Note that $\Phi_{ni}H_{ni}^{\omega}(\omega)\Phi'_{ni}=V_{ni}D_{ni}(\omega)-I_{T}$.
A sufficient condition for $H_{ni}^{\omega}(\omega_{ni}^{*})$ nonsingular
is that $\|V_{ni}\|\|D_{ni}(\omega_{ni}^{*})\|<1$.

Assumption \ref{ass:w}(ii) is an identification condition. It requires
that there is a unique value of $\Lambda_{ni}\omega$ such that $\Pi_{ni}^{\ast}(\omega)$
achieves its saddle point value. Note that $\Pi_{ni}^{\ast}(\omega)$
depends on $\omega$ only through the value of $\Lambda_{ni}\omega$:
if $\omega$ and $\tilde{\omega}$ satisfy $\Lambda_{ni}(\omega-\tilde{\omega})=0$,
then $\omega'\Lambda_{ni}\omega=\tilde{\omega}'\Lambda_{ni}\tilde{\omega}$
and thus $\Pi_{ni}^{\ast}(\omega)=\Pi_{ni}^{\ast}(\tilde{\omega})$.
We assume that $\Pi_{ni}^{\ast}(\omega)$ achieves its saddle point
value at a unique value of $\Lambda_{ni}\omega$ instead of a unique
$\omega$ to account for the fact that $V_{ni}$ may be singular.

\begin{lem}[Uniform LLN for $\Pi_{ni}$]
\label{lem:lln.pai}Suppose that Assumptions \ref{ass:e=000026x}--\ref{ass:discX}
and \ref{ass:w}(i) are satisfied. Conditional on $X$, $\sup_{\omega\in\Omega}|\Pi_{ni}(\omega,\epsilon_{i})-\Pi_{ni}^{\ast}(\omega)|=o_{p}(1)$,
$1\leq i\leq n$.
\end{lem}
\begin{proof}
Defining $\pi_{n,ij}(\omega,\epsilon_{ij})=[U_{n,ij}+\frac{n-1}{n-2}Z_{j}^{\prime}\Phi_{ni}\Lambda_{ni}\omega-\epsilon_{ij}]_{+}$,
we can write 
\[
\Pi_{ni}(\omega,\epsilon_{i})-\Pi_{ni}^{\ast}(\omega)=\frac{1}{n-1}\sum_{j\neq i}(\pi_{n,ij}(\omega,\epsilon_{ij})-\mathbb{E}[\pi_{n,ij}(\omega,\epsilon_{ij})|X]).
\]
By Assumption \ref{ass:w}(i), we have $|Z_{j}^{\prime}\Phi_{ni}\Lambda_{ni}\omega|\leq\sup_{\omega\in\Omega}\|\Phi_{ni}\Lambda_{ni}\omega\Vert\leq M<\infty$.
Hence, for all $\omega\in\Omega$, $\pi_{n,ij}(\omega,\epsilon_{ij})^{2}\leq(|U_{n,ij}-\epsilon_{ij}|+\frac{n-1}{n-2}M)^{2}$,
with $\mathbb{E}[(|U_{n,ij}-\epsilon_{ij}|+\frac{n-1}{n-2}M)^{2}|X]<\infty$.
Also $\pi_{n,ij}(\omega,\epsilon_{ij})$ is continuous in $\omega$
on a compact set $\Omega$. Therefore the conditions of the uniform
LLN are satisfied \citep{Jennrich_1969} and the lemma is proved.
\end{proof}
\begin{lem}[Consistency of $\omega_{ni}(\epsilon_{i})$ for $\omega_{ni}^{*}$]
\label{lem:w.consist}Suppose that Assumptions \ref{ass:e=000026x}--\ref{ass:discX}
and \ref{ass:w}(i)--(ii) are satisfied. Conditional on $X$, $\Lambda_{ni}(\omega_{ni}(\epsilon_{i})-\omega_{ni}^{\ast})=o_{p}(1)$,
$1\leq i\leq n$.
\end{lem}
\begin{proof}
For the components of $\Lambda_{ni}\omega_{ni}(\epsilon_{i})$ corresponding
to a zero eigenvalue in $\Lambda_{ni}$, consistency holds trivially.
For notation ease, we assume $\Lambda_{ni}$ is nonsingular. In this
case, Assumption \ref{ass:w}(ii) holds with $\|\Lambda_{ni}(\omega-\omega_{ni}^{\ast})\|\geq\delta$
replaced by $\|\omega-\omega_{ni}^{\ast}\|\geq\delta$.\footnote{If $\|\omega-\omega_{ni}^{\ast}\|\geq\delta$, then from the inequality
$\|\omega-\omega_{ni}^{\ast}\|\leq\|\Lambda_{ni}^{-1}\|\|\Lambda_{ni}(\omega-\omega_{ni}^{\ast})\|$,
it follows that $\|\Lambda_{ni}(\omega-\omega_{ni}^{\ast})\|\geq\|\Lambda_{ni}^{-1}\|^{-1}\delta$.
Now, choose $\delta$ as the value corresponding to $\|\Lambda_{ni}^{-1}\|^{-1}\delta$
in the original assumption.} Decompose $\omega_{ni}(\epsilon_{i})=(\omega_{ni,+}(\epsilon_{i}),\omega_{ni,-}(\epsilon_{i}))$,
where $\omega{}_{ni,+}(\epsilon_{i})=(\omega{}_{ni,t}(\epsilon_{i}),t\in\mathcal{T}_{ni,+})\in\Omega_{ni,+}$
and $\omega{}_{ni,-}(\epsilon_{i})=(\omega{}_{ni,t}(\epsilon_{i}),t\in\mathcal{T}_{ni,-})\in\Omega_{ni,-}$
represent the components of $\omega_{ni}(\epsilon_{i})$ that correspond
to the positive and negative eigenvalues in $\Lambda_{ni}$, respectively.
Similarly, we decompose $\omega_{ni}^{*}=(\omega_{ni,+}^{*},\omega_{ni,-}^{*})$,
where $\omega_{ni,+}^{*}=(\omega_{ni,t}^{*},t\in\mathcal{T}_{ni,+})\in\Omega_{ni,+}$
and $\omega_{ni,-}^{*}=(\omega_{ni,t}^{*},t\in\mathcal{T}_{ni,-})\in\Omega_{ni,-}$.
It suffices to show that $\omega_{ni,+}(\epsilon_{i})-\omega_{ni,+}^{\ast}=o_{p}(1)$
and $\omega_{ni,-}(\epsilon_{i})-\omega_{ni,-}^{\ast}=o_{p}(1)$.

We start with $\omega_{ni,+}(\epsilon_{i})$. Fix $\delta>0$. Let
$\mathcal{B}_{ni,+}(\omega_{ni,+}^{\ast},\delta)=\{\omega_{+}\in\Omega_{ni,+}:\Vert\omega_{+}-\omega_{ni,+}^{\ast}\Vert<\delta\}$
be a subset of $\Omega_{ni,+}$ containing $\omega_{ni,+}^{\ast}$.
We derive
\begin{eqnarray}
 &  & \Pr(\Vert\omega_{ni,+}(\epsilon_{i})-\omega_{ni,+}^{\ast})\Vert<\delta|X)\nonumber \\
 & \geq & \Pr\left(\left.\Pi_{ni}^{\ast}(\omega_{ni,+}(\epsilon_{i}),\omega_{ni,-}^{*})>\sup_{\omega_{+}\in\Omega_{ni,+}\backslash\mathcal{B}_{ni,+}(\omega_{ni,+}^{\ast},\delta)}\Pi_{ni}^{\ast}(\omega_{+},\omega_{ni,-}^{*})\right\vert X\right).\label{eq:w.consist.pos}
\end{eqnarray}
By the optimality of $\omega_{ni}^{*}$ and $\omega_{ni}(\epsilon_{i})$,
we can derive that $\Pi_{ni}^{\ast}(\omega_{ni}^{*})\leq\Pi_{ni}^{\ast}(\omega_{ni}^{\ast},\omega_{ni,-}(\epsilon_{i}))$
and $\Pi_{ni}(\omega_{ni,+}(\epsilon_{i}),\omega_{ni,-}^{*},\epsilon_{i})\geq\Pi_{ni}(\omega_{ni}(\epsilon_{i}),\epsilon_{i})\geq\Pi_{ni}(\omega_{ni,+}^{\ast},\omega_{ni,-}(\epsilon_{i}),\epsilon_{i})$.
Therefore, 
\begin{eqnarray}
\Pi_{ni}^{\ast}(\omega_{ni,+}(\epsilon_{i}),\omega_{ni,-}^{*})-\Pi_{ni}^{\ast}(\omega_{ni}^{\ast}) & \geq & \Pi_{ni}^{\ast}(\omega_{ni,+}(\epsilon_{i}),\omega_{ni,-}^{*})-\Pi_{ni}^{\ast}(\omega_{ni}^{\ast},\omega_{ni,-}(\epsilon_{i}))\nonumber \\
 & \geq & \Pi_{ni}^{\ast}(\omega_{ni,+}(\epsilon_{i}),\omega_{ni,-}^{*})-\Pi_{ni}(\omega_{ni,+}(\epsilon_{i}),\omega_{ni,-}^{*},\epsilon_{i})\nonumber \\
 &  & +\Pi_{ni}(\omega_{ni}^{\ast},\omega_{ni,-}(\epsilon_{i}),\epsilon_{i})-\Pi_{ni}^{\ast}(\omega_{ni}^{\ast},\omega_{ni,-}(\epsilon_{i}))\nonumber \\
 & \geq & -2\sup_{\omega\in\Omega}|\Pi_{ni}(\omega,\epsilon_{i})-\Pi_{ni}^{\ast}(\omega)|=o_{p}(1),\label{eq:pai_diff.pos}
\end{eqnarray}
by the uniform LLN in Lemma \ref{lem:lln.pai}. From the compactness
of $\Omega_{ni,+}\backslash\mathcal{B}_{ni,+}(\omega_{ni,+}^{\ast},\delta)$,
the continuity of $\Pi_{ni}^{\ast}(\omega)$, and the identification
condition in Assumption \ref{ass:w}(ii),\footnote{Note that $\Vert\omega_{+}-\omega_{ni,+}^{\ast}\Vert\geq\delta$ implies
that $\|\omega-\omega_{ni}^{\ast}\|\geq\delta$ because $\|\omega-\omega_{ni}^{\ast}\|\geq\Vert\omega_{+}-\omega_{ni,+}^{\ast}\Vert$.} we derive that for $n$ sufficiently large, $\sup_{\omega_{+}\in\Omega_{ni,+}\backslash\mathcal{B}_{ni,+}(\omega_{ni,+}^{\ast},\delta)}\Pi_{ni}^{\ast}(\omega_{+},\omega_{ni,-}^{*})=\Pi_{ni}^{\ast}(\bar{\omega}_{+},\omega_{ni,-}^{*})\leq\Pi_{ni}^{\ast}(\omega_{ni}^{\ast})-\xi$
for some $\bar{\omega}_{+}\in\Omega_{ni,+}\backslash\mathcal{B}_{ni,+}(\omega_{ni,+}^{\ast},\delta)$.
Combining the results, we can see that the right-hand side of (\ref{eq:w.consist.pos})
goes to $1$.

As for $\omega_{ni,-}(\epsilon_{i})$, define $\mathcal{B}_{ni,-}(\omega_{ni,-}^{\ast},\delta)=\{\omega_{-}\in\Omega_{ni,-}:\Vert\omega_{-}-\omega_{ni,-}^{\ast}\Vert<\delta\}$
and we have
\begin{eqnarray}
 &  & \Pr(\Vert\omega_{ni,-}(\epsilon_{i})-\omega_{ni,-}^{\ast})\Vert<\delta|X)\nonumber \\
 & \geq & \Pr\left(\left.\Pi_{ni}^{\ast}(\omega_{ni,+}^{*},\omega_{ni,-}(\epsilon_{i}))<\inf_{\omega_{-}\in\Omega_{ni,-}\backslash\mathcal{B}_{ni,-}(\omega_{ni,-}^{\ast},\delta)}\Pi_{ni}^{\ast}(\omega_{ni,+}^{*},\omega_{-})\right\vert X\right).\label{eq:w.consist.neg}
\end{eqnarray}
Similarly to (\ref{eq:pai_diff.pos}), we derive $\Pi_{ni}^{\ast}(\omega_{ni,+}^{*},\omega_{ni,-}(\epsilon_{i}))-\Pi_{ni}^{\ast}(\omega_{ni,+}^{\ast},\omega_{ni,-}^{*})\leq2\sup_{\omega\in\Omega}|\Pi_{ni}(\omega,\epsilon_{i})-\Pi_{ni}^{\ast}(\omega)|=o_{p}(1)$.
Moreover, for $n$ sufficiently large, $\inf_{\omega_{-}\in\Omega_{ni,-}\backslash\mathcal{B}_{ni,-}(\omega_{ni,-}^{\ast},\delta)}\Pi_{ni}^{\ast}(\omega_{ni,+}^{*},\omega_{-})=\Pi_{ni}^{\ast}(\omega_{ni,+}^{*},\bar{\omega}_{-})\geq\Pi_{ni}^{\ast}(\omega_{ni}^{\ast})+\xi$
for some $\bar{\omega}_{-}\in\Omega_{ni,-}\backslash\mathcal{B}_{ni,-}(\omega_{ni,-}^{\ast},\delta)$.
Combining the results shows that the right-hand side of (\ref{eq:w.consist.neg})
goes to $1$.
\end{proof}
\begin{lem}[Asymptotically linear representation of $\omega_{ni}(\epsilon_{i})$]
\label{lem:w.asymlin}Suppose that Assumptions \ref{ass:e=000026x}--\ref{ass:discX}
and \ref{ass:w} are satisfied. Conditional on $X$, $\Lambda_{ni}\omega_{ni}(\epsilon_{i})$
has an asymptotically linear representation
\begin{equation}
\Lambda_{ni}(\omega_{ni}(\epsilon_{i})-\omega_{ni}^{\ast})=\frac{1}{n-1}\sum_{j\neq i}\phi_{n,ij}^{\omega}(\omega_{ni}^{\ast},\epsilon_{ij})+r_{ni}^{\omega}(\epsilon_{i}),\label{eq:w.alr}
\end{equation}
with $r_{ni}^{\omega}(\epsilon_{i})=o_{p}(n^{-1/2})$. In the expression,
$\phi_{n,ij}^{\omega}(\omega_{ni}^{\ast},\epsilon_{ij})\equiv-H_{ni}^{\omega}(\omega_{ni}^{\ast})^{-1}\Lambda_{ni}\phi_{n,ij}^{\pi}(\omega_{ni}^{\ast},\epsilon_{ij})$,
where $\phi_{n,ij}^{\pi}(\omega,\epsilon_{ij})\equiv1\{U_{n,ij}+\frac{n-1}{n-2}Z_{j}^{\prime}\Phi_{ni}\Lambda_{ni}\omega\geq\epsilon_{ij}\}\Phi'_{ni}Z_{j}-\omega$
and $H_{ni}^{\omega}(\omega_{ni}^{\ast})=\frac{1}{n-2}\sum_{j\neq i}f_{\epsilon}(U_{n,ij}+\frac{n-1}{n-2}Z_{j}^{\prime}\Phi_{ni}\Lambda_{ni}\omega_{ni}^{\ast})\Lambda_{ni}\Phi'_{ni}Z_{j}Z_{j}^{\prime}\Phi{}_{ni}-I_{T}$.
\end{lem}
\begin{proof}
$\omega_{ni}(\epsilon_{i})$ satisfies the first-order condition in
Lemma \ref{lem:foc.w}
\begin{equation}
\Gamma_{ni}(\omega,\epsilon_{i})\equiv\Lambda_{ni}\frac{1}{n-1}\sum_{j\neq i}\phi_{n,ij}^{\pi}(\omega,\epsilon_{ij})=0\text{, a.s.}\label{eq:foc.wn}
\end{equation}
Under Assumption \ref{ass:w}(ii), $\omega_{ni}^{\ast}$ is a maximin
solution of $\Pi_{ni}^{\ast}(\omega)$, so $\omega_{ni}^{\ast}$ satisfies
the population counterpart of the first-order condition 
\begin{equation}
\Gamma_{ni}^{\ast}(\omega)\equiv\Lambda_{ni}\frac{1}{n-1}\sum_{j\neq i}\mathbb{E}[\phi_{n,ij}^{\pi}(\omega,\epsilon_{ij})|X]=0,\label{eq:foc.w}
\end{equation}
where $\mathbb{E}[\phi_{n,ij}^{\pi}(\omega,\epsilon_{ij})|X]=F_{\epsilon}(U_{n,ij}+\frac{n-1}{n-2}Z_{j}^{\prime}\Phi_{ni}\Lambda_{ni}\omega)\Phi'_{ni}Z_{j}-\omega$.
View $\Gamma_{ni}^{\ast}(\omega)$ as a function of $\Lambda_{ni}\omega$.
Expanding it at $\Lambda_{ni}\omega_{ni}^{\ast}$ yields
\begin{equation}
\Gamma_{ni}^{\ast}(\omega_{ni}(\epsilon_{i}))=H_{ni}^{\omega}(\omega_{ni}^{\ast})\Lambda_{ni}(\omega_{ni}(\epsilon_{i})-\omega_{ni}^{\ast})+O_{p}(\Vert\Lambda_{ni}(\omega_{ni}(\epsilon_{i})-\omega_{ni}^{\ast})\Vert^{2}).\label{eq:w.gam.taylor}
\end{equation}
For any $\omega\in\Omega$, the inequality $\Vert H_{ni}^{\omega}(\omega_{ni}^{\ast})\Lambda_{ni}(\omega-\omega_{ni}^{\ast})\Vert^{2}\geq c_{n}^{2}\Vert\Lambda_{ni}(\omega-\omega_{ni}^{\ast})\Vert^{2}$
holds, where $c_{n}^{2}=\lambda_{\min}(H_{ni}^{\omega}(\omega_{ni}^{\ast})'H_{ni}^{\omega}(\omega_{ni}^{\ast}))$
is the smallest eigenvalue of the matrix $H_{ni}^{\omega}(\omega_{ni}^{\ast})'H_{ni}^{\omega}(\omega_{ni}^{\ast})$.
Combining this result with Assumption \ref{ass:w}(iii), equation
(\ref{eq:w.gam.taylor}), and the consistency of $\Lambda_{ni}\omega_{ni}(\epsilon_{i})$
in Lemma \ref{lem:w.consist} we obtain
\begin{equation}
\Vert\Gamma_{ni}^{\ast}(\omega_{ni}(\epsilon_{i}))\Vert\geq\Vert\Lambda_{ni}(\omega_{ni}(\epsilon_{i})-\omega_{ni}^{\ast})\Vert(c_{n}+o_{p}(1)).\label{eq:w.gam.diff}
\end{equation}
To derive the convergence rate of $\Lambda_{ni}(\omega_{ni}(\epsilon_{i})-\omega_{ni}^{\ast})$,
it suffices to derive that of $\Gamma_{ni}^{\ast}(\omega_{ni}(\epsilon_{i}))$.

By equations (\ref{eq:foc.wn}) and (\ref{eq:foc.w}), we can write
\begin{align}
 & \Gamma_{ni}^{\ast}(\omega_{ni}(\epsilon_{i}))\nonumber \\
= & -\Gamma_{ni}(\omega_{ni}^{\ast},\epsilon_{i})-(\Gamma_{ni}(\omega_{ni}(\epsilon_{i}),\epsilon_{i})-\Gamma_{ni}^{\ast}(\omega_{ni}(\epsilon_{i}))-(\Gamma_{ni}(\omega_{ni}^{\ast},\epsilon_{i})-\Gamma_{ni}^{\ast}(\omega_{ni}^{\ast}))),\text{ a.s.}\label{eq:w.foc.dcmp}
\end{align}
Define $\phi_{n,ij}^{\gamma}(\omega,\epsilon_{ij})\equiv1\{U_{n,ij}+\frac{n-1}{n-2}Z_{j}^{\prime}\Phi_{ni}\Lambda_{ni}\omega\geq\epsilon_{ij}\}Z_{j}$.
We can write $\Gamma_{ni}(\omega,\epsilon_{i})-\Gamma_{ni}^{\ast}(\omega)=\Lambda_{ni}\Phi'_{ni}\frac{1}{n-1}\sum_{j\neq i}(\phi_{n,ij}^{\gamma}(\omega,\epsilon_{ij})-\mathbb{E}[\phi_{n,ij}^{\gamma}(\omega,\epsilon_{ij})|X])$.
Define the empirical process
\begin{equation}
\mathbb{G}_{n}\phi_{ni}^{\gamma}(\omega,\epsilon_{i})\equiv\frac{1}{\sqrt{n-1}}\sum_{j\neq i}(\phi_{n,ij}^{\gamma}(\omega,\epsilon_{ij})-\mathbb{E}[\phi_{n,ij}^{\gamma}(\omega,\epsilon_{ij})|X]),\text{ }\omega\in\Omega.\label{eq:w.ep.def}
\end{equation}
By Lemma \ref{lem:w.emp}(i) (to be proved later), the last term in
(\ref{eq:w.foc.dcmp}) is negligible:
\begin{align}
 & \Gamma_{ni}(\omega_{ni}(\epsilon_{i}),\epsilon_{i})-\Gamma_{ni}^{\ast}(\omega_{ni}(\epsilon_{i}))-(\Gamma_{ni}(\omega_{ni}^{\ast},\epsilon_{i})-\Gamma_{ni}^{\ast}(\omega_{ni}^{\ast}))\nonumber \\
= & (n-1)^{-1/2}\Lambda_{ni}\Phi'_{ni}\mathbb{G}_{n}(\phi_{ni}^{\gamma}(\omega_{ni}(\epsilon_{i}),\epsilon_{i})-\phi_{ni}^{\gamma}(\omega_{ni}^{\ast},\epsilon_{i}))=o_{p}(n^{-1/2}).\label{eq:w.ep.rate}
\end{align}

The first term on the right-hand side of (\ref{eq:w.foc.dcmp}), $\Gamma_{ni}(\omega_{ni}^{\ast},\epsilon_{i})$,
is a leading term. To derive its rate of convergence, define the random
variable 
\[
Y_{n,ij}\equiv\frac{1}{\sqrt{n-1}}(\phi_{n,ij}^{\gamma}(\omega_{ni}^{\ast},\epsilon_{ij})-\mathbb{E}[\phi_{n,ij}^{\gamma}(\omega_{ni}^{\ast},\epsilon_{ij})|X])\in\mathbb{R}^{T},
\]
and thus $\Gamma_{ni}(\omega_{ni}^{\ast},\epsilon_{i})=\Lambda_{ni}\Phi'_{ni}\frac{1}{\sqrt{n-1}}\sum_{j\neq i}Y_{n,ij}$.
Note that $\{Y_{n,ij},j\neq i\}$ is a triangular array. We apply
the Lindeberg-Feller CLT to show that $\Gamma_{ni}(\omega_{ni}^{\ast},\epsilon_{i})=O_{p}(n^{-1/2}).$
By the Cramer-Wold device it suffices to show that $a^{\prime}\sum_{j\neq i}Y_{n,ij}$
satisfies the Lindeberg condition for any $T\times1$ vector of constants
$a\in\mathbb{R}^{T}$. The Lindeberg condition is that for any $\xi>0$
\begin{equation}
\lim_{n\rightarrow\infty}\sum_{j\neq i}\mathbb{E}\left[\left.\frac{(a^{\prime}Y_{n,ij})^{2}}{a^{\prime}\Sigma_{ni}^{\pi}a}1\{|a^{\prime}Y_{n,ij}|\geq\xi\sqrt{a^{\prime}\Sigma_{ni}^{\pi}a}\}\right\vert X\right]=0,\label{eq:w.lindeberg}
\end{equation}
where
\begin{eqnarray*}
\Sigma_{ni}^{\pi} & = & \sum_{j\neq i}\mathbb{E}[Y_{n,ij}Y_{n,ij}^{\prime}|X]\\
 & = & \frac{1}{n-1}\sum_{j\neq i}F_{\epsilon}\left(U_{n,ij}+\frac{n-1}{n-2}Z_{j}^{\prime}\Phi_{ni}\Lambda_{ni}\omega_{ni}^{\ast}\right)\left(1-F_{\epsilon}\left(U_{n,ij}+\frac{n-1}{n-2}Z_{j}^{\prime}\Phi_{ni}\Lambda_{ni}\omega_{ni}^{\ast}\right)\right)Z_{j}Z_{j}^{\prime}.
\end{eqnarray*}
Observe that the sum in (\ref{eq:w.lindeberg}) is bounded by $\mathbb{E}[(a^{\prime}\Sigma_{ni}^{\pi}a)^{-1}\sum_{j\neq i}(a^{\prime}Y_{n,ij})^{2}1\{\max_{j\neq i}|a^{\prime}Y_{n,ij}|\geq\xi\sqrt{a^{\prime}\Sigma_{ni}^{\pi}a}\}|X]$.
The random variable $(a^{\prime}\Sigma_{ni}^{\pi}a)^{-1}\sum_{j\neq i}(a^{\prime}Y_{n,ij})^{2}$
has an expectation that is bounded uniformly over $n$ and is therefore
$O_{p}(1)$.\footnote{Note that $Y_{n,ij}=\frac{1}{\sqrt{n-1}}(1\{U_{n,ij}+\frac{n-1}{n-2}Z_{j}^{\prime}\Phi_{ni}\Lambda_{ni}\omega\geq\epsilon_{ij}\}-F_{\epsilon}(U_{n,ij}+\frac{n-1}{n-2}Z_{j}^{\prime}\Phi_{ni}\Lambda_{ni}\omega))Z_{j}$.
We can bound it by $\|Y_{n,ij}\|\leq\|Z_{j}\|/\sqrt{n-1}=1/\sqrt{n-1}$.
Moreover, for any $a\neq0$, we have $a^{\prime}\Sigma_{ni}^{\pi}a=\frac{1}{n-1}\sum_{j\neq i}F_{\epsilon}\left(U_{n,ij}+\frac{n-1}{n-2}Z_{j}^{\prime}\Phi_{ni}\Lambda_{ni}\omega_{ni}^{\ast}\right)\left(1-F_{\epsilon}\left(U_{n,ij}+\frac{n-1}{n-2}Z_{j}^{\prime}\Phi_{ni}\Lambda_{ni}\omega_{ni}^{\ast}\right)\right)a'Z_{j}Z_{j}^{\prime}a>0$
because $Z_{j}^{\prime}a=0$ for all $j\neq i$ requires that $a=0$.} Hence, if
\begin{equation}
\frac{\max_{j\neq i}|a^{\prime}Y_{n,ij}|}{\sqrt{a^{\prime}\Sigma_{ni}^{\pi}a}}=o_{p}(1),\label{eq:w.lind.maxY}
\end{equation}
then $(a^{\prime}\Sigma_{ni}^{\pi}a)^{-1}\sum_{j\neq i}|a^{\prime}Y_{n,ij}|^{2}1\{\max_{j\neq i}|a^{\prime}Y_{n,ij}|\geq\xi\sqrt{a^{\prime}\Sigma_{ni}^{\pi}a}\}=O_{p}(1)o_{p}(1)=o_{p}(1)$.
This random variable is bounded by $(a^{\prime}\Sigma_{ni}^{\pi}a)^{-1}\sum_{j\neq i}|a^{\prime}Y_{n,ij}|^{2}$.
We conclude that by dominated convergence the Lindeberg condition
is satisfied if (\ref{eq:w.lind.maxY}) holds.

By Markov's inequality, equation (\ref{eq:w.lind.maxY}) holds if
$\mathbb{E}[\max_{j\neq i}|a^{\prime}Y_{n,ij}|^{2}|X]=o(1)$. By Lemma
2.2.2 in \citet{VW_1996}, we have the bound $\mathbb{E}[\max_{j\neq i}|a^{\prime}Y_{n,ij}|^{2}|X]\leq\Vert\max_{j\neq i}|a^{\prime}Y_{n,ij}|^{2}\Vert_{\psi_{1}|X}\leq K\ln(n+1)\max_{j\neq i}\Vert|a^{\prime}Y_{n,ij}|^{2}\Vert_{\psi_{1}|X}$,
where $K<\infty$ is a constant. Note that the random variable $a^{\prime}Y_{n,ij}$
is bounded by $\Vert a\Vert\Vert Y_{n,ij}\Vert\leq\Vert a\Vert/\sqrt{n-1}<\infty$.
By Hoeffding's inequality for bounded random variables \citep[Theorem 2.8]{BLM_2013},
$\Pr(|a^{\prime}Y_{n,ij}|^{2}\geq t|X)=\Pr(a^{\prime}Y_{n,ij}\geq\sqrt{t}|X)+\Pr(-a^{\prime}Y_{n,ij}\geq\sqrt{t}|X)\leq2\exp(-\frac{(n-1)t}{2\Vert a\Vert^{2}})$.
Hence, by Lemma 2.2.1 in \citet{VW_1996} we can bound $\Vert|a^{\prime}Y_{n,ij}|^{2}\Vert_{\psi_{1}|X}\leq6\Vert a\Vert^{2}/(n-1)$.
Combining these results yields $\mathbb{E}[\max_{j\neq i}|a^{\prime}Y_{n,ij}|^{2}|X]\leq6\Vert a\Vert^{2}K\ln(n+1)/(n-1)=o(1)$,
so the Lindeberg condition holds. We conclude that $\sum_{j\neq i}Y_{n,ij}=O_{p}(1)$
and thus $\Gamma_{ni}(\omega_{ni}^{\ast},\epsilon_{i})=O_{p}(n^{-1/2})$.

Combining equations (\ref{eq:w.gam.diff}), (\ref{eq:w.foc.dcmp})
and (\ref{eq:w.ep.rate}), we obtain $\Vert\Lambda_{ni}(\omega_{ni}(\epsilon_{i})-\omega_{ni}^{\ast})\Vert(c_{n}+o_{p}(1))\leq O_{p}(n^{-1/2})$.
This implies that $\Lambda_{ni}(\omega_{ni}(\epsilon_{i})-\omega_{ni}^{\ast})=O_{p}(n^{-1/2})$.
Combining equations (\ref{eq:w.gam.taylor}), (\ref{eq:w.foc.dcmp})
and (\ref{eq:w.ep.rate}) we derive that $H_{ni}^{\omega}(\omega_{ni}^{\ast})\Lambda_{ni}(\omega_{ni}(\epsilon_{i})-\omega_{ni}^{\ast})=-\Gamma_{ni}(\omega_{ni}^{\ast},\epsilon_{i})+o_{p}(n^{-1/2})$.
By Assumption \ref{ass:w}(iii), the matrix $H_{ni}^{\omega}(\omega_{ni}^{\ast})$
is invertible. Multiplying both sides by the inverse $H_{ni}^{\omega}(\omega_{ni}^{\ast})^{-1}$,
we obtain (\ref{eq:w.alr}).
\end{proof}
\begin{lem}[Uniform Properties of $\omega_{ni}(\epsilon_{i})$]
\label{lem:w.rem}Suppose that Assumptions \ref{ass:e=000026x}--\ref{ass:discX}
and \ref{ass:w} are satisfied. Conditional on $X$, we have (i) $\|\max_{1\leq i\leq n}\Vert\Lambda_{ni}(\omega_{ni}(\epsilon_{i})-\omega_{ni}^{\ast})\Vert^{2}\|_{\psi_{1}|X}=o(n^{-1/2})$,
and (ii) the remainder $r_{ni}^{\omega}(\epsilon_{i})$ in Lemma \ref{lem:w.asymlin}
satisfies $\|\max_{1\leq i\leq n}\Vert r_{ni}^{\omega}(\epsilon_{i})\Vert\|_{\psi_{1}|X}=o(n^{-1/2}).$
\end{lem}
\begin{proof}
Part (i): By Lemma 2.2.2 in \citet{VW_1996} we can bound $\Vert\max_{i}\Vert\Lambda_{ni}(\omega_{ni}(\epsilon_{i})-\omega_{ni}^{\ast})\Vert^{2}\Vert_{\psi_{1}|X}\leq K\ln(n+1)\max_{i}\Vert\Vert\Lambda_{ni}(\omega_{ni}(\epsilon_{i})-\omega_{ni}^{\ast})\Vert^{2}\Vert_{\psi_{1}|X}$,
where $K<\infty$ is a constant. For any random variable $Z\in\mathbb{R}$
and constant $C>0$, we have $\mathbb{E}[\psi_{1}(Z^{2}/C^{2})|X]=\mathbb{E}[\psi_{2}(|Z|/C)|X]$,
where $\psi_{1}(z)=e^{z}-1$ and $\psi_{2}(z)=e^{z^{2}}-1$. This
implies that $\Vert Z^{2}\Vert_{\psi_{1}|X}=\Vert Z\Vert_{\psi_{2}|X}^{2}$
and hence $\Vert\Vert\Lambda_{ni}(\omega_{ni}(\epsilon_{i})-\omega_{ni}^{\ast})\Vert^{2}\Vert_{\psi_{1}|X}=\Vert\Lambda_{ni}(\omega_{ni}(\epsilon_{i})-\omega_{ni}^{\ast})\Vert_{\psi_{2}|X}^{2}$.
Combining these results, we can bound $\Vert\max_{i}\Vert\Lambda_{ni}(\omega_{ni}(\epsilon_{i})-\omega_{ni}^{\ast})\Vert^{2}\Vert_{\psi_{1}|X}$
by $K\ln(n+1)\max_{i}\Vert\Lambda_{ni}(\omega_{ni}(\epsilon_{i})-\omega_{ni}^{\ast})\Vert_{\psi_{2}|X}^{2}$.

To further bound $\Vert\Lambda_{ni}(\omega_{ni}(\epsilon_{i})-\omega_{ni}^{\ast})\Vert_{\psi_{2}|X}$
uniformly in $i$, by comparing equations (\ref{eq:w.gam.taylor})
and (\ref{eq:w.foc.dcmp}) in Lemma \ref{lem:w.asymlin}, we derive
the remainder $r_{ni}^{\omega}(\epsilon_{i})$ in equation (\ref{eq:w.alr})
as
\begin{equation}
r_{ni}^{\omega}(\epsilon_{i})=-H_{ni}^{\omega}(\omega_{ni}^{\ast})^{-1}(O_{p}(\Vert\Lambda_{ni}(\omega_{ni}(\epsilon_{i})-\omega_{ni}^{\ast})\Vert^{2})+\frac{1}{\sqrt{n-1}}\Lambda_{ni}\Phi'_{ni}\mathbb{G}_{n}(\phi_{ni}^{\gamma}(\omega_{ni}(\epsilon_{i}),\epsilon_{i})-\phi_{ni}^{\gamma}(\omega_{ni}^{\ast},\epsilon_{i})).\label{eq:w.rem}
\end{equation}
Substituting $r_{ni}^{\omega}(\epsilon_{i})$ in (\ref{eq:w.alr})
with (\ref{eq:w.rem}), we obtain 
\begin{eqnarray}
 &  & \Vert\Lambda_{ni}(\omega_{ni}(\epsilon_{i})-\omega_{ni}^{\ast})(1+o_{p}(1))\Vert_{\psi_{2}|X}\nonumber \\
 & \leq & \left\Vert \frac{1}{n-1}\sum_{j\neq i}\phi_{n,ij}^{\omega}(\omega_{ni}^{\ast},\epsilon_{ij})\right\Vert _{\psi_{2}|X}\nonumber \\
 &  & +\frac{1}{\sqrt{n-1}}\Vert H_{ni}^{\omega}(\omega_{ni}^{\ast})^{-1}\Lambda_{ni}\Phi'_{ni}\Vert\Vert\mathbb{G}_{n}(\phi_{ni}^{\gamma}(\omega_{ni}(\epsilon_{i}),\epsilon_{i})-\phi_{ni}^{\gamma}(\omega_{ni}^{\ast},\epsilon_{i}))\Vert_{\psi_{2}|X},\label{eq:w.orlicz.bdd}
\end{eqnarray}
where we have used the triangle inequality for the Orlicz norm and
the boundedness of $H_{ni}^{\omega}(\omega_{ni}^{\ast})^{-1}$. Note
that $\Vert\Lambda_{ni}(\omega_{ni}(\epsilon_{i})-\omega_{ni}^{\ast})(1+o_{p}(1))\Vert_{\psi_{2}|X}=\Vert\Lambda_{ni}(\omega_{ni}(\epsilon_{i})-\omega_{ni}^{\ast})\Vert_{\psi_{2}|X}(1+o(1))$.\footnote{For any bounded random variable $Z$ and conditional Orlicz norm $\Vert Z\Vert_{\psi|X}$,
we have $\Vert Zo_{p}(1)\Vert_{\psi|X}=o(\Vert Z\Vert_{\psi|X})$.
This is because for any sequence $\delta_{n}\downarrow0$, if there
were $M<\infty$ such that $\Vert Z\Vert_{\psi|X}\leq M(\Vert Zo_{p}(1)\Vert_{\psi|X}-\delta_{n})$
for $n$ sufficiently large, then since $|Zo_{p}(1)|/\Vert Z\Vert_{\psi|X}=o_{p}(1)$,
we have for sufficiently large $n$, $1<\mathbb{E}[\psi(|Zo_{p}(1)|/(\Vert Zo_{p}(1)\Vert_{\psi|X}-\delta_{n}))|X]\leq\mathbb{E}[\psi(M|Zo_{p}(1)|/\Vert Z\Vert_{\psi|X})|X]\rightarrow0$
by dominated convergence, a contradiction. Therefore, $(\Vert Zo_{p}(1)\Vert_{\psi|X}-\delta_{n})/\Vert Z\Vert_{\psi|X}=o(1)$
and hence $\Vert Zo_{p}(1)\Vert_{\psi|X}=o(\Vert Z\Vert_{\psi|X})$.}

Consider the first term on the right-hand side of (\ref{eq:w.orlicz.bdd}).
Recall that $\phi_{n,ij}^{\omega}(\omega_{ni}^{\ast},\epsilon_{ij})\in\mathbb{R}^{T}$
is the influence function in (\ref{eq:w.alr}). For $1\leq t\leq T,$
let $\phi_{n,ij,t}^{\omega}(\omega_{ni}^{\ast},\epsilon_{ij})$ denote
the $t$th component of $\phi_{n,ij}^{\omega}(\omega_{ni}^{\ast},\epsilon_{ij})$.
Write $\phi_{n,ij}^{\omega}(\omega_{ni}^{\ast},\epsilon_{ij})$ and
$\phi_{n,ij,t}^{\omega}(\omega_{ni}^{\ast},\epsilon_{ij})$ as $\phi_{n,ij}^{\omega}$
and $\phi_{n,ij,t}^{\omega}$. Note that $\Vert\frac{1}{n-1}\sum_{j\neq i}\phi_{n,ij}^{\omega}\Vert=(\sum_{t}(\frac{1}{n-1}\sum_{j\neq i}\phi_{n,ij,t}^{\omega})^{2})^{1/2}\leq\sum_{t}|\frac{1}{n-1}\sum_{j\neq i}\phi_{n,ij,t}^{\omega}|$,
so for any $\kappa>0$, we have $\Pr(\Vert\frac{1}{n-1}\sum_{j\neq i}\phi_{n,ij}^{\omega}\Vert\geq\kappa|X)\leq\Pr(\sum_{t}|\frac{1}{n-1}\sum_{j\neq i}\phi_{n,ij,t}^{\omega}|\geq\kappa|X)\leq\sum_{t}\Pr(|\frac{1}{n-1}\sum_{j\neq i}\phi_{n,ij,t}^{\omega}|\geq\frac{\kappa}{T}|X)$.
It is evident that for any $1\leq t\leq T$ and $1\leq i,j\leq n$,
we can bound $|\phi_{n,ij,t}^{\omega}|<\Vert\phi_{n,ij}^{\omega}\Vert\leq\Vert H_{ni}^{\omega}(\omega_{ni}^{\ast})^{-1}\Vert(\Vert\Lambda_{ni}\Phi'_{ni}\Vert+\Vert\Lambda_{ni}\omega_{ni}^{\ast}\Vert)\leq M<\infty$.
By Hoeffding's inequality for bounded random variables \citep[Theorem 2.8]{BLM_2013},
we have $\Pr(|\frac{1}{n-1}\sum_{j\neq i}\phi_{n,ij,t}^{\omega}|\geq\frac{\kappa}{T}|X)\leq2\exp(-\frac{(n-1)\kappa^{2}}{2M^{2}T^{2}})$
for each $1\leq t\leq T$ and hence $\Pr(\Vert\frac{1}{n-1}\sum_{j\neq i}\phi_{n,ij}^{\omega}\Vert\geq\kappa|X)\leq2T\exp(-\frac{(n-1)\kappa^{2}}{2M^{2}T^{2}})$.
Therefore, by Lemma 2.2.1 in \citet{VW_1996}, we obtain $\|\frac{1}{n-1}\sum_{j\neq i}\phi_{n,ij}^{\omega}(\omega_{ni}^{\ast},\epsilon_{ij})\|_{\psi_{2}|X}\leq\sqrt{2(2T+1)}TM/\sqrt{n-1},$
which is constant across $i$.

By equation (\ref{eq:w.emp.orlicz}) in Lemma \ref{lem:w.emp}(ii)
(to be proved later), we can derive that $\max_{i}\Vert\mathbb{G}_{n}(\phi_{ni}^{\gamma}(\omega_{ni}(\epsilon_{i}),\epsilon_{i})-\phi_{ni}^{\gamma}(\omega_{ni}^{\ast},\epsilon_{i}))\Vert_{\psi_{1}|X}=o(1)$.
Following the proof for equation (\ref{eq:w.emp.orlicz}), we obtain
that $\max_{i}\Vert\mathbb{G}_{n}(\phi_{ni}^{\gamma}(\omega_{ni}(\epsilon_{i}),\epsilon_{i})-\phi_{ni}^{\gamma}(\omega_{ni}^{\ast},\epsilon_{i}))\Vert_{\psi_{2}|X}=o(1)$.
Hence, the second term on the right-hand side of (\ref{eq:w.orlicz.bdd})
is $o(1)/\sqrt{n-1}$ uniformly over $i$. It follows that $\|\max_{i}\Vert\Lambda_{ni}(\omega_{ni}(\epsilon_{i})-\omega_{ni}^{\ast})\Vert^{2}\|_{\psi_{1}|X}\leq K\ln(n+1)((\sqrt{2(2T+1)}TM+o(1))/\sqrt{n-1})^{2}=o(n^{-1/2})$.
Part (i) is proved.

Part (ii): From part (i), we have $\max_{i}\Vert\mathbb{G}_{n}(\phi_{ni}^{\gamma}(\omega_{ni}(\epsilon_{i}),\epsilon_{i})-\phi_{ni}^{\gamma}(\omega_{ni}^{\ast},\epsilon_{i}))\Vert_{\psi_{1}|X}=o(1)$.
Combining this with the statement in part (i) and Assumption \ref{ass:w}(iii),
we derive
\begin{eqnarray*}
 &  & \|\max_{i}\Vert r_{ni}^{\omega}(\epsilon_{i})\Vert\|_{\psi_{1}|X}\\
 & \leq & \max_{i}\Vert H_{ni}^{\omega}(\omega_{ni}^{\ast})^{-1}\Vert(O_{p}(\|\max_{i}\Vert\Lambda_{ni}(\omega_{ni}(\epsilon_{i})-\omega_{ni}^{\ast})\Vert^{2}\|_{\psi_{1}|X})\\
 &  & +(n-1)^{-1/2}\|\Lambda_{ni}\Phi'_{ni}\|\max_{i}\Vert\mathbb{G}_{n}(\phi_{ni}^{\gamma}(\omega_{ni}(\epsilon_{i}),\epsilon_{i})-\phi_{ni}^{\gamma}(\omega_{ni}^{\ast},\epsilon_{i}))\Vert_{\psi_{1}|X})=o(n^{-1/2}).
\end{eqnarray*}
\end{proof}
\begin{lem}[Stochastic Equicontinuity]
\label{lem:w.emp}Suppose that Assumptions \ref{ass:e=000026x}--\ref{ass:discX}
and \ref{ass:w} are satisfied. Conditional on $X$, $\mathbb{G}_{n}\phi_{ni}^{\gamma}(\omega,\epsilon_{i})$
defined in (\ref{eq:w.ep.def}) satisfies that (i) if $\Lambda_{ni}(\omega_{ni}(\epsilon_{i})-\omega_{ni}^{\ast})=o_{p}(1)$,
$\mathbb{G}_{n}(\phi_{ni}^{\gamma}(\omega_{ni}(\epsilon_{i}),\epsilon_{i})-\phi_{ni}^{\gamma}(\omega_{ni}^{\ast},\epsilon_{i}))=o_{p}(1)$,
and (ii) if $\Lambda_{ni}(\omega_{ni}(\epsilon_{i})-\omega_{ni}^{\ast})=O_{p}(n^{-1/2})$,
$\max_{1\leq i\leq n}\Vert\mathbb{G}_{n}(\phi_{ni}^{\gamma}(\omega_{ni}(\epsilon_{i}),\epsilon_{i})-\phi_{ni}^{\gamma}(\omega_{ni}^{\ast},\epsilon_{i}))\Vert=o_{p}(1).$
\end{lem}
\begin{proof}
Part (i): Because $\Lambda_{ni}(\omega_{ni}(\epsilon_{i})-\omega_{ni}^{\ast})=o_{p}(1)$,
we can define $h_{ni}=r_{ni}\Lambda_{ni}(\omega_{ni}(\epsilon_{i})-\omega_{ni}^{\ast})$
for some $r_{ni}\rightarrow\infty$ such that $h_{ni}\in\Omega$ if
$n$ is sufficiently large,\footnote{This implies that $r_{ni}$ diverges more slowly than $\Lambda_{ni}(\omega_{ni}(\epsilon_{i})-\omega_{ni}^{\ast})$
converges to zero.} because by Assumption \ref{ass:w}(i) $\Omega$ contains a compact
neighborhood of $0$.

Recall that $\phi_{n,ij}^{\gamma}(\omega,\epsilon_{ij})=1\{U_{n,ij}+\frac{n-1}{n-2}Z_{j}^{\prime}\Phi_{ni}\Lambda_{ni}\omega\geq\epsilon_{ij}\}Z_{j}$.
View it as a function of $\omega^{\lambda}\equiv\Lambda_{ni}\omega\in\Omega$,
and define $\phi_{n,ij}^{\gamma\lambda}(\omega^{\lambda},\epsilon_{ij})\equiv1\{U_{n,ij}+\frac{n-1}{n-2}Z_{j}^{\prime}\Phi_{ni}\omega^{\lambda}\geq\epsilon_{ij}\}Z_{j}$.
We can write $\mathbb{G}_{n}(\phi_{ni}^{\gamma}(\omega_{ni}(\epsilon_{i}),\epsilon_{i})-\phi_{ni}^{\gamma}(\omega_{ni}^{\ast},\epsilon_{i}))=\mathbb{G}_{n}(\phi_{ni}^{\gamma\lambda}(\omega_{ni}^{\lambda}(\epsilon_{i}),\epsilon_{i})-\phi_{ni}^{\gamma\lambda}(\omega_{ni}^{\ast\lambda},\epsilon_{i}))$,
where we denote $\omega_{ni}^{\lambda}(\epsilon_{i})=\Lambda_{ni}\omega_{ni}(\epsilon_{i})$
and $\omega_{ni}^{\ast\lambda}=\Lambda_{ni}\omega_{ni}^{\ast}$. By
Markov's inequality and the change of variable, it suffices to show
$\mathbb{E}[\sup_{\omega^{\lambda},h\in\Omega}\Vert\mathbb{G}_{n}(\phi_{ni}^{\gamma\lambda}(\omega^{\lambda}+r_{ni}^{-1}h,\epsilon_{i})-\phi_{ni}^{\gamma\lambda}(\omega^{\lambda},\epsilon_{i}))\Vert|X]=o(1)$.
For simplicity, we write $\mathbb{G}_{n}(\phi_{ni}^{\gamma\lambda}(\omega^{\lambda}+r_{ni}^{-1}h,\epsilon_{i})-\phi_{ni}^{\gamma\lambda}(\omega^{\lambda},\epsilon_{i}))$
as $\mathbb{G}_{ni}^{\gamma}(\omega^{\lambda},r_{ni}^{-1}h)$. 

To show this, we need to prove that the empirical process $\mathbb{G}_{ni}^{\gamma}(\omega^{\lambda},r_{ni}^{-1}h)$,
indexed by $\omega^{\lambda},h\in\Omega$, is stochastically equicontinuous.
This is a triangular array with function $\phi_{n,ij}^{\gamma\lambda}$
that varies across $j$, so most of the ready-to-use results for stochastic
equicontinuity \citep{Andrews_1994} are not applicable. Instead,
we apply the maximal inequalities in \citet{VW_1996} to directly
prove the stochastic equicontinuity.

For any $\omega^{\lambda},\tilde{\omega}^{\lambda}\in\Omega$, we
can bound the function $\phi_{n,ij}^{\gamma\lambda}(\omega^{\lambda},\epsilon_{ij})-\phi_{n,ij}^{\gamma\lambda}(\tilde{\omega}^{\lambda},\epsilon_{ij})$
by $|1\{U_{n,ij}+\frac{n-1}{n-2}Z_{j}^{\prime}\Phi_{ni}\omega^{\lambda}\geq\epsilon_{ij}\}-1\{U_{n,ij}+\frac{n-1}{n-2}Z_{j}^{\prime}\Phi_{ni}\tilde{\omega}^{\lambda}\geq\epsilon_{ij}\}|\|Z_{j}\|\leq\eta_{n,ij}(\omega^{\lambda},\tilde{\omega}^{\lambda},\epsilon_{ij})$,
where the bound $\eta_{n,ij}(\omega^{\lambda},\tilde{\omega}^{\lambda},\epsilon_{ij})=1$
if $\epsilon_{ij}$ lies between $U_{n,ij}+\frac{n-1}{n-2}Z_{j}^{\prime}\Phi_{ni}\omega^{\lambda}$
and $U_{n,ij}+\frac{n-1}{n-2}Z_{j}^{\prime}\Phi_{ni}\tilde{\omega}^{\lambda}$,
and $0$ otherwise. For $1\leq t\leq T,$ let $\mathbb{G}_{ni,t}^{\gamma}(\omega^{\lambda},r_{ni}^{-1}h)$
denote the $t$th component of $\mathbb{G}_{ni}^{\gamma}(\omega^{\lambda},r_{ni}^{-1}h)$.
Write $\mathbb{G}_{ni,t}^{\gamma}(\omega^{\lambda},r_{ni}^{-1}h)$
and $\eta_{n,ij}(\omega^{\lambda},\tilde{\omega}^{\lambda},\epsilon_{ij})$
as $\mathbb{G}_{ni,t}^{\gamma}$ and $\eta_{n,ij}$. Define $\Vert\eta_{ni}\Vert_{n}\equiv(\frac{1}{n-1}\sum_{j\neq i}\eta_{n,ij}^{2})^{1/2}\leq1$
as the empirical $L_{2}$ norm of $\eta_{ni}=(\eta_{n,ij},j\neq i)$.
Fix $X$ and $\eta_{ni}$. By Hoeffding's inequality for bounded random
variables \citep[Theorem 2.8]{BLM_2013}, $\Pr(|\mathbb{G}_{ni,t}^{\gamma}|\geq\frac{\kappa}{T}|X,\epsilon_{i})=\Pr(\mathbb{G}_{ni,t}^{\gamma}\geq\frac{\kappa}{T}|X,\eta_{ni})+\Pr(-\mathbb{G}_{ni,t}^{\gamma}\geq\frac{\kappa}{T}|X,\eta_{ni})\leq2\exp(-\frac{2\kappa^{2}}{\Vert\eta_{ni}\Vert_{n}^{2}T^{2}})$
for each $1\leq t\leq T$ and hence $\Pr(\Vert\mathbb{G}_{ni}^{\gamma}\Vert>\kappa|X,\eta_{ni})\leq2T\exp(-\frac{2\kappa^{2}}{\Vert\eta_{ni}\Vert_{n}^{2}T^{2}})$.
It then follows from Lemma 2.2.1 in \citet{VW_1996} that $\|\mathbb{G}_{ni}^{\gamma}\|_{\psi_{2}|X,\eta_{ni}}\leq\sqrt{(2T+1)/2}T\Vert\eta_{ni}\Vert_{n}$.
Note that the conditional $L_{1}$ norm is bounded by a multiple of
the conditional $\psi_{2}$-Orlicz norm. Therefore, by Theorem 2.2.4
in \citet{VW_1996} with $d$ being the empirical $L_{2}$ norm $\|\cdot\|_{n}$
and $\psi=\psi_{2}$, we derive 
\begin{equation}
\mathbb{E}\left[\left.\sup_{\omega^{\lambda},h\in\Omega}\Vert\mathbb{G}_{ni}^{\gamma}(\omega^{\lambda},r_{ni}^{-1}h)\Vert\right\vert X\right]\leq K\mathbb{E}\left[\left.J(1,\mathcal{F}_{ni}(\epsilon_{i}))\sup_{\omega^{\lambda},h\in\Omega}\Vert\eta_{ni}(\omega^{\lambda}+r_{ni}^{-1}h,\omega^{\lambda},\epsilon_{i})\Vert_{n}\right\vert X\right],\label{eq:w.maxineq}
\end{equation}
where $K>0$ is a constant, and $J(1,\mathcal{F}_{ni}(\epsilon_{i}))$
is the uniform entropy integral of the set of triangular arrays $\mathcal{F}_{ni}(\epsilon_{i})\equiv\{(\phi_{n,ij}^{\gamma\lambda}(\omega^{\lambda}+r_{ni}^{-1}h,\epsilon_{ij})-\phi_{n,ij}^{\gamma\lambda}(\omega^{\lambda},\epsilon_{ij}),j\neq i):\omega^{\lambda},h\in\Omega\}$,
i.e., 
\begin{equation}
J(1,\mathcal{F}_{ni}(\epsilon_{i}))\equiv\sup_{\alpha\in\mathbb{R}_{+}^{n-1}}\int_{0}^{1}\sqrt{1+\ln D(\xi\Vert\alpha\odot\bar{\eta}_{ni}(\epsilon_{i})\Vert_{n},\alpha\odot\mathcal{F}_{ni}(\epsilon_{i}),\Vert\cdot\Vert_{n})}d\xi.\label{eq:w.entrp}
\end{equation}
In this expression, $\bar{\eta}_{ni}(\epsilon_{i})\equiv\sup_{\omega^{\lambda},h\in\Omega}\eta_{ni}(\omega^{\lambda}+r_{ni}^{-1}h,\omega^{\lambda},\epsilon_{i})$
is an envelope of $\mathcal{F}_{ni}(\epsilon_{i})$, $\alpha\in\mathbb{R}_{+}^{n-1}$
is a vector of nonnegative constants, $\alpha\odot\bar{\eta}_{ni}(\epsilon_{i})$
is the Hadamard product of $\alpha$ and $\bar{\eta}_{ni}(\epsilon_{i})$,
$\alpha\odot\mathcal{F}_{ni}(\epsilon_{i})$ is the set of Hadamard
products of $\alpha$ and the triangular arrays in $\mathcal{F}_{ni}(\epsilon_{i})$,
and $D(\xi\Vert\alpha\odot\bar{\eta}_{ni}(\epsilon_{i})\Vert_{n},\alpha\odot\mathcal{F}_{ni}(\epsilon_{i}),\Vert\cdot\Vert_{n})$
is the packing number, that is, the maximum number of points in the
set $\alpha\odot\mathcal{F}_{ni}(\epsilon_{i})$ that are separated
by the distance $\xi\Vert\alpha\odot\bar{\eta}_{ni}(\epsilon_{i})\Vert_{n}$
for the norm $\Vert\cdot\Vert_{n}$. The sup outside the integral
is taken over all vectors $\alpha\in\mathbb{R}_{+}^{n-1}$. 

To show that the uniform entropy integral $J(1,\mathcal{F}_{ni}(\epsilon_{i}))$
is finite, consider the indicator function $g_{n,ij}^{\lambda}(\omega^{\lambda},\epsilon_{ij})=1\{U_{n,ij}+\frac{n-1}{n-2}Z_{j}^{\prime}\Phi_{ni}\omega^{\lambda}\geq\epsilon_{ij}\}$,
whose argument involves a linear function of $\omega^{\lambda}$.
We can show that the set $\{(g_{n,ij}^{\lambda}(\omega^{\lambda},\epsilon_{ij}),j\neq i):\omega^{\lambda}\in\Omega\}$
has a pseudo-dimension of at most $T$, so it is Euclidean \citep[Corollary 4.10]{Pollard_1990}.\footnote{To see this, by the definition of pseudo-dimension, it suffices to
show that for each index set $I=\{j_{1},\ldots,j_{T+1}\}\in\{1,\ldots,n\}\backslash\{i\}$
and each point $c\in\mathbb{R}^{T+1}$, there is a subset $J\subseteq I$
such that no $\omega^{\lambda}\in\Omega$ can satisfy the inequalities
$g_{n,ij}(\omega^{\lambda},\epsilon_{ij})>c_{j}$ for $j\in J$ and
$<c_{j}$ for $j\in I\backslash J$. If $c$ has a component $c_{j}$
that lies outside of $(0,1)$, we can choose $J$ such that $j\in J$
if $c_{j}\geq1$ and $j\in I\backslash J$ if $c_{j}\leq0$ so no
$\omega^{\lambda}$ can satisfy the inequalities above. It thus suffices
to consider $c$ with all the components in $(0,1)$ and for such
$c$ the inequalities reduce to $U_{n,ij}+\frac{n-1}{n-2}Z_{j}^{\prime}\Phi_{ni}\omega^{\lambda}\geq\epsilon_{ij}$
for $j\in J$ and $<\epsilon_{ij}$ for $j\in I\backslash J$. Since
$Z_{j}^{\prime}\Phi_{ni}\in\mathbb{R}^{T}$ for all $j$, there exists
a non-zero vector $\tau=(\tau_{1},\ldots,\tau_{T+1})\in\mathbb{R}^{T+1}$
such that $\sum_{t=1}^{T+1}\tau_{t}Z_{j_{t}}^{\prime}\Phi_{ni}=0$,
so $\sum_{t=1}^{T+1}\tau_{t}\frac{n-1}{n-2}Z_{j_{t}}^{\prime}\Phi_{ni}\omega^{\lambda}=0$
for all $\omega^{\lambda}\in\Omega$. We may assume that $\tau_{t}>0$
for at least one $t$. If $\sum_{t=1}^{T+1}\tau_{t}(U_{n,ij_{t}}-\epsilon_{n,ij_{t}})\geq0$,
it is impossible to find a $\omega^{\lambda}\in\Omega$ satisfying
those inequalities for the choice $J=\{j_{t}\in I:\tau_{t}\leq0\}$,
because this would lead to the contradiction $\sum_{t=1}^{T+1}\tau_{t}(U_{n,ij_{t}}-\epsilon_{n,ij_{t}})=\sum_{t=1}^{T+1}\tau_{t}(U_{n,ij_{t}}-\epsilon_{n,ij_{t}})+\sum_{t=1}^{T+1}\tau_{t}\frac{n-1}{n-2}Z_{j_{t}}^{\prime}\Phi_{ni}\omega^{\lambda}=\sum_{t=1}^{T+1}\tau_{t}(U_{n,ij_{t}}+\frac{n-1}{n-2}Z_{j_{t}}^{\prime}\Phi_{ni}\omega^{\lambda}-\epsilon_{n,ij_{t}})<0$.
If $\sum_{t=1}^{T+1}\tau_{t}(U_{n,ij_{t}}-\epsilon_{n,ij_{t}})<0$,
we could choose $J=\{j_{t}\in I:\tau_{t}\geq0\}$ to reach a similar
contradiction.} Note that $\phi_{n,ij}^{\gamma\lambda}(\omega^{\lambda},\epsilon_{ij})=g_{n,ij}^{\lambda}(\omega^{\lambda},\epsilon_{ij})Z_{j}$,
and $Z_{j}$ is a $T\times1$ vector that does not depend on $\omega^{\lambda}$.
By Lemma 5.3 in \citet{Pollard_1990}, the set $\{(\phi_{n,ij}^{\gamma\lambda}(\omega^{\lambda},\epsilon_{ij}),j\neq i):\omega^{\lambda}\in\Omega\}$
is Euclidean. Moreover, the set $\mathcal{F}_{ni}(\epsilon_{i})$
is given by the sum of two sets: $\{(\phi_{n,ij}^{\gamma\lambda}(\omega^{\lambda}+r_{ni}^{-1}h,\epsilon_{ij}),j\neq i):\omega^{\lambda},h\in\Omega\}$
and $\{(-\phi_{n,ij}^{\gamma\lambda}(\omega^{\lambda},\epsilon_{ij}),j\neq i):\omega^{\lambda}\in\Omega\}$.
From the stability result in \citet[Section 5]{Pollard_1990} on the
sum of two sets, $\mathcal{F}_{ni}(\epsilon_{i})$ is Euclidean. Therefore,
$\mathcal{F}_{ni}(\epsilon_{i})$ has a finite uniform entropy integral,
that is, $J(1,\mathcal{F}_{ni}(\epsilon_{i}))\leq\bar{J}$ uniformly
in $\epsilon_{i}$ and $n$, for some $\bar{J}<\infty$.

Next, we analyze the sup term in (\ref{eq:w.maxineq}). By Cauchy-Schwarz
inequality, we have $\mathbb{E}[\sup_{\omega^{\lambda},h\in\Omega}\Vert\eta_{ni}(\omega^{\lambda}+r_{ni}^{-1}h,\omega^{\lambda},\epsilon_{i})\Vert_{n}|X]\leq(\mathbb{E}[\sup_{\omega^{\lambda},h\in\Omega}\frac{1}{n-1}\sum_{j\neq i}\eta_{n,ij}^{2}(\omega^{\lambda}+r_{ni}^{-1}h,\omega^{\lambda},\epsilon_{ij})|X])^{1/2}$.
Consider the empirical process $\mathbb{G}_{n}\eta_{ni}^{2}(\omega^{\lambda}+r_{ni}^{-1}h,\omega^{\lambda},\epsilon_{i})\equiv\frac{1}{\sqrt{n-1}}\sum_{j\neq i}(\eta_{n,ij}^{2}(\omega^{\lambda}+r_{ni}^{-1}h,\omega^{\lambda},\epsilon_{ij})-\mathbb{E}[\eta_{n,ij}^{2}(\omega^{\lambda}+r_{ni}^{-1}h,\omega^{\lambda},\epsilon_{ij})|X])$,
indexed by $\omega^{\lambda},h\in\Omega$. Note that each $\eta_{n,ij}^{2}$
is bounded by $1$. Using an argument similar to the one that leads
to (\ref{eq:w.maxineq}), we derive the upper bound
\begin{equation}
\mathbb{E}\left[\left.\sup_{\omega^{\lambda},h\in\Omega}|\mathbb{G}_{n}\eta_{ni}^{2}(\omega^{\lambda}+r_{ni}^{-1}h,\omega^{\lambda},\epsilon_{i})|\right\vert X\right]\leq K^{\eta}\mathbb{E}[J(1,\mathcal{F}_{ni}^{\eta}(\epsilon_{i}))|X],\label{eq:wbd.maxineq}
\end{equation}
where $K^{\eta}<\infty$ is a constant and $J(1,\mathcal{F}_{ni}^{\eta}(\epsilon_{i}))$
is the uniform entropy integral of the set of triangular arrays $\mathcal{F}_{ni}^{\eta}(\epsilon_{i})\equiv\{(\eta_{n,ij}^{2}(\omega^{\lambda}+r_{ni}^{-1}h,\omega^{\lambda},\epsilon_{ij}),j\neq i):\omega^{\lambda},h\in\Omega\}$.
Similarly to the argument for the set $\mathcal{F}_{ni}(\epsilon_{i})$,
we can show that the set $\mathcal{F}_{ni}^{\eta}(\epsilon_{i})$
has a finite uniform entropy integral $J(1,\mathcal{F}_{ni}^{\eta}(\epsilon_{i}))\leq\bar{J}^{\eta}<\infty$.
From these results we obtain
\begin{eqnarray*}
 &  & \mathbb{E}\left[\left.\sup_{\omega^{\lambda},h\in\Omega}\frac{1}{n-1}\sum_{j\neq i}\eta_{n,ij}^{2}(\omega^{\lambda}+r_{ni}^{-1}h,\omega^{\lambda},\epsilon_{ij})\right|X\right]\\
 &  & -\sup_{\omega^{\lambda},h\in\Omega}\frac{1}{n-1}\sum_{j\neq i}\mathbb{E}[\eta_{n,ij}^{2}(\omega^{\lambda}+r_{ni}^{-1}h,\omega^{\lambda},\epsilon_{ij})|X]\\
 & \leq & \frac{1}{\sqrt{n-1}}\mathbb{E}\left[\left.\sup_{\omega^{\lambda},h\in\Omega}|\mathbb{G}_{n}\eta_{ni}^{2}(\omega^{\lambda}+r_{ni}^{-1}h,\omega^{\lambda},\epsilon_{i})|\right|X\right]\leq\frac{K^{\eta}\bar{J}^{\eta}}{\sqrt{n-1}}.
\end{eqnarray*}
Moreover, by the mean-value theorem, for any $\omega^{\lambda},h\in\Omega$
and any $j\neq i$, we have
\begin{eqnarray*}
 &  & \mathbb{E}[\eta_{n,ij}^{2}(\omega^{\lambda}+r_{ni}^{-1}h,\omega^{\lambda},\epsilon_{ij})|X]\\
 & = & \left|F_{\epsilon}(U_{n,ij}+\frac{n-1}{n-2}Z_{j}^{\prime}\Phi_{ni}(\omega^{\lambda}+r_{ni}^{-1}h))-F_{\epsilon}(U_{n,ij}+\frac{n-1}{n-2}Z_{j}^{\prime}\Phi_{ni}\omega^{\lambda})\right|\Vert Z_{j}\Vert^{2}\\
 & = & r_{ni}^{-1}f_{\epsilon}(U_{n,ij}+\frac{n-1}{n-2}Z_{j}^{\prime}\Phi_{ni}(\omega^{\lambda}+t_{n,ij}r_{ni}^{-1}h))\frac{n-1}{n-2}|Z_{j}^{\prime}\Phi_{ni}h|,
\end{eqnarray*}
for some $t_{n,ij}\in[0,1]$. By the boundedness of $f_{\epsilon}$
under Assumption \ref{ass:e=000026x}(ii) and $\sup_{h\in\Omega}\|h\|<\infty$,
there is a $M<\infty$ such that $\mathbb{E}[\eta_{n,ij}^{2}(\omega^{\lambda}+r_{ni}^{-1}h,\omega^{\lambda},\epsilon_{ij})|X]\leq r_{ni}^{-1}M$
for all $\omega^{\lambda},h\in\Omega$ and all $j$. It follows that
$\mathbb{E}[\sup_{\omega^{\lambda},h\in\Omega}\frac{1}{n-1}\sum_{j\neq i}\eta_{n,ij}^{2}(\omega^{\lambda}+r_{ni}^{-1}h,\omega^{\lambda},\epsilon_{ij})|X]\leq(n-1)^{-1/2}K^{\eta}\bar{J}^{\eta}+r_{ni}^{-1}M$.

Combining the results we conclude that $\mathbb{E}[\sup_{\omega^{\lambda},h\in\Omega}\Vert\mathbb{G}_{ni}^{\gamma}(\omega^{\lambda},r_{ni}^{-1}h)\Vert|X]$
is bounded by $K\bar{J}((n-1)^{-1/2}K^{\eta}\bar{J}^{\eta}+r_{ni}^{-1}M)^{1/2}=o(1)$.
Part (i) is proved.

Part (ii): Because $\Lambda_{ni}(\omega_{ni}(\epsilon_{i})-\omega_{ni}^{\ast})=O_{p}(n^{-1/2})$,
we define $h_{ni}=n^{\kappa}\Lambda_{ni}(\omega_{ni}(\epsilon_{i})-\omega_{ni}^{\ast})$
for $0<\kappa<1/2$ such that $h_{ni}\in\Omega$ if $n$ is sufficiently
large. Recall from part (i) that $\mathbb{G}_{n}(\phi_{ni}^{\gamma}(\omega_{ni}(\epsilon_{i}),\epsilon_{i})-\phi_{ni}^{\gamma}(\omega_{ni}^{\ast},\epsilon_{i}))=\mathbb{G}_{n}(\phi_{ni}^{\gamma\lambda}(\omega_{ni}^{\lambda}(\epsilon_{i}),\epsilon_{i})-\phi_{ni}^{\gamma\lambda}(\omega_{ni}^{\ast\lambda},\epsilon_{i}))$.
By Markov's inequality, it suffices to show that $\mathbb{E}[\max_{i}\sup_{\omega^{\lambda},h\in\Omega}\Vert\mathbb{G}_{n}(\phi_{ni}^{\gamma\lambda}(\omega^{\lambda}+n^{-\kappa}h,\epsilon_{i})-\phi_{ni}^{\gamma\lambda}(\omega^{\lambda},\epsilon_{i}))\Vert|X]=o(1)$.
For simplicity, we write $\mathbb{G}_{n}(\phi_{ni}^{\gamma\lambda}(\omega^{\lambda}+n^{-\kappa}h,\epsilon_{i})-\phi_{ni}^{\gamma\lambda}(\omega^{\lambda},\epsilon_{i}))$
as $\mathbb{G}_{ni}^{\gamma}(\omega^{\lambda},n^{-\kappa}h)$.

By Lemma 2.2.2 in \citet{VW_1996}, we can derive the inequality $\mathbb{E}[\max_{i}\sup_{\omega^{\lambda},h\in\Omega}\Vert\mathbb{G}_{ni}^{\gamma}(\omega^{\lambda},n^{-\kappa}h)\Vert|X]\leq\|\max_{i}\sup_{\omega^{\lambda},h\in\Omega}\Vert\mathbb{G}_{ni}^{\gamma}(\omega^{\lambda},n^{-\kappa}h)\Vert\|_{\psi_{1}|X}\leq K\ln(n+1)\max_{i}\|\sup_{\omega^{\lambda},h\in\Omega}\Vert\mathbb{G}_{ni}^{\gamma}(\omega^{\lambda},n^{-\kappa}h)\Vert\|_{\psi_{1}|X}$
for constant $K<\infty$. Moreover, by Proposition A.1.6 ($p=1$)
and Lemma 2.2.2 in \citet{VW_1996}, for each $i$ we obtain
\begin{eqnarray*}
\left\Vert \sup_{\omega^{\lambda},h\in\Omega}\Vert\mathbb{G}_{ni}^{\gamma}(\omega^{\lambda},n^{-\kappa}h)\Vert\right\Vert _{\psi_{1}|X} & \leq & K_{1}\left(\mathbb{E}\left[\left.\sup_{\omega^{\lambda},h\in\Omega}\Vert\mathbb{G}_{ni}^{\gamma}(\omega^{\lambda},n^{-\kappa}h)\Vert\right|X\right]\right.\\
 &  & +\left.\frac{\ln(n+1)}{\sqrt{n-1}}\max_{j\neq i}\left\Vert \sup_{\omega^{\lambda},h\in\Omega}|\eta_{n,ij}(\omega^{\lambda}+n^{-\kappa}h,\omega^{\lambda},\epsilon_{ij})|\right\Vert _{\psi_{1}|X}\right)
\end{eqnarray*}
for constant $K_{1}<\infty$, and $\eta_{n,ij}$ is defined in part
(i). For the first term on the right-hand side, following part (i)
we derive $\mathbb{E}[\sup_{\omega^{\lambda},h\in\Omega}\Vert\mathbb{G}_{ni}^{\gamma}(\omega^{\lambda},n^{-\kappa}h)\Vert|X]\leq K\bar{J}((n-1)^{-1/2}K^{\eta}\bar{J}^{\eta}+n^{-\kappa}M)^{1/2}$.
The second term on the right-hand side satisfies $\max_{j\neq i}\Vert\sup_{\omega^{\lambda},h\in\Omega}|\eta_{n,ij}(\omega^{\lambda}+n^{-\kappa}h,\omega^{\lambda},\epsilon_{ij})|\Vert_{\psi_{1}|X}\leq1$
by construction.  Therefore, for all $i$ we have the uniform bound
\begin{equation}
\left\Vert \sup_{\omega^{\lambda},h\in\Omega}\Vert\mathbb{G}_{ni}^{\gamma}(\omega^{\lambda},n^{-\kappa}h)\Vert\right\Vert _{\psi_{1}|X}\leq K_{1}\left(K\bar{J}\sqrt{\frac{K^{\eta}\bar{J}^{\eta}}{\sqrt{n-1}}+\frac{M}{n^{\kappa}}}+\frac{\ln(n+1)}{\sqrt{n-1}}\right).\label{eq:w.emp.orlicz}
\end{equation}

Combining the results, we conclude that $\mathbb{E}[\max_{i}\sup_{\omega^{\lambda},h\in\Omega}\Vert\mathbb{G}_{ni}^{\gamma}(\omega^{\lambda},n^{-\kappa}h)\Vert|X]$
is bounded by $K\ln(n+1)K_{1}(K\bar{J}((n-1)^{-1/2}K^{\eta}\bar{J}^{\eta}+n^{-\kappa}M)^{1/2}+(n-1)^{-1/2}\ln(n+1))=o(1)$.
\end{proof}

\subsubsection{\label{online:lemma.theta.asymdist}Asymptotic Distribution of $\hat{\theta}_{n}$}
\begin{lem}[Asymptotic normality of the sample moments]
\label{lem:m.clt}Suppose that Assumptions \ref{ass:e=000026x}--\ref{ass:discX}
and \ref{ass:w} are satisfied. Let $Y_{n}=\frac{1}{\sqrt{n(n-1)}}\sum_{i}\sum_{j\neq i}q_{n,ij}(G_{n,ij}-P_{n,ij}(\theta_{0},p_{n}))$,
where $q_{n,ij}\in\mathbb{R}^{d}$ is a $d\times1$ vector of instruments
that is a function of $X$ and satisfies $\max_{1\leq i,j\leq n}\|q_{n,ij}\|\leq\bar{q}<\infty$.
Define the $d\times1$ vector $\phi_{n,ij}^{y}\equiv q_{n,ij}(g_{n,ij}(\omega_{ni}^{\ast},\epsilon_{ij})-P_{n,ij}^{\ast}(\omega_{ni}^{\ast}))+J_{ni}^{\omega}(\omega_{ni}^{\ast},q_{ni})\phi_{n,ij}^{\omega}(\omega_{ni}^{\ast},\epsilon_{ij})$,
where $g_{n,ij}(\omega,\epsilon_{ij})=1\{U_{n,ij}+\frac{n-1}{n-2}Z_{j}^{\prime}\Phi_{ni}\Lambda_{ni}\omega\geq\epsilon_{ij}\}$
is the indicator function defined in equation (\ref{eq:gij}), $P_{n,ij}^{\ast}(\omega_{ni}^{\ast})=F_{\epsilon}(U_{n,ij}+\frac{n-1}{n-2}Z_{j}^{\prime}\Phi_{ni}\Lambda_{ni}\omega_{ni}^{\ast})$,
$J_{ni}^{\omega}(\omega_{ni}^{\ast},q_{ni})=\frac{1}{n-1}\sum_{j\neq i}q_{n,ij}\nabla_{\omega^{\lambda\prime}}P_{n,ij}^{\ast}(\omega_{ni}^{\ast})$
is the $d\times T$ weighted Jacobian matrix, where $\phi_{n,ij}^{\omega}(\omega_{ni}^{\ast},\epsilon_{ij})\in\mathbb{R}^{T}$
is the influence function defined in Lemma \ref{lem:w.asymlin}. Define
the $d\times d$ variance matrix $\Sigma_{n}=\frac{1}{n(n-1)}\sum_{i}\sum_{j\neq i}\mathbb{E}[\phi_{n,ij}^{y}(\phi_{n,ij}^{y})'|X]$.
Conditional on $X$, $\Sigma_{n}^{-1/2}Y_{n}\overset{d}{\rightarrow}N(0,I_{d})$.
\end{lem}
\begin{proof}
By Theorem \ref{thm:optG}, an observed link $G_{n,ij}$ is given
by the link indicator $g_{n,ij}(\omega,\epsilon_{ij})$ evaluated
at $\omega_{ni}(\epsilon_{i})$, that is, $G_{n,ij}=g_{n,ij}(\omega_{ni}(\epsilon_{i}),\epsilon_{ij})$,
where $\omega_{ni}(\epsilon_{i})$ is a maximin solution of the function
$\Pi_{ni}(\omega)$. Moreover, $P_{n,ij}(\theta_{0},p_{n})=\mathbb{E}[g_{n,ij}(\omega_{ni}(\epsilon_{i}),\epsilon_{ij})|X]$.
Therefore, $Y_{n}$ can be represented as 
\[
Y_{n}=\frac{1}{\sqrt{n(n-1)}}\sum_{i}\sum_{j\neq i}q_{n,ij}(g_{n,ij}(\omega_{ni}(\epsilon_{i}),\epsilon_{ij})-\mathbb{E}[g_{n,ij}(\omega_{ni}(\epsilon_{i}),\epsilon_{ij})|X])\text{.}
\]
The challenge in deriving the asymptotic distribution of $Y_{n}$
lies in the fact that link choices of an individual are correlated
through $\omega_{ni}(\epsilon_{i})$. To account for the correlation,
we decompose $Y_{n}$ into four parts $Y_{n}=T_{1n}+T_{2n}+T_{3n}+T_{4n}$,
where
\begin{eqnarray}
T_{1n} & = & \frac{1}{\sqrt{n(n-1)}}\sum_{i}\sum_{j\neq i}q_{n,ij}(g_{n,ij}(\omega_{ni}^{\ast},\epsilon_{ij})-P_{n,ij}^{\ast}(\omega_{ni}^{\ast}))\nonumber \\
T_{2n} & = & \frac{1}{\sqrt{n(n-1)}}\sum_{i}\sum_{j\neq i}q_{n,ij}(g_{n,ij}(\omega_{ni}(\epsilon_{i}),\epsilon_{ij})-g_{n,ij}(\omega_{ni}^{\ast},\epsilon_{ij})-(P_{n,ij}^{\ast}(\omega_{ni}(\epsilon_{i}))-P_{n,ij}^{\ast}(\omega_{ni}^{\ast})))\nonumber \\
T_{3n} & = & \frac{1}{\sqrt{n(n-1)}}\sum_{i}\sum_{j\neq i}q_{n,ij}(P_{n,ij}^{\ast}(\omega_{ni}(\epsilon_{i}))-\mathbb{E}[P_{n,ij}^{\ast}(\omega_{ni}(\epsilon_{i}))|X])\nonumber \\
T_{4n} & = & \frac{1}{\sqrt{n(n-1)}}\sum_{i}\sum_{j\neq i}q_{n,ij}(\mathbb{E}[P_{n,ij}^{\ast}(\omega_{ni}(\epsilon_{i}))|X]-\mathbb{E}[g_{n,ij}(\omega_{ni}(\epsilon_{i}),\epsilon_{ij})|X]).\label{eq:Y.dcmp}
\end{eqnarray}
The four terms in the decomposition can be interpreted as follows.
The first term $T_{1n}$ is the sample moment if we replace $\omega_{ni}(\epsilon_{i})$
by its limit $\omega_{ni}^{\ast}$. This substitution removes the
correlation between the link choices of an individual. The second
term $T_{2n}$ is the difference between the dependent sample moment
and the independent one in $T_{1n}$. The fact that this term is shown
to be negligible indicates that the correlation between link choices
vanishes as $n$ grows large. The sampling variation in $\omega_{ni}(\epsilon_{i})$
is captured by the third term $T_{3n}$ which contributes to the asymptotic
variance of the moment function. Finally, the fourth term $T_{4n}$
satisfies $T_{4n}=-\mathbb{E}[T_{2n}|X]$ and hence is asymptotically
negligible. 

Let us now examine the four terms in (\ref{eq:Y.dcmp}).

\textbf{Step 1:} $T_{1n}$. The term $T_{1n}$ is a normalized sum
of link indicators that are evaluated at $\omega_{ni}^{\ast}$ rather
than $\omega_{ni}(\epsilon_{i})$ and thus are independent. This is
a leading term in $Y_{n}$ and has an asymptotically normal distribution
because the CLT applies. It captures the sampling variation in link
choices due to $\epsilon_{ij}$.

\textbf{Step 2:} $T_{2n}$. We show that $T_{2n}$ is $o_{p}(1)$.
For each $i$, define the empirical process $\mathbb{G}_{n}q_{ni}g_{ni}(\omega,\epsilon_{i})=\frac{1}{\sqrt{n-1}}\sum_{j\neq i}q_{n,ij}(g_{n,ij}(\omega,\epsilon_{ij})-P_{n,ij}^{\ast}(\omega))$,
$\omega\in\Omega$. Then $T_{2n}$ is a normalized average of these
empirical processes for all $i$, 
\[
T_{2n}=\frac{1}{\sqrt{n}}\sum_{i}\mathbb{G}_{n}q_{ni}(g_{ni}(\omega_{ni}(\epsilon_{i}),\epsilon_{i})-g_{ni}(\omega_{ni}^{\ast},\epsilon_{i})).
\]
While each empirical process in $T_{2n}$ is $o_{p}(1)$ by establishing
stochastic equicontinuity, we cannot directly invoke a stochastic
equicontinuity argument to show that their normalized average $T_{2n}$
is $o_{p}(1)$. Instead, we use an maximal inequality to derive a
uniform bound on the $L_{2}$ norm of each empirical process.

Note that each $\mathbb{G}_{n}q_{ni}(g_{ni}(\omega_{ni}(\epsilon_{i}),\epsilon_{i})-g_{ni}(\omega_{ni}^{\ast},\epsilon_{i}))$
only involves $\epsilon_{i}$, so given $X$ they are independent
across $i$. Moreover, by Lemma \ref{lem:w.asymlin} we have $\Lambda_{ni}(\omega_{ni}(\epsilon_{i})-\omega_{ni}^{\ast})=O_{p}(n^{-1/2})$,
so if we define $h_{ni}=n^{\kappa}\Lambda_{ni}(\omega_{ni}(\epsilon_{i})-\omega_{ni}^{\ast})$
for $0<\kappa<1/2$, then $h_{ni}\in\Omega$ if $n$ is sufficiently
large, because by Assumption \ref{ass:w}(i) $\Omega$ contains a
compact neighborhood of $0$. Further, view $g_{n,ij}(\omega,\epsilon_{ij})$
as a function of $\omega^{\lambda}=\Lambda_{ni}\omega\in\Omega$,
and define $g_{n,ij}^{\lambda}(\omega^{\lambda},\epsilon_{ij})\equiv1\{U_{n,ij}+\frac{n-1}{n-2}Z_{j}^{\prime}\Phi_{ni}\omega^{\lambda}\geq\epsilon_{ij}\}$.
Then $g_{ni}(\omega_{ni}(\epsilon_{i}),\epsilon_{i})-g_{ni}(\omega_{ni}^{\ast},\epsilon_{i})=g_{ni}^{\lambda}(\omega_{ni}^{\lambda}(\epsilon_{i}),\epsilon_{i})-g_{ni}^{\lambda}(\omega_{ni}^{\ast\lambda},\epsilon_{i})$,
where $\omega_{ni}^{\lambda}(\epsilon_{i})=\Lambda_{ni}\omega_{ni}(\epsilon_{i})$
and $\omega_{ni}^{\ast\lambda}=\Lambda_{ni}\omega_{ni}^{\ast}$. Combining
these results we obtain the bound
\begin{equation}
\mathbb{E}[\Vert T_{2n}\Vert^{2}|X]\leq n^{-1}\sum_{i}\mathbb{E}\left[\left.\sup_{\omega^{\lambda},h\in\Omega}\Vert\mathbb{G}_{n}q_{ni}(g_{ni}^{\lambda}(\omega^{\lambda}+n^{-\kappa}h,\epsilon_{i})-g_{ni}^{\lambda}(\omega^{\lambda},\epsilon_{i}))\Vert^{2}\right|X\right].\label{eq:T2.Ebdd}
\end{equation}

Below we derive a bound on each term in the summation in (\ref{eq:T2.Ebdd})
that is uniform in $i$. Observe that for any $\omega^{\lambda},\tilde{\omega}^{\lambda}\in\Omega$,
the function $q_{n,ij}(g_{n,ij}^{\lambda}(\omega^{\lambda},\epsilon_{ij})-g_{n,ij}^{\lambda}(\tilde{\omega}^{\lambda},\epsilon_{ij}))$
can be bounded by $\Vert q_{n,ij}\Vert|1\{U_{n,ij}+\frac{n-1}{n-2}Z_{j}^{\prime}\Phi_{ni}\omega^{\lambda}\geq\epsilon_{ij}\}-1\{U_{n,ij}+\frac{n-1}{n-2}Z_{j}^{\prime}\Phi_{ni}\tilde{\omega}^{\lambda}\geq\epsilon_{ij}\}|\leq\eta_{n,ij}(\omega^{\lambda},\tilde{\omega}^{\lambda},\epsilon_{ij})$,
where $\eta_{n,ij}(\omega^{\lambda},\tilde{\omega}^{\lambda},\epsilon_{ij})=\Vert q_{n,ij}\Vert$
if $\epsilon_{ij}$ lies between $U_{n,ij}+\frac{n-1}{n-2}Z_{j}^{\prime}\Phi_{ni}\omega^{\lambda}$
and $U_{n,ij}+\frac{n-1}{n-2}Z_{j}^{\prime}\Phi_{ni}\tilde{\omega}^{\lambda}$,
and $0$ otherwise. Similarly to the proof of equation (\ref{eq:w.maxineq}),
we derive for each $i$ that
\begin{eqnarray}
 &  & \mathbb{E}\left[\left.\sup_{\omega^{\lambda},h\in\Omega}\Vert\mathbb{G}_{n}q_{ni}(g_{ni}^{\lambda}(\omega^{\lambda}+n^{-\kappa}h,\epsilon_{i})-g_{ni}^{\lambda}(\omega^{\lambda},\epsilon_{i}))\Vert^{2}\right|X\right]\nonumber \\
 & \leq & K\mathbb{E}\left[\left.J(1,\mathcal{F}_{ni}(\epsilon_{i}))^{2}\sup_{\omega^{\lambda},,h\in\Omega}\Vert\eta_{ni}(\omega^{\lambda}+n^{-\kappa}h,\omega^{\lambda},\epsilon_{i})\Vert_{n}^{2}\right|X\right]\label{eq:T2.maxineq}
\end{eqnarray}
for constant $K<\infty$, where $\Vert\eta_{ni}\Vert_{n}=(\frac{1}{n-1}\sum_{j\neq i}\eta_{n,ij}^{2})^{1/2}$,
and $J(1,\mathcal{F}_{ni}(\epsilon_{i}))$ is the uniform entropy
integral defined as in (\ref{eq:w.entrp}) for the set of triangular
arrays $\mathcal{F}_{ni}(\epsilon_{i})=\{(q_{n,ij}(g_{n,ij}^{\lambda}(\omega^{\lambda}+n^{-\kappa}h,\epsilon_{ij})-g_{n,ij}^{\lambda}(\omega^{\lambda},\epsilon_{ij})),j\neq i):\omega^{\lambda},h\in\Omega\}$.
Similarly as in Lemma \ref{lem:w.emp}, we can show that $J(1,\mathcal{F}_{ni}(\epsilon_{i}))\leq\bar{J}$
for $\bar{J}<\infty$.

We follow the proof of Lemma \ref{lem:w.emp} to bound the sup term
in (\ref{eq:T2.maxineq}). Recall that $\|\eta_{n,ij}\|^{2}\leq\max_{1\leq i,j\leq n}\Vert q_{n,ij}\Vert^{2}\leq\bar{q}^{2}<\infty$.
Similarly as in Lemma \ref{lem:w.emp}, we can show that the set of
triangular arrays $\{(\eta_{n,ij}^{2}(\omega^{\lambda}+n^{-\kappa}h,\omega^{\lambda},\epsilon_{ij}),j\neq i):\omega^{\lambda},h\in\Omega\}$
has a finite uniform entropy integral bounded by $\bar{J}^{\eta}<\infty$.
Analogous to the proof of equation (\ref{eq:w.maxineq}), we can derive
$\mathbb{E}\left[\sup_{\omega^{\lambda},h\in\Omega}|\mathbb{G}_{n}\eta_{ni}^{2}(\omega^{\lambda}+n^{-\kappa}h,\omega^{\lambda},\epsilon_{ij})||X\right]\leq K^{\eta}\bar{J}^{\eta}\bar{q}^{2}$
with constant $K^{\eta}<\infty$. Hence,
\begin{eqnarray*}
 &  & \mathbb{E}\left[\left.\sup_{\omega^{\lambda},h\in\Omega}\frac{1}{n-1}\sum_{j\neq i}\eta_{n,ij}^{2}(\omega^{\lambda}+n^{-\kappa}h,\omega^{\lambda},\epsilon_{ij})\right|X\right]\\
 &  & -\sup_{\omega^{\lambda},h\in\Omega}\frac{1}{n-1}\sum_{j\neq i}\mathbb{E}[\eta_{n,ij}^{2}(\omega^{\lambda}+n^{-\kappa}h,\omega^{\lambda},\epsilon_{ij})|X]\\
 & \leq & \frac{1}{\sqrt{n-1}}\mathbb{E}\left[\left.\sup_{\omega^{\lambda},h\in\Omega}|\mathbb{G}_{n}\eta_{ni}^{2}(\omega^{\lambda}+n^{-\kappa}h,\omega^{\lambda},\epsilon_{i})|\right|X\right]\leq\frac{K^{\eta}\bar{J}^{\eta}\bar{q}^{2}}{\sqrt{n-1}}.
\end{eqnarray*}
Moreover, by the mean-value theorem, for any $\omega^{\lambda},h\in\Omega$
and any $j\neq i$, we have 
\begin{eqnarray*}
 &  & \mathbb{E}[\eta_{n,ij}^{2}(\omega^{\lambda}+n^{-\kappa}h,\omega^{\lambda},\epsilon_{ij})|X]\\
 & = & \left|F_{\epsilon}(U_{n,ij}+\frac{n-1}{n-2}Z_{j}^{\prime}\Phi_{ni}(\omega^{\lambda}+n^{-\kappa}h))-F_{\epsilon}(U_{n,ij}+\frac{n-1}{n-2}Z_{j}^{\prime}\Phi_{ni}\omega^{\lambda})\right|\Vert q_{n,ij}\Vert^{2}\\
 & = & n^{-\kappa}f_{\epsilon}(U_{n,ij}+\frac{n-1}{n-2}Z_{j}^{\prime}\Phi_{ni}(\omega^{\lambda}+t_{n,ij}n^{-\kappa}h))\frac{n-1}{n-2}|Z_{j}^{\prime}\Phi_{ni}h|\Vert q_{n,ij}\Vert^{2}
\end{eqnarray*}
for some $t_{n,ij}\in[0,1]$. By the boundedness of $f_{\epsilon}$
under Assumption \ref{ass:e=000026x}(ii), $\sup_{h\in\Omega}\Vert h\Vert<\infty$,
and $\Vert q_{n,ij}\Vert^{2}\leq\bar{q}^{2}<\infty$, there is a $M<\infty$
such that $\mathbb{E}[\eta_{n,ij}^{2}(\omega^{\lambda}+n^{-\kappa}h,\omega^{\lambda},\epsilon_{ij})|X]\leq n^{-\kappa}M$
for all $\omega^{\lambda},h\in\Omega$ and all $i,j$. Hence, $\sup_{\omega^{\lambda},h\in\Omega}\frac{1}{n-1}\sum_{j\neq i}\mathbb{E}[\eta_{n,ij}^{2}(\omega^{\lambda}+n^{-\kappa}h,\omega^{\lambda},\epsilon_{ij})|X]\leq n^{-\kappa}M$.
From these results we derive $\mathbb{E}[\sup_{\omega^{\lambda},h\in\Omega}\frac{1}{n-1}\sum_{j\neq i}\eta_{n,ij}^{2}(\omega^{\lambda}+n^{-\kappa}h,\omega^{\lambda},\epsilon_{ij})|X]\leq(n-1)^{-1/2}K^{\eta}\bar{J}^{\eta}\bar{q}^{2}+n^{-\kappa}M$
for each $i$. Note that the bound is constant across $i$. Combining
the results we can bound $\mathbb{E}[\Vert T_{2n}\Vert^{2}|X]$ by
$K\bar{J}^{2}((n-1)^{-1/2}K^{\eta}\bar{J}^{\eta}\bar{q}^{2}+n^{-\kappa}M)=o(1)$
and hence $T_{2n}=o_{p}(1)$ by Markov's inequality.

\textbf{Step 3:} $T_{3n}$. View $P_{n,ij}^{\ast}(\omega)$ as a function
of $\Lambda_{ni}\omega$. By Taylor expansion, we have
\[
P_{n,ij}^{\ast}(\omega_{ni}(\epsilon_{i}))=P_{n,ij}^{\ast}(\omega_{ni}^{\ast})+\nabla_{\omega^{\lambda\prime}}P_{n,ij}^{\ast}(\omega_{ni}^{\ast})\Lambda_{ni}(\omega_{ni}(\epsilon_{i})-\omega_{ni}^{\ast})+O_{p}(\Vert\Lambda_{ni}(\omega_{ni}(\epsilon_{i})-\omega_{ni}^{\ast})\Vert^{2}),
\]
where $\nabla_{\omega^{\lambda\prime}}P_{n,ij}^{\ast}(\omega_{ni}^{\ast})=\frac{n-1}{n-2}f_{\epsilon}(U_{n,ij}+\frac{n-1}{n-2}Z_{j}^{\prime}\Phi_{ni}\Lambda_{ni}\omega_{ni}^{\ast})Z_{j}^{\prime}\Phi_{ni}$
is the derivative of $P_{n,ij}^{\ast}(\omega)$ with respect to $\Lambda_{ni}\omega$
at $\Lambda_{ni}\omega_{ni}^{\ast}$. Lemma \ref{lem:w.asymlin} shows
the asymptotically linear representation $\Lambda_{ni}(\omega_{ni}(\epsilon_{i})-\omega_{ni}^{\ast})=\frac{1}{n-1}\sum_{j\neq i}\phi_{n,ij}^{\omega}(\omega_{ni}^{\ast},\epsilon_{ij})+r_{ni}^{\omega}(\epsilon_{i})$.
Let $\bar{q}_{ni}=\frac{1}{n-1}\sum_{j\neq i}q_{n,ij}$ be the $d_{\theta}\times1$
vector of instruments averaged over $j$. By the asymptotically linear
representation, we can decompose $T_{3n}$ into three parts $T_{3n}=T_{3n}^{l}+(r_{1n}-\mathbb{E}[r_{1n}|X])+(r_{2n}-\mathbb{E}[r_{2n}|X])$,
where
\begin{eqnarray*}
T_{3n}^{l} & = & \frac{1}{\sqrt{n(n-1)}}\sum_{i}\sum_{j\neq i}J_{ni}^{\omega}(\omega_{ni}^{\ast},q_{ni})\phi_{n,ij}^{\omega}(\omega_{ni}^{\ast},\epsilon_{ij})\\
r_{1n} & = & \sqrt{\frac{n-1}{n}}\sum_{i}J_{ni}^{\omega}(\omega_{ni}^{\ast},q_{ni})r_{ni}^{\omega}(\epsilon_{i})\\
r_{2n} & = & \sqrt{\frac{n-1}{n}}\sum_{i}\bar{q}_{ni}O_{p}(\Vert\Lambda_{ni}(\omega_{ni}(\epsilon_{i})-\omega_{ni}^{\ast})\Vert^{2}).
\end{eqnarray*}
$T_{3n}^{l}$ is a leading term that contributes to the asymptotic
distribution of $Y_{n}$. It captures the sampling variation due to
$\omega_{ni}(\epsilon_{i})$. We will combine it with $T_{1n}$ to
derive the asymptotic distribution of $Y_{n}$. Below we show that
the two centered remainders $r_{1n}-\mathbb{E}[r_{1n}|X]$ and $r_{2n}-\mathbb{E}[r_{2n}|X]$
are both $o_{p}(1)$.

Given $X$, each $r_{ni}^{\omega}(\epsilon_{i})$ only depends on
$\epsilon_{i}$, so they are independent across $i$ (Assumption \ref{ass:e=000026x}(i)).
Hence, 
\begin{eqnarray*}
 &  & \mathbb{E}[\Vert r_{1n}-\mathbb{E}[r_{1n}|X]\Vert^{2}|X]\\
 & = & \frac{n-1}{n}\sum_{i}\mathbb{E}[\Vert J_{ni}^{\omega}(\omega_{ni}^{\ast},q_{ni})(r_{ni}^{\omega}(\epsilon_{i})-\mathbb{E}[r_{ni}^{\omega}(\epsilon_{i})|X])\Vert^{2}|X]\\
 & \leq & (n-1)\max_{i}\Vert J_{ni}^{\omega}(\omega_{ni}^{\ast},q_{ni})\Vert^{2}\max_{i}\mathbb{E}[\Vert r_{ni}^{\omega}(\epsilon_{i})-\mathbb{E}[r_{ni}^{\omega}(\epsilon_{i})|X]\Vert^{2}|X].
\end{eqnarray*}
For any random variable $Z$, recall that $\mathbb{E}[|Z||X]\leq\Vert Z\Vert_{\psi_{1}|X}$
and $\mathbb{E}[Z^{2}|X]\leq4\Vert Z\Vert_{\psi_{1}|X}^{2}$. From
$\Vert\max_{i}\Vert r_{ni}^{\omega}(\epsilon_{i})\Vert\Vert_{\psi_{1}|X}=o(n^{-1/2})$
(Lemma \ref{lem:w.rem}(ii)) we thus derive that $\mathbb{E}[\max_{i}\Vert r_{ni}^{\omega}(\epsilon_{i})\Vert|X]=o(n^{-1/2})$
and $\mathbb{E}[\max_{i}\Vert r_{ni}^{\omega}(\epsilon_{i})\Vert^{2}|X]=o(n^{-1})$.
Therefore, $\max_{i}\mathbb{E}[\Vert r_{ni}^{\omega}(\epsilon_{i})-\mathbb{E}[r_{ni}^{\omega}(\epsilon_{i})|X]\Vert^{2}|X]\leq\max_{i}\mathbb{E}[(\Vert r_{ni}^{\omega}(\epsilon_{i})\Vert+\Vert\mathbb{E}[r_{ni}^{\omega}(\epsilon_{i})|X]\Vert)^{2}|X]\leq\mathbb{E}[\max_{i}\Vert r_{ni}^{\omega}(\epsilon_{i})\Vert^{2}|X]+3(\mathbb{E}[\max_{i}\Vert r_{ni}^{\omega}(\epsilon_{i})\Vert|X])^{2}=o(n^{-1})$.
Because $J_{ni}^{\omega}(\omega_{ni}^{\ast},q_{ni})$ is bounded uniformly
in $i$, we obtain $\mathbb{E}[\Vert r_{1n}-\mathbb{E}[r_{1n}|X]\Vert^{2}|X]=o(1)$
and thus $r_{1n}-\mathbb{E}[r_{1n}|X]=o_{p}(1)$ by Markov's inequality. 

Similarly, with $O(\Vert\Lambda_{ni}(\omega_{ni}(\epsilon_{i})-\omega_{ni}^{\ast})\Vert^{2})$
in place of $r_{ni}^{\omega}(\epsilon_{i})$ and $\bar{q}_{ni}$ in
place of $J_{ni}^{\omega}(\omega_{ni}^{\ast},q_{ni})$ and by $\Vert\max_{i}\Vert\Lambda_{ni}(\omega_{ni}(\epsilon_{i})-\omega_{ni}^{\ast})\Vert^{2}\Vert_{\psi_{1}|X}=o(n^{-1/2})$
(Lemma \ref{lem:w.rem}(i)), we can show that $r_{2n}-\mathbb{E}[r_{2n}|X]=o_{p}(1)$.

\textbf{Step 4:} $T_{4n}$. Recall that $T_{4n}=-\mathbb{E}[T_{2n}|X]$.
In Step 2 we showed that $\mathbb{E}[\Vert T_{2n}\mathbb{\Vert}^{2}|X]=o(1)$.
Because $\Vert\mathbb{E}[T_{2n}|X]\Vert\leq\mathbb{E}[\Vert T_{2n}\Vert|X]\leq(\mathbb{E}[\Vert T_{2n}\mathbb{\Vert}^{2}|X])^{1/2}$,
we have $T_{4n}=o(1)$.

Combing the four steps, we derive that $Y_{ni}=T_{1n}+T_{3n}^{l}+o_{p}(1)=\sum_{i}\phi_{ni}^{y}+o_{p}(1)$,
where for each $i$, $\phi_{ni}^{y}=\frac{1}{\sqrt{n(n-1)}}\sum_{j\neq i}\phi_{n,ij}^{y}$.
Given $X$, $\phi_{ni}^{y}$, $i=1,\ldots,n$, are independent but
not identically distributed. We apply the Lindeberg-Feller CLT to
derive the asymptotic distribution of $\sum_{i}\phi_{ni}^{y}$. Note
that $\mathbb{E}[\phi_{ni}^{y}|X]=0$ for all $i$. By the Cramer-Wold
device it suffices to show that $a^{\prime}\sum_{i}\phi_{ni}^{y}$
satisfies the Lindeberg condition for any $d_{\theta}\times1$ vector
of constants $a\in\mathbb{R}^{d_{\theta}}$. The Lindeberg condition
is that for any $\xi>0$ 
\begin{equation}
\lim_{n\rightarrow\infty}\frac{1}{a^{\prime}\Sigma_{n}a}\sum_{i}\mathbb{E}[(a^{\prime}\phi_{ni}^{y})^{2}1\{|a^{\prime}\phi_{ni}^{y}|\geq\xi\sqrt{a^{\prime}\Sigma_{n}a}\}|X]=0,\label{eq:lindeberg}
\end{equation}
where $\Sigma_{n}=\sum_{i}\mathbb{E}[\phi_{ni}^{y}(\phi_{ni}^{y})^{\prime}|X]=\frac{1}{n(n-1)}\sum_{i}\sum_{j\neq i}\mathbb{E}[\phi_{n,ij}^{y}(\phi_{n,ij}^{y})'|X]$.
Following the argument in the proof of Lemma \ref{lem:w.asymlin},
the Lindeberg condition holds if 
\begin{equation}
\frac{\max_{i}|a^{\prime}\phi_{ni}^{y}|}{\sqrt{a^{\prime}\Sigma_{n}a}}=o_{p}(1).\label{eq:max_aY}
\end{equation}
By Markov's inequality, equation (\ref{eq:max_aY}) holds if $\mathbb{E}[\max_{i}(a^{\prime}\phi_{ni}^{y})^{2}|X]=o(1)$.
By Lemma 2.2.2 in \citet{VW_1996}, $\mathbb{E}[\max_{i}(a^{\prime}\phi_{ni}^{y})^{2}|X]\leq\Vert\max_{i}(a^{\prime}\phi_{ni}^{y})^{2}\Vert_{\psi_{1}|X}\leq K\ln(n+1)\max_{i}\Vert(a^{\prime}\phi_{ni}^{y})^{2}\Vert_{\psi_{1}|X}$
with constant $K<\infty$. Moreover, note that $a^{\prime}\phi_{ni}^{y}=(n(n-1))^{-1/2}\sum_{j\neq i}a^{\prime}\phi_{n,ij}^{y}$,
and $|a^{\prime}\phi_{n,ij}^{y}|\leq\Vert a\Vert(2\Vert q_{n,ij}\Vert+\Vert J_{ni}^{\omega}(\omega_{ni}^{\ast},q_{ni})\Vert\Vert\phi_{n,ij}^{\omega}(\omega_{ni}^{\ast},\epsilon_{ij})\Vert)\leq M_{n}<\infty$.
By Hoeffding's inequality for bounded random variables \citep[Theorem 2.8]{BLM_2013},
we derive that $\Pr((a^{\prime}\phi_{ni}^{y})^{2}\geq t|X)=\Pr(a^{\prime}\phi_{ni}^{y}\geq\sqrt{t}|X)+\Pr(-a^{\prime}\phi_{ni}^{y}\geq\sqrt{t}|X)\leq2\exp(-\frac{nt}{2M_{n}^{2}})$.
It follows from Lemma 2.2.1 in \citet{VW_1996} that $\Vert(a^{\prime}\phi_{ni}^{y})^{2}\Vert_{\psi_{1}|X}\leq\frac{6}{n}M_{n}^{2}$.
Each $\mathbb{E}[\phi_{n,ij}^{y}(\phi_{n,ij}^{y})'|X]$ is positive
definite, so $a^{\prime}\Sigma_{n}a>0$. Combining these results yields
$\mathbb{E}[\max_{i}(a^{\prime}\phi_{ni}^{y})^{2}|X]$ by $6KM_{n}^{2}\ln(n+1)/n=o(1)$,
so the Lindeberg condition holds.

By the Lindeberg-Feller CLT, $(a^{\prime}\Sigma_{n}a)^{-1/2}a^{\prime}\sum_{i}\phi_{ni}^{y}\overset{d}{\rightarrow}N(0,1)$.
Because $\Sigma_{n}$ is positive definite, there is a nonsingular
symmetric matrix $\Sigma_{n}^{1/2}$ such that $\Sigma_{n}^{1/2}\Sigma_{n}^{1/2}=\Sigma_{n}$.
Let $\tilde{a}=\Sigma_{n}^{1/2}a$, then $a^{\prime}\sum_{i}\phi_{ni}^{y}=\tilde{a}^{\prime}\Sigma_{n}^{-1/2}\sum_{i}\phi_{ni}^{y}$
and $a^{\prime}\Sigma_{n}a=\tilde{a}^{\prime}\Sigma_{n}^{-1/2}\Sigma_{n}\Sigma_{n}^{-1/2}\tilde{a}=\tilde{a}^{\prime}\tilde{a}$.
Note that $\Sigma_{n}$ is nonsingular, so $\tilde{a}$ is also an
arbitrary vector in $\mathbb{R}^{d_{\theta}}$. The previous result
implies that $\tilde{a}^{\prime}\Sigma_{n}^{-1/2}\sum_{i}\phi_{ni}^{y}\overset{d}{\rightarrow}N(0,\tilde{a}^{\prime}\tilde{a})$.
By the Cramer-Wold device, $\Sigma_{n}^{-1/2}\sum_{i}\phi_{ni}^{y}\overset{d}{\rightarrow}N(0,I_{d_{\theta}})$.
Because $Y_{n}=\sum_{i}\phi_{ni}^{y}+o_{p}(1)$, we conclude that
by Slutsky's theorem $Y_{n}$ has the asymptotic distribution $\Sigma_{n}^{-1/2}Y_{n}\overset{d}{\rightarrow}N(0,I_{d_{\theta}})$.
\end{proof}

\subsection{Lemmas for Section \ref{sec:extension}}
\begin{lem}[Consistency of $\omega_{ni}^{\phi\lambda}(\epsilon_{i})$ for $\omega_{i}^{\phi\lambda*}$]
\label{lem:wlim.consist}Suppose that Assumptions \ref{ass:e=000026x}--\ref{ass:discX}
and \ref{ass:lim} are satisfied. Conditional on $X_{i}$, we have
$\omega_{ni}^{\phi\lambda}(\epsilon_{i})-\omega_{i}^{\phi\lambda*}=o_{p}(1)$.
\end{lem}
\begin{proof}
The finite-$n$ first-order condition in (\ref{eq:foc.maxmin}) in
Lemma \ref{lem:foc.w} (multiplied by $\Phi_{ni}$) implies that $\omega_{ni}^{\phi\lambda}(\epsilon_{i})=\Phi_{ni}\Lambda_{ni}\omega_{ni}(\epsilon_{i})$
solves the equation
\begin{equation}
\Gamma_{ni}(\omega^{\phi\lambda};\epsilon_{i},X)\equiv V_{ni}\frac{1}{n-1}\sum_{j\neq i}1\left\{ U_{n,ij}+\frac{n-1}{n-2}Z_{j}^{\prime}\omega^{\phi\lambda}\geq\epsilon_{ij}\right\} Z_{j}-\omega^{\phi\lambda}=0,\text{ a.s.}\label{eq:foc.PLomega}
\end{equation}
 Moreover, the limiting first-order condition in (\ref{eq:foc.maxmin.lim})
(multiplied by $\Phi_{i}^{*}$) implies that $\omega_{i}^{\phi\lambda*}=\Phi_{i}^{*}\Lambda_{i}^{*}\omega_{i}^{*}$
solves the equation
\begin{equation}
\Gamma_{i}^{*}(\omega^{\phi\lambda};X_{i})\equiv V_{i}^{*}\mathbb{E}[F_{\epsilon}(U_{ij}^{*}+Z_{j}^{\prime}\omega^{\phi\lambda})Z_{j}|X_{i}]-\omega^{\phi\lambda}=0.\label{eq:foc.PLomega.lim}
\end{equation}
Note that $\|\Gamma_{i}^{*}(\omega_{ni}^{\phi\lambda}(\epsilon_{i});X_{i})\|\leq\|\Gamma_{ni}(\omega_{ni}^{\phi\lambda}(\epsilon_{i});\epsilon_{i},X)-\Gamma_{i}^{*}(\omega_{ni}^{\phi\lambda}(\epsilon_{i});X_{i})\|+\|\Gamma_{ni}(\omega_{ni}^{\phi\lambda}(\epsilon_{i});\epsilon_{i},X)\|\leq\sup_{\omega^{\phi\lambda}\in\Omega}\|\Gamma_{ni}(\omega^{\phi\lambda};\epsilon_{i},X)-\Gamma_{i}^{*}(\omega^{\phi\lambda};X_{i})\|$
almost surely. Because the function $\Gamma_{i}^{*}(\omega^{\phi\lambda};X_{i})$
is continuous in $\omega^{\phi\lambda}$ (Assumption \ref{ass:e=000026x}(ii))
on a compact set $\Omega$, by the uniqueness of $\omega_{i}^{\phi\lambda*}$
(Assumption \ref{ass:lim}(ii)), for any $\delta>0$, there exists
$\kappa(\delta)\equiv\inf_{\omega^{\phi\lambda}\in\Omega:\|\omega^{\phi\lambda}-\omega_{i}^{\phi\lambda*}\|\geq\delta}\|\Gamma_{i}^{*}(\omega^{\phi\lambda};X_{i})\|>0$
such that $\|\Gamma_{i}^{*}(\omega_{ni}^{\phi\lambda}(\epsilon_{i});X_{i})\|<\kappa(\delta)\Rightarrow\|\omega_{ni}^{\phi\lambda}(\epsilon_{i})-\omega_{i}^{\phi\lambda*}\|<\delta$.
Hence, $\Pr(\|\omega_{ni}^{\phi\lambda}(\epsilon_{i})-\omega_{i}^{\phi\lambda*}\|<\delta)|X_{i})\geq\Pr(\sup_{\omega^{\phi\lambda}\in\Omega}\|\Gamma_{ni}(\omega^{\phi\lambda};\epsilon_{i},X)-\Gamma_{i}^{*}(\omega^{\phi\lambda};X_{i})\|<\kappa(\delta)|X_{i})$.
It suffices to show that $\sup_{\omega^{\phi\lambda}\in\Omega}\|\Gamma_{ni}(\omega^{\phi\lambda};\epsilon_{i},X)-\Gamma_{i}^{*}(\omega^{\phi\lambda};X_{i})\|=o_{p}(1)$.
By the triangle inequality
\begin{eqnarray}
 &  & \sup_{\omega^{\phi\lambda}\in\Omega}\|\Gamma_{ni}(\omega^{\phi\lambda};\epsilon_{i},X)-\Gamma_{i}^{*}(\omega^{\phi\lambda};X_{i})\|\nonumber \\
 & \leq & \|V_{ni}\|\sup_{\omega^{\phi\lambda}\in\Omega}\left\Vert \frac{1}{n-1}\sum_{j\neq i}\left(1\left\{ U_{n,ij}+\frac{n-1}{n-2}Z_{j}^{\prime}\omega^{\phi\lambda}\geq\epsilon_{ij}\right\} -1\{U_{ij}^{*}+Z_{j}^{\prime}\omega^{\phi\lambda}\geq\epsilon_{ij}\}\right)Z_{j}\right\Vert \nonumber \\
 &  & +\|V_{ni}\|\sup_{\omega^{\phi\lambda}\in\Omega}\left\Vert \frac{1}{n-1}\sum_{j\neq i}(1\{U_{ij}^{*}+Z_{j}^{\prime}\omega^{\phi\lambda}\geq\epsilon_{ij}\}Z_{j}-\mathbb{E}[F_{\epsilon}(U_{ij}^{*}+Z_{j}^{\prime}\omega^{\phi\lambda})Z_{j}|X_{i}])\right\Vert \nonumber \\
 &  & +\|V_{ni}-V_{i}^{*}\|\sup_{\omega^{\phi\lambda}\in\Omega}\mathbb{E}[F_{\epsilon}(U_{ij}^{*}+Z_{j}^{\prime}\omega^{\phi\lambda})Z_{j}|X_{i}]\label{eq:foc.diff}
\end{eqnarray}
For any $\xi_{n}>0$, we can bound $(1\{U_{n,ij}+\frac{n-1}{n-2}Z_{j}^{\prime}\omega^{\phi\lambda}\geq\epsilon_{ij}\}-1\{U_{ij}^{*}+Z_{j}^{\prime}\omega^{\phi\lambda}\geq\epsilon_{ij}\})Z_{j}$
by $1\{\max_{j\neq i}|U_{n,ij}-U_{ij}^{*}|+\frac{1}{n-2}\sup_{\omega^{\phi\lambda}\in\Omega}\|\omega^{\phi\lambda}\|>\xi_{n}\}+1\{|U_{ij}^{*}+Z_{j}^{\prime}\omega^{\phi\lambda}-\epsilon_{ij}|\leq\xi_{n}\}$.
It follows from $\max_{j\neq i}|U_{n,ij}-U_{ij}^{*}|=o_{p}(1)$ and
$\sup_{\omega^{\phi\lambda}\in\Omega}\|\omega^{\phi\lambda}\|<\infty$
that $1\{\max_{j\neq i}|U_{n,ij}-U_{ij}^{*}|+\frac{1}{n-2}\sup_{\omega^{\phi\lambda}\in\Omega}\|\omega^{\phi\lambda}\|>\xi_{n}\}=o_{p}(1)$.
Moreover, because $1\{|U_{ij}^{*}+Z_{j}^{\prime}\omega^{\phi\lambda}-\epsilon_{ij}|\leq\xi_{n}\}$
is continuous in $\omega^{\phi\lambda}$ with probability one, by
the uniform LLN in \citet[Lemma 2.4]{Newey_McFadden_1994} we have
$\sup_{\omega^{\phi\lambda}\in\Omega}|\frac{1}{n-1}\sum_{j\neq i}1\{|U_{ij}^{*}+Z_{j}^{\prime}\omega^{\phi\lambda}-\epsilon_{ij}|\leq\xi_{n}\}-\mathbb{E}[1\{|U_{ij}^{*}+Z_{j}^{\prime}\omega^{\phi\lambda}-\epsilon_{ij}|\leq\xi_{n}\}|X_{i}]|=o_{p}(1)$.
By the mean-value theorem, $\mathbb{E}[1\{|U_{ij}^{*}+Z_{j}^{\prime}\omega^{\phi\lambda}-\epsilon_{ij}|\leq\xi_{n}\}|X_{i}]=\mathbb{E}[F_{\epsilon}(U_{ij}^{*}+Z_{j}^{\prime}\omega^{\phi\lambda}+\xi_{n})-F_{\epsilon}(U_{ij}^{*}+Z_{j}^{\prime}\omega^{\phi\lambda}-\xi_{n})|X_{i}]=2\mathbb{E}[f_{\epsilon}(U_{ij}^{*}+Z_{j}^{\prime}\omega^{\phi\lambda}+t_{ij}\xi_{n})|X_{i}]\xi_{n}\leq C\xi_{n}$
for some $-1\leq t_{ij}\leq1$. By choosing $\xi_{n}\downarrow0$
as $n\rightarrow\infty$ we can make $\mathbb{E}[1\{|U_{ij}^{*}+Z_{j}^{\prime}\omega^{\phi\lambda}-\epsilon_{ij}|\leq\xi_{n}\}|X_{i}]$
arbitrarily small. Combining these results shows that the first term
on the right-hand side of (\ref{eq:foc.diff}) is $o_{p}(1)$. Applying
the uniform LLN in \citet[Lemma 2.4]{Newey_McFadden_1994} again we
can show that the second term on the right-hand side of (\ref{eq:foc.diff})
is $o_{p}(1)$. The last term in (\ref{eq:foc.diff}) is $o_{p}(1)$
because $\|V_{ni}-V_{i}^{*}\|=o_{p}(1)$. The lemma is proved.

\end{proof}

\begin{lem}[Example \ref{ex:U=000026V.consist}]
\label{lem:U=000026V.consist}Under Assumption \ref{ass:lim}(iii),
Assumption \ref{ass:lim}(iv) is satisfied for the specification in
(\ref{eq:Eu})--(\ref{eq:Ev}) and $U^{*}(X_{i},X_{j},p)$ and $V^{*}(X_{i},p)$
defined in Example \ref{ex:U=000026V.consist}.
\end{lem}
\begin{proof}
By definition
\begin{eqnarray*}
U_{n,ij}(X,p)-U^{*}(X_{i},X_{j},p) & = & \frac{1}{n-2}\sum_{k\neq i,j}(p(X_{k},X_{j})-\mathbb{E}[p(X_{k},X_{j})|X_{j}])\beta_{5}\\
 &  & +\frac{1}{n-2}\sum_{k\neq i,j}(p(X_{j},X_{k})-\mathbb{E}[p(X_{j},X_{k})|X_{j}])\beta_{6}\\
 &  & -\frac{1}{2(n-2)}Z_{j}^{\prime}V_{ni}(X,p)Z_{j}.
\end{eqnarray*}
Conditional on$X_{j}$, both $p(X_{k},X_{j})$ and $p(X_{j},X_{k})$
are i.i.d. across $k$ (Assumption \ref{ass:lim}(iii)). Hence, by
the LLN the first two terms on the right-hand side are $o_{p}(1)$.
The last term on the right-hand side is $o_{p}(1)$ by the boundedness
of $V_{ni}$. Because $X_{j}$ takes $T$ values, we obtain $\max_{j\neq i}|U_{n,ij}(X,p)-U^{*}(X_{i},X_{j},p)|=o_{p}(1)$.
Moreover, for any $1\leq s,t\leq T$,
\begin{eqnarray*}
V_{ni,st}(X,p)-V_{st}^{*}(p) & = & \frac{1}{n-3}\sum_{l\neq i,j,k}(p(x_{s},X_{l})p(X_{l},x_{t})+p(x_{t},X_{l})p(X_{l},x_{s})\\
 &  & -\mathbb{E}[p(x_{s},X_{l})p(X_{l},x_{t})+p(x_{t},X_{l})p(X_{l},x_{s})])\gamma_{2}(x_{s},x_{t})
\end{eqnarray*}
By Assumption \ref{ass:lim}(iii), $p(x_{s},X_{l})p(X_{l},x_{t})+p(x_{t},X_{l})p(X_{l},x_{s})$
is i.i.d. across $l$. By the LLN, we obtain $|V_{ni,st}(X,p)-V_{st}(p)|=o_{p}(1)$
and hence $\|V_{ni}(X,p)-V^{*}(X_{i},p)\|=o_{p}(1)$.
\end{proof}

\end{document}